\documentclass[showpacs,preprintnumbers,amsmath,amssymb]{revtex4}

\newcommand{\bea}{\begin{eqnarray}}
\newcommand{\eea}{\end{eqnarray}}

\begin{document}
%%%%%%%%%%%%%%%%%%%%%%%%%%%%%%%%%%%%%%%%%%%%%%%%%%%%%%%%%%%%%%%

%%%%%%%%%%%%%%%%%%%%%%%%%%%%%%%%%%%%%%%%%%%%%%%%%%%%%%%%%%%%%%%
\title{Second-order perturbations of cosmological fluids: \\
       Relativistic effects of pressure, multi-component,
       curvature, and rotation}
\author{Jai-chan Hwang}
 \email{jchan@knu.ac.kr}
 \affiliation{Department of Astronomy and Atmospheric Sciences,
              Kyungpook National University, Taegu, Korea}
\author{Hyerim Noh}
 \email{hr@kasi.re.kr}
 \affiliation{Korea Astronomy and Space Science Institute,
              Daejon, Korea}

%%%%%%%%%%%%%%%%%%%%%%%%%%%%%%%%%%%%%%%%%%%%%%%%%%%%%%%%%%%%%%%
\date{\today}

%%%%%%%%%%%%%%%%%%%%%%%%%%%%%%%%%%%%%%%%%%%%%%%%%%%%%%%%%%%%%%%
\begin{abstract}
We present general relativistic correction terms appearing in
Newton's gravity to the second-order perturbations of cosmological
fluids. In our previous work we have shown that to the second-order
perturbations, the density and velocity perturbation equations of
general relativistic zero-pressure, irrotational, single-component
fluid in a flat background {\it coincide exactly} with the ones
known in Newton's theory. We also have shown the effect of
gravitational waves to the second-order, and pure general
relativistic correction terms appearing in the third-order
perturbations. Here, we present results of second-order
perturbations relaxing all the assumptions made in our previous
works. We derive the general relativistic correction terms arising
due to (i) pressure, (ii) multi-component, (iii) background
curvature, and (iv) rotation. In case of multi-component
zero-pressure, irrotational fluids under the flat background, we
effectively do not have relativistic correction terms, thus the
relativistic result again {\it coincides} with the Newtonian ones.
In the other three cases we generally have pure general relativistic
correction terms. In case of pressure, the relativistic corrections
appear even in the level of background and linear perturbation
equations. In the presence of background curvature, or rotation,
pure relativistic correction terms directly appear in the Newtonian
equations of motion of density and velocity perturbations to the
second order. In the small-scale limit (far inside the horizon),
relativistic equations including the rotation {\it coincide} with
the ones in Newton's gravity. All equations in this work include the
cosmological constant in the background world model.  We also
present the case of multiple minimally coupled scalar fields, and
properly derive the large-scale conservation properties of curvature
perturbation variable in various temporal gauge conditions to the
second order.
\end{abstract}
%%%%%%%%%%%%%%%%%%%%%%%%%%%%%%%%%%%%%%%%%%%%%%%%%%%%%%%%%%%%%%%
\noindent \pacs{PACS numbers: 04.50.+h, 04.62.+v, 98.80.-k}

%%%%%%%%%%%%%%%%%%%%%%%%%%%%%%%%%%%%%%%%%%%%%%%%%%%%%%%%%%%%%%%
\maketitle

%%%%%%%%%%%%%%%%%%%%%%%%%%%%%%%%%%%%%%%%%%%%%%%%%%%%%%%%%%%%%%%
\tableofcontents

%%%%%%%%%%%%%%%%%%%%%%%%%%%%%%%%%%%%%%%%%%%%%%%%%%%%%%%%%%%%%%%
%
% Introduction
%
%%%%%%%%%%%%%%%%%%%%%%%%%%%%%%%%%%%%%%%%%%%%%%%%%%%%%%%%%%%%%%%
\section{Introduction}
                                          \label{sec:Introduction}

Large amount of cosmological data on the large-scale structures and
motions of galaxies \cite{2dF,SDSS}, and the temperature and
polarization anisotropies of cosmic microwave background radiation
\cite{COBE,WMAP} have been accumulating recently. In current
standard cosmological scenario such structures are explained as
small (linear) or large (nonlinear) deviations from spatially
homogeneous and isotropic Friedmann background world model. In order
to explain these data theoretically, researchers rely on the linear
perturbation theory based on relativistic gravity, and quasi-linear
perturbation theories and nonlinear simulations based on Newton's
gravity. To the linear order in perturbation the general
relativistic result was first derived by Lifshitz in 1946
\cite{Lifshitz-1946}, and later shown to coincide with the Newtonian
result in a zero-pressure medium \cite{Bonnor-1957}. The same is
also known to be true for the background world model. That is, the
general relativistic result was first derived by Friedmann in 1922
\cite{Friedmann-1922}, and later shown to coincide with Newtonian
result in a zero-pressure medium \cite{Milne-1934}.

The observed large-scale distribution of galaxies shows that in the
largest observed scale (say, larger than several hundred mega-parsec
scale) the distribution may not be inconsistent with the linear
assumption around the Friedmann background. However, as the scale
becomes smaller the distribution apparently shows quasi-linear to
fully nonlinear structures. The fully nonlinear processes occur in
small scale where the relativistic effects characterized by $GM/rc^2
\sim v^2/c^2$ are quite small. If we could ignore such relativistic
effects, Newton's gravity would be sufficient to handle the relevant
nonlinear processes. If we need to consider the weakly relativistic
correction terms in fully nonlinear stage, instead of the
relativistic perturbation approach which can handle the fully
relativistic processes under weakly nonlinear assumption, we can use
the post-Newtonian approximation developed in the context of
cosmology in \cite{PN-2005}.

For structures in the quasi-linear evolution phase, previous
researches were based on Newton's gravity especially assuming the
single component zero-pressure fluid without rotational perturbation
\cite{quasilinear}. In our previous works in \cite{NL,second-order}
we have shown that, in the single component zero-pressure fluid
without rotational perturbation, cosmological scalar-type
perturbation equations in a spatially flat background coincide
exactly with the Newtonian ones up to the second order in
perturbation. In \cite{NL,second-order} we also have shown the
contribution of gravitational wave perturbations to the hydrodynamic
parts in the second-order perturbations. In Newton's gravity the
hydrodynamic equations of zero-pressure fluid contain only the
quadratic order nonlinearity. In \cite{third-order-PRD} we presented
pure general relativistic correction terms appearing in the
third-order perturbation, and showed that all third-order correction
terms are $10^{-5}$ times smaller than the second-order
relativistic/Newtonian terms, and independent of the horizon scale.

In this work we will take into account of the pure general
relativistic effects appearing in the second-order perturbations
which were ignored in our previous work in \cite{second-order}. We
will consider general relativistic effects of (i) pressure, (ii)
multi-component, (iii) background curvature, (iv) rotation in
cosmological fluids to the second-order perturbations. As results we
will show that, although in \cite{second-order} we have shown the
exact relativistic/Newtonian correspondence to the second-order
perturbations by ignoring the above four conditions and the
gravitational waves, as we take these four effects into account we
often encounter pure general relativistic effects appearing in the
corresponding Newtonian equations even to the second order in
perturbations. Our results will show that the relativistic/Newtonian
correspondence continues even in the multi-component situation
assuming zero-pressure irrotational fluid in a flat background, but
in the presence of the cosmological constant. This is a practically
useful result because the matter content of present universe is
dominated by collisionless dark matter and baryon both of which
practically have zero-pressure. In the other three cases, relaxing
any of the assumptions about pressure, rotation, and background
curvature generally leads to pure general relativistic correction
terms to the second order. We will present such correction terms in
the context of Newtonian hydrodynamics. One additional
relativistic/Newtonian correspondence occurs in the case of rotation
in small-scale (sub-horizon-scale) limit which is another
practically important result. This correspondence allows us to use
the Newtonian equations safely in such a small-scale limit even in
the presence of rotational perturbation to the second order.

In Sec.\ \ref{sec:Newtonian} we summarize Newtonian hydrodynamic
perturbation equations valid to fully nonlinear order. In Sec.\
\ref{sec:previous_work} we briefly summarize our previous result of
relativistic/Newtonian correspondence to the second order, and pure
general relativistic correction terms appearing in the third order.
In Sec.\ \ref{sec:NL-eqs} we present parts of the covariant and the
ADM (Arnowitt-Deser-Misner) equations which are valid in
multi-component situation. In Sec.\ \ref{sec:pert-eqs} we present
the basic perturbation equations valid to second order. In \cite{NL}
the basic set of equations was presented using fluid quantities
based on the normal-frame four-vector. The fluid quantities in the
present work are based on the energy-frame four-vector, and in this
section we present the basic equations using such fluid quantities.
In Secs.\ \ref{sec:pressure}-\ref{sec:rotation} we analyse the
effects of the pressure, the multi-components, the background
curvature, and the rotational perturbation, respectively. In Sec.\
\ref{sec:fields} we present equations in the scalar fields and
generalized gravity theories using the energy-frame fluid
quantities. In Sec.\ \ref{sec:conservations} we properly derive
conservation properties of curvature perturbation in various
temporal gauge (hypersurface) conditions to the second order in
perturbations. Section \ref{sec:Discussion} is a discussion. In this
work we follow notations used in \cite{NL,second-order}. We set $c
\equiv 1$, but when we compare with Newtonian case we often recover
the speed of light $c$.

%%%%%%%%%%%%%%%%%%%%%%%%%%%%%%%%%%%%%%%%%%%%%%%%%%%%%%%%%%%%%%%
%
% Newtonian perturbations
%
%%%%%%%%%%%%%%%%%%%%%%%%%%%%%%%%%%%%%%%%%%%%%%%%%%%%%%%%%%%%%%
\section{Newtonian nonlinear perturbations}
                                            \label{sec:Newtonian}

In order to compare properly the relativistic results with the
Newtonian ones, in this section we summarize the Newtonian
cosmological perturbation theory in fully nonlinear context. We
consider multi-component fluids in the presence of isotropic
pressure. In case of $n$-fluids with the mass densities $\varrho_i$,
the pressures $p_i$, the velocities ${\bf v}_i$ ($i = 1,2, \dots
n$), and the gravitational potential $\Phi$, we have \bea
   & & \dot \varrho_i + \nabla \cdot ( \varrho_i {\bf v}_i ) = 0,
   \label{mass-conservation-i} \\
   & & \dot {\bf v}_i + {\bf v}_i \cdot \nabla {\bf v}_i
       = - {1 \over \varrho_i} \nabla p_i - \nabla \Phi,
   \label{Euler-eq-i} \\
   & & \nabla^2 \Phi = 4 \pi G \sum_{j = 1}^n \varrho_j.
   \label{Poisson-eq-i-N}
\eea {\it Assuming} the presence of spatially homogeneous and
isotropic but temporally dynamic background, we introduce fully
nonlinear perturbations as \bea
   & & \varrho_i = \bar \varrho_i + \delta \varrho_i, \quad
       p_i = \bar p_i + \delta p_i, \quad
       {\bf v}_i = H {\bf r} + {\bf u}_i, \quad
       \Phi = \bar \Phi + \delta \Phi,
\eea where $H \equiv \dot a/a$, and $a(t)$ is a cosmic scale factor.
We move to the comoving coordinate ${\bf x}$ where \bea
   & & {\bf r} \equiv a (t) {\bf x},
\eea thus \bea
   & & \nabla = \nabla_{\bf r} = {1 \over a} \nabla_{\bf x},
   \nonumber \\
   & & {\partial \over \partial t}
       = {\partial \over \partial t} \Big|_{\bf r}
       = {\partial \over \partial t} \Big|_{\bf x}
       + \left( {\partial \over \partial t} \Big|_{\bf r} {\bf x} \right)
       \cdot \nabla_{\bf x}
       = {\partial \over \partial t} \Big|_{\bf x}
       - H {\bf x} \cdot \nabla_{\bf x}.
\eea In the following we neglect the subindex ${\bf x}$. To the
background order we have \bea
   & & \dot \varrho_i + 3 H \varrho_i = 0, \quad
       {\ddot a \over a} = - {4 \pi G \over 3} \sum_j \varrho_j,
       \quad
       H^2 = {8 \pi G \over 3} \sum_j \varrho_j + {2 E \over a^2},
   \label{BG-N-eqs}
\eea where $E$ is an integration constant which can be interpreted
as the specific total energy in Newton's gravity; in Einstein's
gravity we have $2 E = - Kc^2$ where $K$ can be normalized to be the
sign of spatial curvature. To the perturbed order we have
\cite{Peebles-1980} \bea
   & & \dot \delta_i + {1 \over a} \nabla \cdot {\bf u}_i
       = - {1 \over a} \nabla \cdot \left( \delta_i {\bf u}_i \right),
   \label{dot-delta-eq-N-i} \\
   & & \dot {\bf u}_i
       + H {\bf u}_i
       + {1 \over a} {\bf u}_i \cdot \nabla {\bf u}_i
       = - {1 \over a \bar \varrho_i} {\nabla \delta p_i \over 1 +
       \delta_i}
       - {1 \over a} \nabla \delta \Phi,
   \label{dot-v-eq-N-i} \\
   & & {1 \over a^2} \nabla^2 \delta \Phi
       = 4 \pi G \sum_{j} \bar \varrho_j \delta_j.
\eea By introducing the expansion $\theta_i$ and the rotation
$\overrightarrow \omega_i$ of each component as \bea
   & & \theta_i \equiv {1 \over a} \nabla \cdot {\bf u}_i, \quad
       \overrightarrow \omega_i
       \equiv {1 \over a} \nabla \times {\bf u}_i,
   \label{theta-omega-N}
\eea Eq.\ (\ref{dot-v-eq-N-i}) gives \bea
   & & \dot \theta_i + 2 H \theta_i
       + 4 \pi G \sum_j \bar \varrho_j \delta_j
       =
       - {1 \over a^2} \nabla \cdot \left( {\bf u}_i \cdot
       \nabla {\bf u}_i \right)
       - {1 \over a^2 \bar \varrho_i} \nabla \cdot
       \left( {\nabla \delta p_i \over 1 + \delta_i} \right),
   \label{dot-velocity-eq-N-i} \\
   & & \dot {\overrightarrow \omega}_i + 2 H \overrightarrow \omega_i
       =
       - {1 \over a^2} \nabla \times \left( {\bf u}_i \cdot
       \nabla {\bf u}_i \right)
       + {1 \over a^2 \bar \varrho_i}
       { (\nabla \delta_i) \times \nabla \delta p_i
       \over (1 + \delta_i)^2}.
   \label{dot-rotation-eq-N-i}
\eea By introducing decomposition of perturbed velocity into the
potential- and transverse parts as  \bea
   & & {\bf u}_i \equiv \nabla u_i + {\bf u}^{(v)}_i, \quad
       \nabla \cdot {\bf u}^{(v)}_i \equiv 0; \quad
       \theta_i = {\Delta \over a} u_i, \quad
       \overrightarrow \omega_i = {1 \over a} \nabla \times {\bf
       u}^{(v)}_i,
\eea instead of Eq.\ (\ref{dot-rotation-eq-N-i}) we have \bea
   & & \dot {\bf u}^{(v)}_i
       + H {\bf u}^{(v)}_i
       = - {1 \over a} \left[ {\bf u}_i \cdot \nabla {\bf u}_i
       + {1 \over \bar \varrho_i} {\nabla \delta p_i \over 1 +
       \delta_i}
       - \nabla \Delta^{-1} \nabla \cdot
       \left( {\bf u}_i \cdot \nabla {\bf u}_i
       + {1 \over \bar \varrho_i} {\nabla \delta p_i \over 1 +
       \delta_i}
       \right) \right].
   \label{dot-rotation-eq-N-i2}
\eea We note that the pure $u_i$ contributions in the
right-hand-side of Eq.\ (\ref{dot-rotation-eq-N-i}) or Eq.\
(\ref{dot-rotation-eq-N-i2}) vanish. Thus, under vanishing pressure,
pure irrotational perturbation {\it cannot} generate the rotational
perturbation. Equation (\ref{dot-rotation-eq-N-i}) shows that
presence of pressure perturbation oblique (i.e., non-parallel) to
the density perturbation can generate rotational perturbation.
Combining Eqs.\ (\ref{dot-delta-eq-N-i}),(\ref{dot-velocity-eq-N-i})
we can derive \bea
   & & \ddot \delta_i + 2 H \dot \delta_i
       - 4 \pi G \sum_j \bar \varrho_j \delta_j
       =
       - {1 \over a^2} \left[ a \nabla \cdot \left( \delta_i {\bf u}_i
       \right) \right]^\cdot
       + {1 \over a^2} \nabla \cdot \left( {\bf u}_i \cdot
       \nabla {\bf u}_i \right)
       + {1 \over a^2 \bar \varrho_i} \nabla \cdot
       \left( {\nabla \delta p_i \over 1 + \delta_i} \right).
   \label{ddot-delta-eq-N-i}
\eea Equations (\ref{dot-delta-eq-N-i})-(\ref{ddot-delta-eq-N-i})
are valid to fully nonlinear order. Notice that for vanishing
pressure these equations have only quadratic order nonlinearity in
perturbations.

%%%%%%%%%%%%%%%%%%%%%%%%%%%%%%%%%%%%%%%%%%%%%%%%%%%%%%%%%%%%%%%
%
% Previous work
%
%%%%%%%%%%%%%%%%%%%%%%%%%%%%%%%%%%%%%%%%%%%%%%%%%%%%%%%%%%%%%%%
\section{Summary of previous work}
                                            \label{sec:previous_work}

In \cite{NL,second-order} we have derived second-order perturbation
equations valid for the single component, irrotational, and
zero-pressure medium in zero-curvature background. These are \bea
   & & \dot \delta
       + {1 \over a} \nabla \cdot {\bf u}
       = - {1 \over a} \nabla \cdot \left( \delta {\bf u} \right),
   \label{E-conserv-eq5} \\
   & & {1 \over a} \nabla \cdot \left( \dot {\bf u} + H {\bf u}
       \right)
       + 4 \pi G \varrho \delta
       = - {1 \over a^2} \nabla \cdot \left( {\bf u} \cdot \nabla
       {\bf u} \right)
       - \dot C^{(t)\alpha\beta}
       \left( {2 \over a} u_{\alpha|\beta}
       + \dot C^{(t)}_{\alpha\beta} \right),
   \label{Raychaudhury-eq5} \\
   & & \ddot \delta
       + 2 H \dot \delta
       - 4 \pi G \varrho \delta
       = {1 \over a^2}
       \nabla \cdot \left( {\bf u} \cdot \nabla {\bf u} \right)
       - {1 \over a^2} \left[ a
       \nabla \cdot \left( \delta {\bf u} \right) \right]^\cdot
       + \dot C^{(t)\alpha\beta}
       \left( {2 \over a} u_{\alpha|\beta}
       + \dot C^{(t)}_{\alpha\beta} \right).
   \label{ddot-delta-eq5}
\eea Except for the presence of tensor-type perturbation, Eqs.\
(\ref{E-conserv-eq5})-(\ref{ddot-delta-eq5}) are exactly the same as
the ones known in the Newtonian theory. We note that these equations
are valid in the presence of $\Lambda$. To the linear order, these
are valid in the presence of general $K$, see Sec.\
\ref{sec:curvature}. We have correctly identified the relativistic
density and velocity perturbation variables which correspond to the
Newtonian counterparts to the second order. In the relativistic
context, our $\delta$ and ${\bf u}$ are the density perturbation and
(related to) the perturbed expansion scalar, respectively, in the
comoving gauge; the variables are equivalently gauge-invariant.
However, we were not able to identify relativistic variable which
corresponds to the Newtonian gravitational potential to the second
order; this is understandable if we consider the factor of two
difference between Einstein's (post-Newtonian) and Newton's gravity
theories in predicting the light bending under the gravitational
field. To the linear order the spatial curvature perturbation in the
zero-shear gauge can be identified as the perturbed Newtonian
potential \cite{Harrison-1967,Bardeen-1980}. Equations
(\ref{E-conserv-eq5})-(\ref{ddot-delta-eq5}) include effects of
gravitational waves to the density and velocity perturbations.
Equations of the gravitational waves can be found in
\cite{second-order}.

To the third order, we have \cite{third-order-PRD} \bea
   & & \dot \delta + {1 \over a} \nabla \cdot {\bf u}
       = - {1 \over a} \nabla \cdot \left( \delta {\bf u} \right)
       + {1 \over a} \left[ 2 \varphi {\bf u}
       - \nabla \left( \Delta^{-1} X \right) \right] \cdot \nabla \delta,
   \label{delta-eq-3rd} \\
   & & {1 \over a} \nabla \cdot \left( \dot {\bf u}
       + {\dot a \over a} {\bf u} \right)
       + 4 \pi G \mu \delta
       = - {1 \over a^2} \nabla \cdot \left( {\bf u}
       \cdot \nabla {\bf u} \right)
   \nonumber \\
   & & \qquad
       - {2 \over 3 a^2} \varphi
       {\bf u} \cdot \nabla \left( \nabla \cdot {\bf u} \right)
       + {4 \over a^2} \nabla \cdot \left[ \varphi
       \left( {\bf u} \cdot \nabla {\bf u}
       - {1 \over 3} {\bf u} \nabla \cdot {\bf u} \right) \right]
       - {\Delta \over a^2}
       \left[ {\bf u} \cdot \nabla \left( \Delta^{-1} X \right) \right]
       + {1 \over a^2} {\bf u} \cdot \nabla X
       + {2 \over 3a^2} X \nabla \cdot {\bf u},
   \label{u-eq-3rd} \\
   & & \ddot \delta + 2 {\dot a \over a} \dot \delta
       - 4 \pi G \mu \delta
       = - {1 \over a^2} {\partial \over \partial t}
       \left[ a \nabla \cdot \left( \delta {\bf u} \right) \right]
       + {1 \over a^2} \nabla \cdot \left( {\bf u}
       \cdot \nabla {\bf u} \right)
       + {1 \over a^2} {\partial \over \partial t}
       \left\{ a \left[ 2 \varphi {\bf u}
       - \nabla \left( \Delta^{-1} X \right) \right] \cdot \nabla \delta
       \right\}
   \nonumber \\
   & & \qquad
       + {2 \over 3 a^2} \varphi
       {\bf u} \cdot \nabla \left( \nabla \cdot {\bf u} \right)
       - {4 \over a^2} \nabla \cdot \left[ \varphi
       \left( {\bf u} \cdot \nabla {\bf u}
       - {1 \over 3} {\bf u} \nabla \cdot {\bf u} \right) \right]
       + {\Delta \over a^2}
       \left[ {\bf u} \cdot \nabla \left( \Delta^{-1} X \right) \right]
       - {1 \over a^2} {\bf u} \cdot \nabla X
       - {2 \over 3a^2} X \nabla \cdot {\bf u},
   \label{density-eq-3rd}
\eea where \bea
   X
   &\equiv&
       2 \varphi \nabla \cdot {\bf u}
       - {\bf u} \cdot \nabla \varphi
       + {3 \over 2} \Delta^{-1} \nabla \cdot
       \left[ {\bf u} \cdot \nabla \left( \nabla \varphi \right)
       + {\bf u} \Delta \varphi \right].
   \label{X-eq-3rd}
\eea In these equations we ignored the role of tensor-type
perturbation; for a complete set of equations, see
\cite{third-order-PRD}. The variable $\varphi$ is a perturbed-order
metric (spatial curvature) variable in the comoving gauge condition,
see later.

All the third-order correction terms in Eqs.\
(\ref{delta-eq-3rd})-(\ref{density-eq-3rd}) are simply of
$\varphi$-order higher than the second-order relativistic/Newtonian
terms. Thus, the pure general relativistic effects are at least
$\varphi$-order higher than the relativistic/Newtonian ones in the
second order equations. Thus, we only need the behavior of $\varphi$
to the linear order which is related to the other hydrodynamic
variables as \bea
   & & \varphi = - \delta \Phi
       + \dot a \Delta^{-1} \nabla \cdot {\bf u}.
\eea It also satisfies \cite{Hwang-Noh-Newtonian} \bea
   & & \dot \varphi = 0,
\eea thus $\varphi = C ({\bf x})$ with {\it no} decaying mode; this
is true considering the presence of the cosmological constant, see
\cite{Hwang-Noh-Newtonian}.

%%%%%%%%%%%%%%%%%%%%%%%%%%%%%%%%%%%%%%%%%%%%%%%%%%%%%%%%%%%%%%%
%
% Fully nonlinear equations
%
%%%%%%%%%%%%%%%%%%%%%%%%%%%%%%%%%%%%%%%%%%%%%%%%%%%%%%%%%%%%%%%
\section{Relativistic fully nonlinear equations}
                                                 \label{sec:NL-eqs}

In this section, for convenience, we present some additional
covariant or ADM equations not available in \cite{NL}. In the
multi-component situation, we have \bea
   & & \tilde T_{ab} \equiv \sum_j \tilde T_{(j)ab}.
\eea The energy-momentum conservation gives \bea
   & & \tilde T_{(i)a;b}^{\;\;\;\;b}
       = \tilde I_{(i)a}, \quad
       \sum_j \tilde I_{(j)a} = 0.
   \label{EM-conservation}
\eea Tildes indicate the covariant quantities.

%%%%%%%%%%%%%%%%%%%%%%%%%%%%%%%%%%%%%%%%%%%%%%%%%%%%%%%%%%%%%%
\subsection{Covariant equations}

We introduce the fluid quantities as \bea
   & & \tilde T_{(i)ab}
       \equiv \tilde \mu_{(i)} \tilde u_{(i)a} \tilde u_{(i)b}
       + \tilde p_{(i)} \left( \tilde g_{ab}
       + \tilde u_{(i)a} \tilde u_{(i)b} \right)
       + \tilde q_{(i)a} \tilde u_{(i)b}
       + \tilde q_{(i)b} \tilde u_{(i)a}
       + \tilde \pi_{(i)ab},
   \label{Tab_i}
\eea where \bea
   & & \tilde u_{(i)}^a \tilde u_{(i)a} \equiv -1, \quad
       \tilde u_{(i)}^a \tilde q_{(i)a} \equiv 0
       \equiv \tilde u_{(i)}^b \tilde \pi_{(i)ab}, \quad
       \tilde \pi_{(i)a}^{\;\;\;\;a} \equiv 0.
\eea The fluid quantities of each component are based on the fluid
four-vector $\tilde u_{(i)a}$ as \bea
   & & \tilde \mu_{(i)} \equiv \tilde T_{(i)ab} \tilde u_{(i)}^a
       \tilde u_{(i)}^b, \quad
       \tilde p_{(i)} \equiv {1 \over 3} \tilde T_{(i)ab} \tilde
       h_{(i)}^{ab}, \quad
       \tilde q_{(i)a} \equiv - \tilde T_{(i)cd} \tilde u_{(i)}^c
       \tilde h_{(i)a}^{\;\;\;\; d}, \quad
       \tilde \pi_{(i)ab} \equiv \tilde T_{(i)cd} \tilde
       h_{(i)a}^{\;\;\;\; c} \tilde h_{(i)b}^{\;\;\;\; d}
       - \tilde p_{(i)} \tilde h_{(i)ab},
\eea where $\tilde h_{(i)ab} \equiv \tilde g_{ab} + \tilde u_{(i)a}
\tilde u_{(i)b}$. Equation (\ref{EM-conservation}) gives \bea
   & & \tilde {\dot {\tilde \mu}}_{(i)}
       + \left( \tilde \mu_{(i)} + \tilde p_{(i)} \right)
       \tilde \theta_{(i)}
       + \tilde q^a_{(i);a}
       + \tilde q_{(i)}^a \tilde a_{(i)a}
       + \tilde \pi^{ab}_{(i)} \tilde \sigma_{(i)ab}
       = - \tilde u^a_{(i)} \tilde I_{(i)a},
   \label{E-conserv-cov-i} \\
   & & \left( \tilde \mu_{(i)} + \tilde p_{(i)} \right)
       \tilde a_{(i)a}
       + \tilde h_{(i)a}^{\;\;\;\;b}
       \left( \tilde p_{(i),b}
       + \tilde {\dot {\tilde q}}_{(i)b}
       + \tilde \pi_{(i)b;c}^{\;\;\;\;c} \right)
       + \tilde q_{(i)}^b
       \left( \tilde \omega_{(i)ab}
       + \tilde \sigma_{(i)ab}
       + {4 \over 3} \tilde \theta_{(i)} \tilde h_{(i)ab} \right)
       = \tilde h_{(i)a}^{\;\;\;\;b} \tilde I_{(i)b},
   \label{Mom-conserv-cov-i}
\eea where the kinematic quantities are also based on the fluid
four-vector $\tilde u_{(i)a}$ as \bea
   & & \tilde h_{(i)a}^{\;\;\;\;c} \tilde h_{(i)b}^{\;\;\;\;d}
       \tilde u_{(i)c;d}
       = \tilde h_{(i)[a}^{\;\;\;\;c} \tilde h_{(i)b]}^{\;\;\;\;d}
       \tilde u_{(i)c;d}
       + \tilde h_{(i)(a}^{\;\;\;\;c} \tilde h_{(i)b)}^{\;\;\;\;d}
       \tilde u_{(i)c;d}
       \equiv \tilde \omega_{(i)ab} + \tilde \theta_{(i)ab}
       = \tilde u_{(i)a;b} + \tilde a_{(i)a} \tilde u_{(i)b},
   \nonumber \\
   & &
       \tilde \theta_{(i)} \equiv u_{(i);a}^a, \quad
       \tilde \sigma_{(i)ab} \equiv \tilde \theta_{(i)ab}
       - {1 \over 3} \tilde \theta_{(i)} \tilde h_{(i)ab}, \quad
       \tilde a_{(i)a} \equiv \tilde u_{(i)a;b} \tilde u_{(i)}^b
       \equiv \tilde {\dot {\tilde u}}_{(i)a}, \quad
       \tilde {\dot {\tilde \mu}}_{(i)} \equiv
       \tilde \mu_{(i),a} \tilde u^a_{(i)}.
   \label{kinematic_quantities-i}
\eea In the multi-component situation, we can derive the
corresponding equation of Raychaudhury equation for the individual
component. From $\tilde u_{(i)a;bc} - \tilde u_{(i)a;cb} \equiv
\tilde u_{(i)d} \tilde R^d_{\;\; abc}$ we can derive \bea
   & & \tilde {\dot {\tilde \theta}}_{(i)}
       + {1 \over 3} \tilde \theta^2_{(i)}
       - \tilde a^a_{(i);a}
       + \tilde \sigma^{ab}_{(i)} \tilde \sigma_{(i)ab}
       - \tilde \omega^{ab}_{(i)} \tilde \omega_{(i)ab}
       = 4 \pi G \left( \tilde \mu - 3 \tilde p
       - 2 \tilde T_{ab} \tilde u^a_{(i)} \tilde u^b_{(i)} \right)
       + \Lambda.
   \label{Raychaudhury-eq-cov-i}
\eea In the energy-frame we take $\tilde q_{(i)a} \equiv 0$ for each
component of the fluids without losing any physical degree of
freedom.

In a single component situation, taking the energy-frame, the energy
conservation equation, the momentum conservation equation, and the
Raychaudhury equation are \cite{covariant} \bea
   & & \tilde {\dot {\tilde \mu}}
       + \left( \tilde \mu + \tilde p \right) \tilde \theta
       + \tilde \pi^{ab} \tilde \sigma_{ab}
       = 0,
   \label{E-conserv-cov} \\
   & & \left( \tilde \mu + \tilde p \right) \tilde a_a
       + \tilde h^b_a \left( \tilde p_{,b}
       + \tilde \pi^c_{b;c} \right)
       = 0,
   \label{Mom-conserv-cov} \\
   & & \tilde {\dot {\tilde \theta}}
       + {1 \over 3} \tilde \theta^2
       - \tilde a^a_{\;\; ;a}
       + \tilde \sigma^{ab} \tilde \sigma_{ab}
       - \tilde \omega^{ab} \tilde \omega_{ab}
       + 4 \pi G \left( \tilde \mu + 3 \tilde p \right)
       - \Lambda
       = 0.
   \label{Raychaudhury-eq-cov}
\eea By combining Eqs.\
(\ref{E-conserv-cov})-(\ref{Raychaudhury-eq-cov}) we can derive \bea
   & & \left( { \tilde {\dot {\tilde \mu}}
       + \tilde \pi^{ab} \tilde \sigma_{ab}
       \over \tilde \mu + \tilde p }
       \right)^{\tilde \cdot}
       - {1 \over 3} \left( { \tilde {\dot {\tilde \mu}}
       + \tilde \pi^{ab} \tilde \sigma_{ab}
       \over \tilde \mu + \tilde p } \right)^2
       = 4 \pi G \left( \tilde \mu + 3 \tilde p \right)
       - \Lambda
       + \tilde \sigma^{ab} \tilde \sigma_{ab}
       - \tilde \omega^{ab} \tilde \omega_{ab}
       + \left[ { \tilde h^{ab} ( \tilde p_{,b}
       + \tilde \pi^c_{b;c} ) \over
       \tilde \mu + \tilde p } \right]_{;a}.
   \label{covariant-density-eq}
\eea This equation was derived in Eq.\ (88) of \cite{HV-1990}, see
also \cite{Jackson-1972}. In the multi-component case, in the
energy-frame, combining Eqs.\
(\ref{E-conserv-cov-i})-(\ref{Raychaudhury-eq-cov-i}) we can derive
\bea
   & & \left( { \tilde {\dot {\tilde \mu}}_{(i)}
       + \tilde \pi^{ab}_{(i)} \tilde \sigma_{(i)ab}
       + \tilde u^a_{(i)} \tilde I_{(i)a}
       \over \tilde \mu_{(i)} + \tilde p_{(i)} }
       \right)^{\tilde \cdot}
       - {1 \over 3} \left( { \tilde {\dot {\tilde \mu}}_{(i)}
       + \tilde \pi^{ab}_{(i)} \tilde \sigma_{(i)ab}
       + \tilde u^a_{(i)} \tilde I_{(i)a}
       \over
       \tilde \mu_{(i)} + \tilde p_{(i)} } \right)^2
   \nonumber \\
   & & \qquad
       = - 4 \pi G \left( \tilde \mu - 3 \tilde p
       - 2 \tilde T_{ab} \tilde u^a_{(i)} \tilde u^b_{(i)} \right)
       - \Lambda
       + \tilde \sigma^{ab}_{(i)} \tilde \sigma_{(i)ab}
       - \tilde \omega^{ab}_{(i)} \tilde \omega_{(i)ab}
       + \left[ { \tilde h^{ab}_{(i)} ( \tilde p_{(i),b}
       + \tilde \pi^{\;\;\;\;c}_{(i)b;c} - \tilde I_{(i)b} ) \over
       \tilde \mu_{(i)} + \tilde p_{(i)} } \right]_{;a}.
   \label{covariant-density-eq-i}
\eea

In \cite{Langlois-Vernizzi} Langlois and Vernizzi derived a simple
covariant relation which leads to one of the conserved variable in
the large-scale limit. These authors introduced \bea
   & & \tilde \zeta_a \equiv \tilde h_a^b \left[
       \tilde \alpha_{,b}
       + {\tilde \mu_{,b} \over 3 (\tilde \mu + \tilde p)} \right], \quad
       \tilde {\dot {\tilde \alpha}}
       \equiv \tilde \alpha_{,a} \tilde u^a
       \equiv {1 \over 3} \tilde \theta.
\eea Using only the energy conservation in Eq.\
(\ref{E-conserv-cov}) we can derive \bea
   & & {\pounds_{\tilde u}} \tilde \zeta_a
       = - {\tilde \theta \over 3 (\tilde \mu + \tilde p)}
       \left[ \tilde p_{,a}
       + {\tilde {\dot {\tilde p}} \over (\tilde \mu + \tilde p)
       \tilde \theta} \tilde \mu_{,a} \right]
       - \left[ {\tilde u_a \tilde \pi^{bc} \tilde \sigma_{bc} \over
       3 (\tilde \mu + \tilde p)} \right]^{\tilde \cdot}
       - {\left( \tilde \pi^{bc} \tilde \sigma_{bc} \right)_{,a} \over
       3 (\tilde \mu + \tilde p)}
       + {\tilde \mu_{,a} \tilde \pi^{bc} \tilde \sigma_{bc} \over
       3 (\tilde \mu + \tilde p)^2},
   \label{LV-relation1}
\eea where ${\pounds_{\tilde u}}$ is a Lie derivative along $\tilde
u_a$ with ${\pounds_{\tilde u}} \tilde \zeta_a \equiv \tilde
\zeta_{a;b} \tilde u^b + \tilde \zeta^b \tilde u_{b;a}$. Thus, for
vanishing anisotropic pressure, we have the Langlois-Vernizzi
relation \cite{Langlois-Vernizzi} \bea
   & & {\pounds_{\tilde u}} \tilde \zeta_a
       = - {\tilde \theta \over 3 (\tilde \mu + \tilde p)}
       \tilde h_a^b \left( \tilde p_{,b}
       - {\tilde {\dot {\tilde p}} \over \tilde {\dot {\tilde \mu}}}
       \tilde \mu_{,b} \right).
\eea These equations are valid in a single component fluid, or in
multiple component fluids for the collective fluid variables. We can
easily extend the relation to the individual fluid component as
follows. We introduce \bea
   & & \tilde \zeta_{(i)a} \equiv \tilde h_{(i)a}^{\;\;\;\;b} \left[
       \tilde \alpha_{(i),b}
       + {\tilde \mu_{(i),b} \over 3 (\tilde \mu_{(i)} + \tilde p_{(i)})} \right], \quad
       \tilde {\dot {\tilde \alpha}}_{(i)}
       \equiv \tilde \alpha_{(i),a} \tilde u_{(i)}^a
       \equiv { 1\over 3} \tilde \theta_{(i)}.
\eea Using only Eq.\ (\ref{E-conserv-cov-i}) we can derive \bea
   & & {\pounds_{\tilde u_{(i)}}} \tilde \zeta_{(i)a}
       = - {\tilde \theta_{(i)} \over 3 (\tilde \mu_{(i)} + \tilde p_{(i)})}
       \left[ \tilde p_{(i),a}
       + {\tilde {\dot {\tilde p}}_{(i)} \over (\tilde \mu_{(i)} + \tilde p_{(i)})
       \tilde \theta_{(i)}} \tilde \mu_{(i),a} \right]
   \nonumber \\
   & & \qquad
       - \left[ {\tilde u_{(i)a} \left( \tilde \pi_{(i)}^{bc} \tilde \sigma_{(i)bc}
       + \tilde u_{(i)}^b \tilde I_{(i)b} \right) \over
       3 (\tilde \mu_{(i)} + \tilde p_{(i)})} \right]^{\tilde \cdot}
       - {\left( \tilde \pi_{(i)}^{bc} \tilde \sigma_{(i)bc}
       + \tilde u_{(i)}^b \tilde I_{(i)b} \right)_{,a} \over
       3 (\tilde \mu_{(i)} + \tilde p_{(i)})}
       + {\tilde \mu_{(i),a} \left( \tilde \pi_{(i)}^{bc} \tilde \sigma_{(i)bc}
       + \tilde u_{(i)}^b \tilde I_{(i)b} \right) \over
       3 (\tilde \mu_{(i)} + \tilde p_{(i)})^2}.
\eea Thus, for vanishing anisotropic pressure and direct
interactions among fluids, i.e., $\tilde \pi_{(i)ab} = 0 = \tilde
I_{(i)a}$, we have \bea
   & & {\pounds_{\tilde u_{(i)}}} \tilde \zeta_{(i)a}
       = - {\tilde \theta_{(i)} \over 3 (\tilde \mu + \tilde p)}
       \tilde h_{(i)a}^{\;\;\;\;b} \left( \tilde p_{(i),b}
       - {\tilde {\dot {\tilde p}}_{(i)} \over \tilde {\dot {\tilde \mu}}_{(i)}}
       \tilde \mu_{(i),b} \right).
\eea Application of these compact relations to large-scale
conservation properties to the second order will be studied in Sec.\
\ref{sec:conservations}.

%%%%%%%%%%%%%%%%%%%%%%%%%%%%%%%%%%%%%%%%%%%%%%%%%%%%%%%%%%%%%%
\subsection{ADM equations}

The ADM formulation \cite{ADM} is presented in Eqs.\
(2)-(13),(47),(48) of \cite{NL}. Interpretation of the ADM fluid
quantities in Eqs.\ (45),(46) of \cite{NL} was based on the
normal-frame fluid quantities; for relations to the energy-frame
fluid quantities, see Eq.\ (\ref{ADM-fluid-quantities}) below. The
ADM fluid quantities of individual component are introduced as \bea
   & & E_{(i)} \equiv \tilde n_a \tilde n_b \tilde T^{ab}_{(i)}
       = N^2 \tilde T_{(i)}^{00}, \quad
       J_{(i)\alpha} \equiv - \tilde n_b \tilde
       T_{(i)\alpha}^{\;\;\;\;b}
       = N \tilde T_{(i)\alpha}^{\;\;\;\;0},
   \nonumber \\
   & &
       S_{(i)\alpha\beta} \equiv \tilde T_{(i)\alpha\beta}, \quad
       S_{(i)} \equiv h^{\alpha\beta} S_{(i)\alpha\beta}, \quad
       \bar S_{(i)\alpha\beta} \equiv S_{(i)\alpha\beta}
       -{1 \over 3} h_{\alpha\beta} S_{(i)}.
   \label{ADM-fluid}
\eea  Equation (\ref{EM-conservation}) gives Eqs.\
(12),(13),(47),(48) in \cite{NL}. Equations (10),(12),(47) in
\cite{NL} can be arranged as \bea
   & & {1 \over N} \left( \partial_0
       - N^\alpha \partial_\alpha \right) K^\alpha_\alpha
       - {1 \over 3} \left( K^\alpha_\alpha \right)^2
       = 4 \pi G \left( E + S \right) - \Lambda
       + \bar K^{\alpha\beta} \bar K_{\alpha\beta}
       - {1 \over N} N^{:\alpha}_{\;\;\;\alpha},
   \label{ADM-1} \\
   & & K^\alpha_\alpha = \left( E + {1 \over 3} S \right)^{-1}
       \left[ {1 \over N} \left( \partial_0
       - N^\alpha \partial_\alpha \right) E
       + {1 \over N^2} \left( N^2 J^\alpha \right)_{:\alpha}
       - \bar S^{\alpha\beta} \bar K_{\alpha\beta} \right],
   \label{ADM-2} \\
   & & K^\alpha_\alpha = \left( E_{(i)} + {1 \over 3} S_{(i)} \right)^{-1}
       \left[ {1 \over N} \left( \partial_0
       - N^\alpha \partial_\alpha \right) E_{(i)}
       + {1 \over N^2} \left( N^2 J^\alpha_{(i)} \right)_{:\alpha}
       - \bar S_{(i)}^{\alpha\beta} \bar K_{\alpha\beta}
       + {1 \over N} \left( \tilde I_{(i)0}
       - \tilde I_{(i)\alpha} N^\alpha \right) \right].
   \label{ADM-3}
\eea Momentum conservation equations for the collective and
individual components can be found in Eqs.\ (13),(48) of \cite{NL}.
By combining Eqs.\ (\ref{ADM-1})-(\ref{ADM-3}) we can derive the ADM
counterpart of the density perturbation equations in Eqs.\
(\ref{covariant-density-eq}),(\ref{covariant-density-eq-i}).

%%%%%%%%%%%%%%%%%%%%%%%%%%%%%%%%%%%%%%%%%%%%%%%%%%%%%%%%%%%%%%%
%
% Second-order perturbations
%
%%%%%%%%%%%%%%%%%%%%%%%%%%%%%%%%%%%%%%%%%%%%%%%%%%%%%%%%%%%%%%
\section{Second-order perturbations}
                                           \label{sec:pert-eqs}

We use a metric convention in Eq.\ (49) of \cite{NL} \bea
   & & \tilde g_{00}
       \equiv - a^2 \left( 1 + 2 A  \right), \quad
       \tilde g_{0\alpha} \equiv - a^2 B_\alpha, \quad
       \tilde g_{\alpha\beta}
       \equiv a^2 \left( g^{(3)}_{\alpha\beta} + 2 C_{\alpha\beta}
       \right).
\eea The subindex $0$ indicates the conformal time $\eta$ with $a d
\eta \equiv c dt$. To the second-order in perturbation we introduce
the fluid four-vector of individual component as \bea
   & & \tilde u_{(i)}^\alpha \equiv {1 \over a} V_{(i)}^\alpha,
       \quad
       \tilde u_{(i)}^0
       = {1 \over a} \left( 1 - A + {3 \over 2} A^2
       + {1 \over 2} V_{(i)}^\alpha V_{(i)\alpha}
       - B^\alpha V_{(i)\alpha} \right);
   \nonumber \\
   & & \tilde u_{(i)\alpha}
       = a \left( V_{(i)\alpha}
       - B_\alpha
       + A B_\alpha
       + 2 V_{(i)}^\beta C_{\alpha\beta} \right), \quad
       \tilde u_{(i)0}
       = - a \left( 1 + A
       - {1 \over 2} A^2
       + {1 \over 2} V_{(i)}^\alpha V_{(i)\alpha} \right).
   \label{u-second-order}
\eea In this definition of fluid four-vector we follow the notation
in Eq.\ (53) of \cite{NL}. If we introduce $\tilde u_{(i)}^\alpha
\equiv \bar V_{(i)}^\alpha \tilde u_{(i)}^0$, we have
$V_{(i)}^\alpha = ( 1 - A ) \bar V_{(i)}^\alpha$.  The fluid
quantities of individual component are introduced as \bea
   & & \tilde \mu_{(i)} \equiv \mu_{(i)} + \delta \mu_{(i)}, \quad
       \tilde p_{(i)} \equiv p_{(i)} + \delta p_{(i)}, \quad
       \tilde \pi_{(i)\alpha\beta} \equiv a^2 \Pi_{(i)\alpha\beta},
       \quad
       \tilde \pi_{(i)0\alpha} = - a^2 \Pi_{(i)\alpha\beta}
       V_{(i)}^\beta, \quad
       \tilde \pi_{(i)00} = 0,
\eea where from $\tilde \pi^{\;\;\;\; a}_{(i)a} \equiv 0$ we have
\bea
   & & \Pi_{(i)\alpha}^{\;\;\;\;\alpha}
       - 2 C^{\alpha\beta} \Pi_{(i)\alpha\beta} = 0.
\eea The energy-momentum tensor of individual component in the
energy-frame follows from Eq.\ (\ref{Tab_i}) as \bea
   & & \tilde T_{(i)0}^{\;\;\;\; 0}
       = - \mu_{(i)} - \delta \mu_{(i)}
       - \left( \mu_{(i)} + p_{(i)} \right)
       \left( V_{(i)}^\alpha - B^\alpha \right) V_{(i)\alpha},
   \nonumber \\
   & & \tilde T_{(i) \alpha}^{\;\;\;\; 0}
       = \left( \mu_{(i)} + p_{(i)} \right)
       \left( V_{(i)\alpha} - B_\alpha
       - A V_{(i)\alpha} + 2 A B_\alpha
       + 2 V_{(i)}^\beta C_{\alpha\beta} \right)
       + \left( \delta \mu_{(i)} + \delta p_{(i)} \right)
       \left( V_{(i)\alpha} - B_\alpha \right)
   \nonumber \\
   & & \qquad
       + \Pi_{(i)\alpha\beta} \left( V_{(i)}^\beta - B^\beta
       \right),
   \nonumber \\
   & & \tilde T_{(i) \beta}^{\;\;\;\; \alpha}
       = \left( p_{(i)} + \delta p_{(i)} \right) \delta^\alpha_\beta
       + \left( \mu_{(i)} + p_{(i)} \right) V^\alpha_{(i)}
       \left( V_{(i)\beta} - B_\beta \right)
       + \Pi_{(i)\beta}^{\;\;\;\;\alpha}
       - 2 C^{\alpha\gamma} \Pi_{(i)\beta\gamma}.
   \label{Tab-pert}
\eea Using $\tilde T^a_b = \sum_j \tilde T_{(j)b}^{\;\;\;\;a}$, and
the total fluid quantities in Eq.\ (82) of \cite{NL} we have \bea
   & & \mu = \sum_j \mu_{(j)}, \quad
       p = \sum_j p_{(j)},
\eea for the background order fluid quantities, and \bea
   & & \delta \mu
       = \sum_j \left[
       \delta \mu_{(j)}
       + \left( \mu_{(j)} + p_{(j)} \right) \left( V_{(j)}^\alpha -
       B^\alpha \right)
       \left( V_{(j)\alpha} - V_\alpha \right) \right],
   \nonumber \\
   & & \delta p
       = \sum_j \left[
       \delta p_{(j)}
       + {1 \over 3} \left( \mu_{(j)} + p_{(j)} \right)
       \left( V_{(j)}^\alpha - B^\alpha \right)
       \left( V_{(j)\alpha} - V_\alpha \right) \right],
   \nonumber \\
   & & \left( \mu + p \right) V_\alpha
       = \sum_j \left[ \left( \mu_{(j)} + p_{(j)} \right) V_{(j)\alpha}
       + \left( \delta \mu_{(j)} + \delta p_{(j)} \right)
       \left( V_{(j)\alpha} - V_\alpha \right)
       + \Pi^{\;\;\;\;\beta}_{(j)\alpha}
       \left( V_{(j)\beta} - V_\beta \right) \right],
   \nonumber \\
   & & \Pi^\alpha_\beta
       = \sum_j \left\{ \Pi^{\;\;\;\;\alpha}_{(j)\beta}
       + \left( \mu_{(j)} + p_{(j)} \right)  \left[
       \left( v_{(j)}^\alpha - B^\alpha \right)
       \left( V_{(j)\beta} - V_\beta \right)
       - {1 \over 3} \delta^\alpha_\beta
       \left( V_{(j)}^\gamma - B^\gamma \right)
       \left( V_{(j)\gamma} - V_\gamma \right) \right] \right\},
   \label{sum-1}
\eea for perturbed order fluid quantities to the second-order. From
Eq.\ (\ref{ADM-fluid}) we have the ADM fluid quantities based on the
energy-frame fluid quantities \bea
   & & E_{(i)} = \mu_{(i)} + \delta \mu_{(i)}
       + \left( \mu_{(i)} + p_{(i)} \right)
       \left( V_{(i)}^\alpha - B^\alpha \right)
       \left( V_{(i)\alpha} - B_\alpha \right),
   \nonumber \\
   & & J_{(i)\alpha}
       = a \Big[ \left( \mu_{(i)} + p_{(i)} \right)
       \left( V_{(i)\alpha} - B_\alpha
       + A B_\alpha
       + 2 V^\beta_{(i)} C_{\alpha\beta} \right)
       + \left( \delta \mu_{(i)} + \delta p_{(i)} \right)
       \left( V_{(i)\alpha} - B_\alpha \right)
       + \Pi_{(i)\alpha\beta} \left( V_{(i)}^\beta - B^\beta \right)
       \Big],
   \nonumber \\
   & & S_{(i)}
       = 3 \left( p_{(i)} + \delta p_{(i)} \right)
       + \left( \mu_{(i)} + p_{(i)} \right)
       \left( V_{(i)}^\alpha - B^\alpha \right)
       \left( V_{(i)\alpha} - B_\alpha \right),
   \nonumber \\
   & & \bar S_{(i)\alpha\beta}
       = a^2 \left\{
       \Pi_{(i)\alpha\beta}
       + \left( \mu_{(i)} + p_{(i)} \right)
       \left[
       \left( V_{(i)\alpha} - B_\alpha \right)
       \left( V_{(i)\beta} - B_\beta \right)
       - {1 \over 3} g^{(3)}_{\alpha\beta}
       \left( V_{(i)}^\gamma - B^\gamma \right)
       \left( V_{(i)\gamma} - B_\gamma \right) \right]
       \right\}.
   \label{ADM-fluid-quantities}
\eea We can compare Eq.\ (\ref{ADM-fluid-quantities}) with the ADM
fluid quantities based on the normal-frame fluid quantities in Eq.\
(76) of \cite{NL}; in a single component case we simply delete $(i)$
subindices.

In \cite{NL} the fluid quantities are based on the normal-frame
vector. By taking $\tilde u_{(i)\alpha} \equiv 0$, $\tilde u_{(i)a}$
becomes the normal-frame vector $\tilde n_a$, see Eq.\ (54) in
\cite{NL}. Based on the normal-frame vector, to the second order,
the fluid quantities have contributions due to the frame choice: for
example, even in the zero-pressure fluid, the perturbed pressure
based on the normal-frame does not necessarily vanish to the
second-order, see Eq.\ (\ref{normal-to-energy-frame}) below. By
comparing Eq.\ (\ref{ADM-fluid-quantities}) with Eq.\ (76) of
\cite{NL} we have \bea
   & & \delta \mu_{(i)}^N
       = \delta \mu_{(i)}
       + \left( \mu_{(i)} + p_{(i)} \right)
       \left( V_{(i)}^\alpha - B^\alpha \right)
       \left( V_{(i)\alpha} - B_\alpha \right),
   \nonumber \\
   & & \delta p_{(i)}^N
       = \delta p_{(i)}
       + {1 \over 3} \left( \mu_{(i)} + p_{(i)} \right)
       \left( V_{(i)}^\alpha - B^\alpha \right)
       \left( V_{(i)\alpha} - B_\alpha \right),
   \nonumber \\
   & & Q_{(i)\alpha}
       = \left( \mu_{(i)} + p_{(i)} \right)
       \left( V_{(i)\alpha} - B_\alpha
       + A B_\alpha
       + 2 V_{(i)}^\beta C_{\alpha\beta} \right)
       + \left( \delta \mu_{(i)} + \delta p_{(i)} \right)
       \left( V_{(i)\alpha} - B_\alpha \right)
       + \Pi_{(i)\alpha\beta} \left( V_{(i)}^\beta - B^\beta
       \right),
   \nonumber \\
   & & \Pi_{(i)\alpha\beta}^N
       = \Pi_{(i)\alpha\beta}
       + \left( \mu_{(i)} + p_{(i)} \right) \left[
       \left( V_{(i)\alpha} - B_\alpha \right)
       \left( V_{(i)\beta} - B_\beta \right)
       - {1 \over 3} g^{(3)}_{\alpha\beta}
       \left( V_{(i)}^\gamma - B^\gamma \right)
       \left( V_{(i)\gamma} - B_\gamma \right) \right].
   \label{normal-to-energy-frame}
\eea For the total fluid quantities the relations between the two
frames are presented in Eq.\ (87) of \cite{NL}. Thus, by replacing
all fluid quantities in Eqs.\ (99)-(107) of \cite{NL} using Eq.\
(\ref{normal-to-energy-frame}) and Eq.\ (87) of \cite{NL} we have
the equations in the energy frame. Using $Q_{(i)\alpha}$ in Eq.\
(\ref{normal-to-energy-frame}), Eq.\ (\ref{Tab-pert}) gives \bea
   & & \tilde T_{(i) \alpha}^{\;\;\;\; 0}
       = \left( 1 - A \right) Q_{(i)\alpha}.
\eea As the fluid four-velocity of $i$-th component we can use
either $Q_{(i)\alpha}$ or $V_{(i)\alpha} - B_\alpha$ related by Eq.\
(\ref{normal-to-energy-frame}).

The kinematic quantities in the energy-frame are presented in Eqs.\
(63)-(66) of \cite{NL}. In Eq.\ (\ref{kinematic_quantities-i}) we
introduced kinematic quantities for the individual component. To the
second order we can show \bea
   & & \tilde \theta_{(i)}
       = 3 {a^\prime \over a^2}
       - 3 {a^\prime \over a^2} A
       + {1 \over a} C^{\alpha\prime}_\alpha
       + {1 \over a} V^\alpha_{(i)|\alpha}
       + {9 \over 2} {a^\prime \over a^2} A^2
       - 3 {a^\prime \over a^2} B^\alpha V_{(i)\alpha}
       - {1 \over a} A C^{\alpha\prime}_\alpha
       + {1 \over a} V^\alpha_{(i)} \left( A_{,\alpha}
       + C^\beta_{\beta|\alpha} \right)
   \nonumber \\
   & & \qquad
       + {3 \over 2} {a^\prime \over a^2} V^\alpha_{(i)} V_{(i)\alpha}
       + {1 \over a} \left( V^\alpha_{(i)} - B^\alpha \right)
       \left( V_{(i)\alpha} - B_\alpha \right)^\prime
       - 2 {1 \over a} C^{\alpha\beta} C_{\alpha\beta}^\prime,
   \nonumber \\
   & & \tilde \sigma_{(i)\alpha\beta}
       = a \Big[ V_{(i)(\alpha|\beta)}
       + C_{\alpha\beta}^\prime
       - A C_{\alpha\beta}^\prime
       + \left( V_{(i)(\alpha} - B_{(\alpha} \right)
       \left( V_{(i)\beta)} - B_{\beta)} \right)^\prime
       + V_{(i)(\alpha} A_{,\beta)}
       + V_{(i)}^\gamma C_{\alpha\beta|\gamma}
       + 2 V^\gamma_{(i)|(\alpha} C_{(\beta)\gamma}
   \nonumber \\
   & & \qquad
       - {2 \over 3} C_{\alpha\beta}
       \left( C_\gamma^{\gamma\prime}
       + V^\gamma_{(i)|\gamma} \right) \Big]
   \nonumber \\
   & & \qquad
       - {1 \over 3} g^{(3)}_{\alpha\beta} a \left[
       V^\gamma_{(i)|\gamma}
       + C_\gamma^{\gamma\prime}
       - A C_\gamma^{\gamma\prime}
       + \left( V_{(i)}^\gamma - B^\gamma \right)
       \left( V_{(i)\gamma} - B_\gamma \right)^\prime
       + V_{(i)}^\gamma A_{,\gamma}
       + V_{(i)}^\gamma C^\delta_{\delta|\gamma}
       - 2 C^{\gamma\delta} C_{\gamma\delta}^\prime \right],
   \nonumber \\
   & & \tilde \omega_{(i)\alpha\beta}
       = a \left( V_{(i)[\alpha} - B_{[\alpha}
       + A B_{[\alpha}
       + 2 V_{(i)}^\gamma C_{\gamma[\alpha} \right)_{|\beta]}
       - a \left( V_{(i)[\alpha} - B_{[\alpha} \right)
       \left[ \left( V_{(i)\beta]} - B_{\beta]} \right)^\prime
       + A_{,\beta]} \right],
   \nonumber \\
   & & \tilde a_{(i)\alpha}
       = A_{,\alpha}
       + {1 \over a} \left[ a \left( V_{(i)\alpha} - B_\alpha
       + A B_\alpha
       + 2 V_{(i)}^\beta C_{\alpha\beta}  \right) \right]^\prime
       - 2 A A_{,\alpha}
       - A {1 \over a} \left[ a \left( V_{(i)\alpha} - B_\alpha
       \right) \right]^\prime
   \nonumber \\
   & & \qquad
       + V_{(i)}^\beta \left[ \left( V_{(i)\alpha} - B_\alpha \right)_{|\beta}
       + B_{\beta|\alpha} \right],
   \nonumber \\
   & & \tilde \sigma_{(i)\alpha0} = - V_{(i)}^\beta \tilde
       \sigma_{(i)\alpha\beta}, \quad
       \tilde \sigma_{(i)00} = 0; \quad
       \tilde \omega_{(i)\alpha0} = - V_{(i)}^\beta \tilde
       \omega_{(i)\alpha\beta}, \quad
       \tilde \omega_{(i)00} = 0; \quad
       \tilde a_{(i)0} = - V_{(i)}^\alpha \tilde a_{(i)\alpha},
   \label{kinematic-quantities-pert}
\eea where a prime indicates the time derivative based on $\eta$.

The gauge transformation properties of the fluid quantities are
presented in Eqs.\ (232)-(235) of \cite{NL} for the normal-frame,
and Eqs.\ (238) of \cite{NL} for the energy-frame. A prescription to
get the gauge transformation properties for individual fluid
quantities is also presented below Eq.\ (235) of \cite{NL}. Under
the gauge transformation we have \bea
   & & \delta \hat \mu_{(i)}
       = \delta \mu_{(i)}
       - \left( \mu_{(i)}^\prime + \delta \mu_{(i)}^\prime \right)
       \xi^0
       - \delta \mu_{(i),\alpha} \xi^\alpha
       + {1 \over 2} \mu_{(i)}^{\prime\prime} \xi^0 \xi^0
       + \mu_{(i)}^\prime \left( \xi^0 \xi^{0\prime}
       + \xi^\alpha \xi^0_{\;\;,\alpha} \right),
   \nonumber \\
   & & \delta \hat p_{(i)}
       = \delta p_{(i)}
       - \left( p_{(i)}^\prime + \delta p_{(i)}^\prime \right)
       \xi^0
       - \delta p_{(i),\alpha} \xi^\alpha
       + {1 \over 2} p_{(i)}^{\prime\prime} \xi^0 \xi^0
       + p_{(i)}^\prime \left( \xi^0 \xi^{0\prime}
       + \xi^\alpha \xi^0_{\;\;,\alpha} \right),
   \nonumber \\
   & & \hat V_{(i)\alpha} - \hat B_\alpha
       + \hat A \hat B_\alpha + 2 \hat V_{(i)}^\beta \hat
       C_{\alpha\beta}
       = V_{(i)\alpha} - B_\alpha
       + A B_\alpha + 2 V_{(i)}^\beta C_{\alpha\beta}
       + \xi^0_{\;\;,\alpha}
       - \left( V_{(i)\alpha} - B_\alpha \right)^\prime \xi^0
       - {a^\prime \over a} \left( V_{(i)\alpha} - B_\alpha \right)
       \xi^0
   \nonumber \\
   & & \qquad
       - \left( V_{(i)\beta} - B_\beta \right)
       \xi^\beta_{\;\;,\alpha}
       - \left( V_{(i)\alpha} - B_\alpha \right)_{,\beta} \xi^\beta
       + \left( A - \xi^{0\prime} - {a^\prime \over a} \xi^0 \right)
       \xi^0_{\;\;,\alpha}
       - \xi^\beta_{\;\;,\alpha} \xi^0_{\;\;,\beta}
       - \xi^0 \xi^{0\prime}_{\;\;,\alpha}
       - \xi^\beta \xi^0_{\;\;,\alpha\beta},
   \nonumber \\
   & & \hat \Pi_{(i)\alpha\beta}
       = \Pi_{(i)\alpha\beta}
       - \left( \Pi^\prime_{(i)\alpha\beta}
       + 2 {a^\prime \over a} \Pi_{(i)\alpha\beta} \right) \xi^0
       - \Pi_{(i)\alpha\beta,\gamma} \xi^\gamma
       - 2 \Pi_{(i)\gamma(\alpha} \xi^\gamma_{\;\;,\beta)}.
   \label{GT}
\eea

%%%%%%%%%%%%%%%%%%%%%%%%%%%%%%%%%%%%%%%%%%%%%%%%%%%%%%%%%%%%%%
\subsection{Basic equations in the energy-frame}

The basic set of equations with fluid quantities based on the
normal-frame is presented in Eqs.\ (99)-(107) of \cite{NL}. By using
Eq.\ (\ref{normal-to-energy-frame}) we can recover the equations
with fluid quantities based on the energy frame. For convenience, in
the following we present the complete set of equations with fluid
quantities in the energy frame. These equations are written without
taking any gauge conditions yet, thus in a sort of gauge-ready form.
To the linear order this method was suggested by Bardeen
\cite{Bardeen-1988,H-PRW-1991}.

\noindent Definition of $\delta K$: \bea
   & & \bar K + 3 H
       + \delta K - 3 H A + \dot C^\alpha_\alpha
       + {1\over a} B^\alpha_{\;\;|\alpha}
   \nonumber \\
   & & \qquad
       = - A \left( {9 \over 2} H A - \dot C^\alpha_\alpha
       - {1 \over a} B^\alpha_{\;\;|\alpha} \right)
       + {3 \over 2} H B^\alpha B_\alpha
       + {1\over a} B^\alpha \left( 2 C^\beta_{\alpha|\beta}
       - C^\beta_{\beta|\alpha} \right)
       + 2 C^{\alpha\beta} \left( \dot C_{\alpha\beta}
       + {1\over a} B_{\alpha|\beta} \right)
       \equiv n_0.
   \label{def-of-kappa}
\eea Energy constraint equation: \bea
   & & 16 \pi G \mu + 2 \Lambda - 6 H^2 - {1\over a^2} R^{(3)}
       + 16 \pi G \delta \mu + 4 H \delta K
       - {1\over a^2} \left(
       2 C^{\beta|\alpha}_{\alpha\;\;\;\;\beta}
       - 2 C^{\alpha|\beta}_{\alpha\;\;\;\;\beta}
       - {2 \over 3} R^{(3)} C^\alpha_\alpha \right)
   \nonumber \\
   & & \qquad
       = {2 \over 3} \delta K^2
       - 16 \pi G \left( \mu + p \right)
       \left( V^\alpha - B^\alpha \right)
       \left( V_\alpha - B_\alpha \right)
       - \left( \dot C_{\alpha\beta} + {1\over a} B_{(\alpha|\beta)} \right)
       \left( \dot C^{\alpha\beta} + {1\over a} B^{\alpha|\beta} \right)
       + {1\over 3} \left( \dot C^\alpha_\alpha
       + {1\over a} B^\alpha_{\;\;|\alpha} \right)^2
   \nonumber \\
   & & \qquad
       + {1 \over a^2} \Bigg[
       4 C^{\alpha\beta} \left( - C^\gamma_{\alpha|\beta\gamma}
       - C^{\gamma}_{\alpha|\gamma\beta}
       + C^{\;\;\;\;|\gamma}_{\alpha\beta\;\;\;\gamma}
       + C^\gamma_{\gamma|\alpha\beta} \right)
       + {4 \over 3} R^{(3)} C^\alpha_\gamma C^\gamma_\alpha
   \nonumber \\
   & & \qquad
       - \left( 2 C^\gamma_{\beta|\gamma} - C^\gamma_{\gamma|\beta} \right)
       \left( 2 C^{\alpha\beta}_{\;\;\;\;|\alpha} - C^{\alpha|\beta}_\alpha
       \right)
       + C^{\alpha\beta|\gamma} \left( 3 C_{\alpha\beta|\gamma}
       - 2 C_{\alpha\gamma|\beta} \right)
       \Bigg]
       \equiv n_1.
   \label{E-constraint}
\eea Momentum constraint equation: \bea
   & & \left[ \dot C^\beta_\alpha + {1\over 2a} \left( B^\beta_{\;\;|\alpha}
       + B_\alpha^{\;\;|\beta} \right) \right]_{|\beta}
       - {1\over 3} \left( \dot C^\gamma_\gamma
       + {1\over a} B^\gamma_{\;\;|\gamma} \right)_{,\alpha}
       + {2\over 3} \delta K_{,\alpha}
       + 8 \pi G a \left( \mu + p \right) \left( V_\alpha -B_\alpha
       + A B_\alpha + 2 V^\beta C_{\alpha\beta} \right)
   \nonumber \\
   & & \qquad
       = - {2\over 3} A \delta K_{,\alpha}
       - 8 \pi G a \left[
       \left( \delta \mu + \delta p \right) \left( V_\alpha
       -B_\alpha \right)
       + \left( \mu + p \right) A \left( V_\alpha - B_\alpha \right)
       + \Pi_{\alpha\beta} \left( V^\beta - B^\beta \right) \right]
   \nonumber \\
   & & \qquad
       + A_{,\beta} \left[ \dot C^\beta_\alpha
       + {1\over 2a} \left( B^\beta_{\;\;|\alpha}
       + B_\alpha^{\;\;|\beta} \right) \right]
       + \left( 2 C^\beta_{\gamma|\beta}
       - C^\beta_{\beta|\gamma} \right)
       \left[ \dot C^\gamma_\alpha
       + {1\over 2a} \left( B_\alpha^{\;\;|\gamma}
       + B^\gamma_{\;\;|\alpha} \right) \right]
       + 2 C^{\beta\gamma} \left( \dot C_{\alpha\gamma}
       + {1\over a} B_{(\alpha|\gamma)} \right)_{|\beta}
   \nonumber \\
   & & \qquad
       + {1\over a} \left[ B_\gamma \left( C^{\beta\gamma}_{\;\;\;\;\;|\alpha}
       + C^{\gamma|\beta}_\alpha
       - C^{\beta|\gamma}_\alpha \right) \right]_{|\beta}
       + {1\over 3} C^\gamma_{\beta|\alpha} \left(
       \dot C^\beta_\gamma + {1\over a} B^\beta_{\;\;|\gamma} \right)
   \nonumber \\
   & & \qquad
       - {1\over 3} \left\{
       A_{,\alpha} \left( \dot C^\gamma_\gamma
       + {1\over a} B^\gamma_{\;\;|\gamma} \right)
       + 2 C^{\gamma\delta} \left( \dot C_{\gamma\delta}
       + {1\over a} B_{\gamma|\delta} \right)_{|\alpha}
       + {1\over a}\left[ B^\delta \left( 2 C^\gamma_{\delta|\gamma}
       - C^\gamma_{\gamma|\delta} \right) \right]_{|\alpha} \right\}
       \equiv n_{2\alpha}.
   \label{Mom-constraint}
\eea Trace of the ADM propagation equation: \bea
   & & - \left[ 3 \dot H + 3 H^2
       + 4 \pi G \left( \mu + 3 p \right)
       - \Lambda \right]
       + \delta \dot K + 2 H \delta K
       - 4 \pi G \left( \delta \mu + 3 \delta p
       \right)
       + \left( 3 \dot H + {\Delta \over a^2} \right) A
   \nonumber \\
   & & \qquad
       = A \delta \dot K
       - {1 \over a} \delta K_{,\alpha} B^\alpha
       + {1 \over 3} \delta K^2
       + 8 \pi G \left( \mu + p \right)
       \left( V^\alpha - B^\alpha \right)
       \left( V_\alpha - B_\alpha \right)
   \nonumber \\
   & & \qquad
       + {3 \over 2} \dot H \left( 3 A^2 - B^\alpha B_\alpha \right)
       + {1 \over a^2} \left[ 2 A \Delta A
       + A^{,\alpha} A_{,\alpha}
       - {1 \over 2} \Delta \left( B^\alpha  B_\alpha \right)
       + A^{,\alpha} \left( 2 C^\beta_{\alpha|\beta}
       - C^\beta_{\beta|\alpha} \right)
       + 2 C^{\alpha\beta} A_{,\alpha|\beta} \right]
   \nonumber \\
   & & \qquad
       + \left( \dot C^{\alpha\beta}
       + {1 \over a} B^{(\alpha|\beta)} \right)
       \left( \dot C_{\alpha\beta} + {1 \over a} B_{\alpha|\beta}
       \right)
       - {1 \over 3} \left( \dot C^\alpha_\alpha
       + {1 \over a} B^\alpha_{\;\;|\alpha} \right)^2
       \equiv n_3.
   \label{Raychaudhury-eq}
\eea Tracefree ADM propagation equation: \bea
   & &
       \left[ \dot C^\alpha_\beta + {1\over 2a} \left( B^\alpha_{\;\;|\beta}
       + B_\beta^{\;\;|\alpha} \right) \right]^\cdot
       + 3 H
       \left[ \dot C^\alpha_\beta + {1\over 2a} \left( B^\alpha_{\;\;|\beta}
       + B_\beta^{\;\;|\alpha} \right) \right]
       - {1\over a^2} A^{|\alpha}_{\;\;\;\;\beta}
   \nonumber \\
   & & \qquad
       - {1\over 3} \delta^\alpha_\beta \left[
       \left( \dot C^\gamma_\gamma + {1\over a} B^\gamma_{\;\;|\gamma}
       \right)^\cdot
       + 3 H
       \left( \dot C^\gamma_\gamma + {1\over a} B^\gamma_{\;\;|\gamma} \right)
       - {1\over a^2} A^{|\gamma}_{\;\;\;\;\gamma} \right]
   \nonumber \\
   & & \qquad
       + {1 \over a^2} \left[
       C^{\alpha\gamma}_{\;\;\;\;\;|\beta\gamma}
       + C^{\gamma|\alpha}_{\beta\;\;\;\;\gamma}
       - C^{\alpha|\gamma}_{\beta\;\;\;\;\gamma}
       - C^{\gamma|\alpha}_{\gamma\;\;\;\;\beta}
       - {2 \over 3} R^{(3)} C^\alpha_\beta
       - {1\over 3} \delta^\alpha_\beta \left(
       2 C^{\delta|\gamma}_{\gamma\;\;\;\;\delta}
       - 2 C^{\gamma|\delta}_{\gamma\;\;\;\;\delta}
       - {2 \over 3} R^{(3)} C^\gamma_\gamma \right) \right]
       - 8 \pi G \Pi^\alpha_\beta
   \nonumber \\
   & & \qquad
       =
       \left\{
       \left[ \dot C^\alpha_\beta + {1\over 2a} \left( B^\alpha_{\;\;|\beta}
       + B_\beta^{\;\;|\alpha} \right) \right] A
       + 2 C^{\alpha\gamma} \left( \dot C_{\beta\gamma}
       + {1\over a} B_{(\beta|\gamma)} \right)
       + {1\over a}B_\gamma \left( C^{\alpha\gamma}_{\;\;\;\;\;|\beta}
       + C^{\gamma|\alpha}_\beta - C^{\alpha|\gamma}_\beta \right)
       \right\}^\cdot
   \nonumber \\
   & & \qquad
       + 3 H \left\{
       \left[ \dot C^\alpha_\beta + {1\over 2a} \left( B^\alpha_{\;\;|\beta}
       + B_\beta^{\;\;|\alpha} \right) \right] A
       + 2 C^{\alpha\gamma} \left( \dot C_{\beta\gamma}
       + {1\over a} B_{(\beta|\gamma)} \right)
       + {1\over a}B_\gamma \left( C^{\alpha\gamma}_{\;\;\;\;\;|\beta}
       + C^{\gamma|\alpha}_\beta - C^{\alpha|\gamma}_\beta \right)
       \right\}
   \nonumber \\
   & & \qquad
       + \left[ \dot C^\alpha_\beta + {1\over 2a} \left( B^\alpha_{\;\;|\beta}
       + B_\beta^{\;\;|\alpha} \right) \right]^\cdot A
       - {1\over a} \left[ \dot C^\alpha_\beta + {1\over 2a}
       \left( B^\alpha_{\;\;\;|\beta} + B^{\;\;|\alpha}_\beta \right)
       \right]_{|\gamma} B^\gamma
       + \delta K
       \left[ \dot C^\alpha_\beta + {1\over 2a} \left( B^\alpha_{\;\;|\beta}
       + B_\beta^{\;\;|\alpha} \right) \right]
   \nonumber \\
   & & \qquad
       + {1 \over a^2} \left[ - A A^{|\alpha}_{\;\;\;\;\beta}
       + {1 \over 2} \left( - A^2 + B^\gamma B_\gamma
       \right)^{|\alpha}_{\;\;\;\;\beta}
       - 2 C^{\alpha\gamma} A_{,\beta|\gamma}
       - \left( C^{\alpha\gamma}_{\;\;\;\;\;|\beta}
       + C^{\gamma|\alpha}_\beta - C^{\alpha|\gamma}_\beta \right)
       A_{,\gamma} \right]
   \nonumber \\
   & & \qquad
       - {1 \over 3} \delta^\alpha_\beta \Bigg\{
       \left[
       \left( \dot C^\gamma_\gamma + {1\over a} B^\gamma_{\;\;|\gamma}
       \right) A
       + 2 C^{\gamma\delta} \left( \dot C_{\gamma\delta}
       + {1\over a} B_{\gamma|\delta} \right)
       + {1\over a} B^\delta \left( 2 C^\gamma_{\delta|\gamma}
       - C^\gamma_{\gamma|\delta} \right) \right]^\cdot
   \nonumber \\
   & & \qquad
       + 3 H \left[
       \left( \dot C^\gamma_\gamma
       + {1\over a} B^\gamma_{\;\;|\gamma} \right) A
       + 2 C^{\gamma\delta} \left( \dot C_{\gamma\delta}
       + {1\over a} B_{\gamma|\delta} \right)
       + {1\over a} B^\delta \left( 2 C^\gamma_{\delta|\gamma}
       - C^\gamma_{\gamma|\delta} \right) \right]
   \nonumber \\
   & & \qquad
       + \left( \dot C^\gamma_\gamma + {1\over a} B^\gamma_{\;\;|\gamma}
       \right)^\cdot A
       - {1\over a} \left( \dot C^\gamma_\gamma
       + {1\over a} B^\gamma_{\;\;|\gamma} \right)_{|\delta} B^\delta
       + \delta K \left( \dot C^\gamma_\gamma
       + {1\over a} B^\gamma_{\;\;|\gamma} \right)
   \nonumber \\
   & & \qquad
       + {1 \over a^2} \left[ - A A^{|\gamma}_{\;\;\;\;\gamma}
       + {1 \over 2} \left( - A^2 + B^\delta B_\delta
       \right)^{|\gamma}_{\;\;\;\;\gamma}
       - 2 C^{\gamma\delta} A_{,\gamma|\delta}
       - \left( 2 C^{\gamma\delta}_{\;\;\;\;|\gamma}
       - C^{\gamma|\delta}_\gamma \right) A_{,\delta} \right]
       \Bigg\}
   \nonumber \\
   & & \qquad
       + {1 \over a} B^\alpha_{\;\;|\gamma}
       \left[ \dot C^\gamma_\beta + {1\over 2a} \left( B^\gamma_{\;\;|\beta}
       + B_\beta^{\;\;|\gamma} \right) \right]
       - {1 \over a} B^\gamma_{\;\;|\beta} \left[ \dot C^\alpha_\gamma
       + {1 \over 2a} \left( B^\alpha_{\;\;|\gamma}
       + B_\gamma^{\;\;|\alpha} \right) \right]
   \nonumber \\
   & & \qquad
       + {1 \over a^2} \Bigg\{
       2 C^{\gamma\delta} \left( C^\alpha_{\delta|\beta\gamma}
       + C^{\;\;\;\;|\alpha}_{\delta\beta\;\;\;\;\gamma}
       - C^\alpha_{\beta|\delta\gamma}
       - C^{\;\;\;\;|\alpha}_{\delta\gamma\;\;\;\;\beta} \right)
       + 2 C^{\alpha\gamma} \left(
       C^\delta_{\gamma|\beta\delta}
       + C^\delta_{\beta|\gamma\delta}
       - C^{\;\;\;\;|\delta}_{\beta\gamma\;\;\;\delta}
       - C^\delta_{\delta|\gamma\beta} \right)
       - {4 \over 3} R^{(3)} C^\alpha_\gamma C^\gamma_\beta
   \nonumber \\
   & & \qquad
       + \left( 2 C^\gamma_{\delta|\gamma}
       - C^\gamma_{\gamma|\delta} \right)
       \left( C^{\alpha\delta}_{\;\;\;\;\;|\beta}
       + C^{\delta|\alpha}_\beta
       - C^{\alpha|\delta}_\beta \right)
       - C_{\gamma\delta|\beta} C^{\gamma\delta|\alpha}
       + 2 C^{\alpha\gamma|\delta} \left( C_{\beta\delta|\gamma}
       - C_{\beta\gamma|\delta} \right)
   \nonumber \\
   & & \qquad
       - {1 \over 3} \delta^\alpha_\beta \Bigg[
       4 C^{\gamma\delta} \left( C^\epsilon_{\gamma|\delta\epsilon}
       + C^{\epsilon}_{\gamma|\epsilon\delta}
       - C^{\;\;\;\;|\epsilon}_{\gamma\delta\;\;\;\;\epsilon}
       - C^\epsilon_{\epsilon|\gamma\delta} \right)
       - {4 \over 3} R^{(3)} C^\delta_\gamma C^\gamma_\delta
   \nonumber \\
   & & \qquad
       + \left( 2 C^\epsilon_{\delta|\epsilon}
       - C^\epsilon_{\epsilon|\delta} \right)
       \left( 2 C^{\gamma\delta}_{\;\;\;\;|\gamma}
       - C^{\gamma|\delta}_\gamma
       \right)
       + C^{\gamma\delta|\epsilon} \left(
       2 C_{\gamma\epsilon|\delta} - 3 C_{\gamma\delta|\epsilon} \right)
       \Bigg] \Bigg\}
   \nonumber \\
   & & \qquad
       -16 \pi G C^{\alpha\gamma} \Pi_{\beta\gamma}
       + 8 \pi G \left( \mu + p \right) \left[
       \left( V^\alpha - B^\alpha \right) \left( V_\beta - B_\beta
       \right)
       - {1 \over 3} \delta^\alpha_\beta \left( V^\gamma -
       B^\gamma \right)
       \left( V_\gamma - B_\gamma \right) \right]
       \equiv n_{4\beta}^{\;\;\alpha}.
   \label{tracefree-ADM-propagation}
\eea Energy conservation equation: \bea
   & & \left[ \dot \mu
       + 3 H \left( \mu + p \right) \right]
       + \delta \dot \mu
       + 3 H \left( \delta \mu + \delta p \right)
       - \left( \mu + p \right) \left( \delta K - 3 H A
       \right)
       + {1 \over a} \left( \mu + p \right)
       \left[ V^\alpha - B^\alpha
       + A B^\alpha
       + 2 V^\beta C^\alpha_\beta \right]_{|\alpha}
   \nonumber \\
   & & \qquad
       =
       - {1 \over a} \delta \mu_{,\alpha} B^\alpha
       + \left( \delta \mu + \delta p \right)
       \left( \delta K - 3 H A \right)
       + \left( \mu + p \right) A \delta K
       + {3 \over 2} H \left( \mu + p \right)
       \left( A^2 - B^\alpha B_\alpha \right)
   \nonumber \\
   & & \qquad
       - {1 \over a^4} \left[ a^4 \left( \mu + p \right)
       \left( V^\alpha - B^\alpha \right)
       \left( V_\alpha - B_\alpha \right) \right]^\cdot
       - {1 \over a}
       \left[
       \left( \delta \mu + \delta p \right)
       \left( V^\alpha - B^\alpha \right)
       + \Pi^{\alpha\beta}
       \left( V_{\beta} - B_\beta \right) \right]_{|\alpha}
   \nonumber \\
   & & \qquad
       + {1 \over a} \left( \mu + p \right)
       \left\{
       - A \left( V^\alpha - B^\alpha \right)_{|\alpha}
       - 2 A_{,\alpha} \left( V^\alpha - B^\alpha \right)
       + 2 \left[ C^{\alpha\beta}
       \left( V_{\beta} - B_\beta \right) \right]_{|\alpha}
       - C_{\alpha|\beta}^\alpha
       \left( V^{\beta} - B^\beta \right) \right\}
   \nonumber \\
   & & \qquad
       - \Pi^{\alpha\beta} \left( \dot C_{\alpha\beta}
       + {1 \over a} B_{\alpha|\beta} \right)
       \equiv n_5.
   \label{E-conservation}
\eea Momentum conservation equation: \bea
   & & {1 \over a^4} \left[ a^4 \left( \mu + p \right)
       \left( V_{\alpha} - B_\alpha
       + A B_\alpha
       + 2 V^\beta C_{\alpha\beta} \right)
       \right]^\cdot
       + {1 \over a} \left( \mu + p \right) A_{,\alpha}
       + {1 \over a} \left(
       \delta p_{,\alpha}
       + \Pi_{\alpha|\beta}^{\beta} \right)
   \nonumber \\
   & & \qquad
       = \left( \mu + p \right)
       \left( \delta K - 3 H A \right)
       \left( V_{\alpha} - B_\alpha \right)
       - {1 \over a^4} \left\{ a^4
       \left[
       \left( \delta \mu + \delta p \right)
       \left( V_{\alpha} - B_\alpha \right)
       + \Pi_{\alpha}^{\beta}
       \left( V_{\beta} - B_\beta \right)
       \right] \right\}^\cdot
   \nonumber \\
   & & \qquad
       + {1 \over a} \Bigg\{
       - \left( \delta p_{,\alpha}
       + \Pi_{\alpha|\beta}^{\beta} \right) A
       - \left( \delta \mu + \delta p \right) A_{,\alpha}
   \nonumber \\
   & & \qquad
       - \left( \mu + p \right)
       \left[ - A A_{,\alpha}
       + B_{\beta|\alpha} V^{\beta}
       + \left( V_{\alpha} - B_\alpha \right)_{|\beta} V^\beta
       + \left( V_{\alpha} - B_\alpha \right)
       \left( V^{\beta} - B^\beta \right)_{|\beta}
       \right]
   \nonumber \\
   & & \qquad
       + 2 \left( C^{\beta\gamma} \Pi_{\alpha\gamma}
       \right)_{|\beta}
       - C^\beta_{\beta|\gamma} \Pi_{\alpha}^{\gamma}
       + C^\gamma_{\beta|\alpha} \Pi_{\gamma}^{\beta}
       - A_{,\beta} \Pi_{\alpha}^{\beta}
       \Bigg\}
       \equiv n_{6\alpha}.
   \label{Mom-conservation}
\eea In the multi-component situation we additionally have the
energy and the momentum conservation of individual component. Using
the energy-frame fluid quantities, Eqs.\ (106),(107) in \cite{NL}
become \bea
   & & \Big[ \dot \mu_{(i)}
       + 3 H \left( \mu_{(i)} + p_{(i)} \right)
       + {1 \over a} I_{(i)0} \Big]
       + \delta \dot \mu_{(i)}
       + 3 H \left( \delta \mu_{(i)} + \delta p_{(i)} \right)
       - \left( \mu_{(i)} + p_{(i)} \right) \left( \delta K - 3 H A
       \right)
   \nonumber \\
   & & \qquad
       + {1 \over a} \left( \mu_{(i)} + p_{(i)} \right)
       \left[ V_{(i)}^\alpha - B^\alpha
       + A B^\alpha
       + 2 V_{(i)}^\beta C^\alpha_\beta \right]_{|\alpha}
       + {1 \over a} \delta I_{(i)0}
   \nonumber \\
   & & \qquad
       =
       - {1 \over a} \delta \mu_{(i),\alpha} B^\alpha
       + \left( \delta \mu_{(i)} + \delta p_{(i)} \right)
       \left( \delta K - 3 H A \right)
       + \left( \mu_{(i)} + p_{(i)} \right) A \delta K
       + {3 \over 2} H \left( \mu_{(i)} + p_{(i)} \right)
       \left( A^2 - B^\alpha B_\alpha \right)
   \nonumber \\
   & & \qquad
       - {1 \over a^4} \left[ a^4 \left( \mu_{(i)} + p_{(i)} \right)
       \left( V_{(i)}^\alpha - B^\alpha \right)
       \left( V_{(i)\alpha} - B_\alpha \right)
       \right]^\cdot
       - {1 \over a}
       \left[
       \left( \delta \mu_{(i)} + \delta p_{(i)} \right)
       \left( V_{(i)}^\alpha - B^\alpha \right)
       + \Pi_{(i)}^{\alpha\beta}
       \left( V_{(i)\beta} - B_\beta \right) \right]_{|\alpha}
   \nonumber \\
   & & \qquad
       + {1 \over a} \left( \mu_{(i)} + p_{(i)} \right)
       \left\{
       - A \left( V_{(i)}^\alpha - B^\alpha \right)_{|\alpha}
       - 2 A_{,\alpha} \left( V_{(i)}^\alpha - B^\alpha \right)
       + 2 \left[ C^{\alpha\beta}
       \left( V_{(i)\beta} - B_\beta \right) \right]_{|\alpha}
       - C^{\alpha|\beta}_\alpha
       \left( V_{(i)\beta} - B_\beta \right) \right\}
   \nonumber \\
   & & \qquad
       - \Pi_{(i)}^{\alpha\beta} \left( \dot C_{\alpha\beta}
       + {1 \over a} B_{\alpha|\beta} \right)
       - {1 \over a} \delta I_{(i)\alpha} B^\alpha
       \equiv n_{(i)5},
   \label{E-conservation-i} \\
   & & {1 \over a^4} \left[ a^4 \left( \mu_{(i)} + p_{(i)} \right)
       \left( V_{(i)\alpha} - B_\alpha
       + A B_\alpha
       + 2 V_{(i)}^\beta C_{\alpha\beta} \right)
       \right]^\cdot
       + {1 \over a} \left( \mu_{(i)} + p_{(i)} \right) A_{,\alpha}
       + {1 \over a} \left(
       \delta p_{(i),\alpha}
       + \Pi_{(i)\alpha|\beta}^{\;\;\;\;\beta}
       - \delta I_{(i)\alpha} \right)
   \nonumber \\
   & & \qquad
       = \left( \mu_{(i)} + p_{(i)} \right)
       \left( \delta K - 3 H A \right)
       \left( V_{(i)\alpha} - B_\alpha \right)
       - {1 \over a^4} \left\{ a^4
       \left[
       \left( \delta \mu_{(i)} + \delta p_{(i)} \right)
       \left( V_{(i)\alpha} - B_\alpha \right)
       + \Pi_{(i)\alpha}^{\;\;\;\;\beta}
       \left( V_{(i)\beta} - B_\beta \right)
       \right] \right\}^\cdot
   \nonumber \\
   & & \qquad
       + {1 \over a} \Bigg\{
       - \left( \delta p_{(i),\alpha}
       + \Pi_{(i)\alpha|\beta}^{\;\;\;\;\beta}
       - \delta I_{(i)\alpha} \right) A
       - \left( \delta \mu_{(i)} + \delta p_{(i)} \right) A_{,\alpha}
   \nonumber \\
   & & \qquad
       - \left( \mu_{(i)} + p_{(i)} \right)
       \left[ - A A_{,\alpha}
       + B_{\beta|\alpha} V_{(i)}^{\beta}
       + \left( V_{(i)\alpha} - B_\alpha \right)_{|\beta} V_{(i)}^\beta
       + \left( V_{(i)\alpha} - B_\alpha \right)
       \left( V_{(i)}^{\beta} - B^\beta \right)_{|\beta}
       \right]
   \nonumber \\
   & & \qquad
       + 2 \left( C^{\beta\gamma} \Pi_{(i)\alpha\gamma}
       \right)_{|\beta}
       - C^\beta_{\beta|\gamma} \Pi_{(i)\alpha}^{\;\;\;\;\gamma}
       + C^\gamma_{\beta|\alpha} \Pi_{(i)\gamma}^{\;\;\;\;\beta}
       - A_{,\beta} \Pi_{(i)\alpha}^{\;\;\;\;\beta}
       \Bigg\}
       \equiv n_{(i)6\alpha}.
   \label{Mom-conservation-i}
\eea By removing indices indicating the components in Eqs.\
(\ref{E-conservation-i}),(\ref{Mom-conservation-i}) we recover
equations for the collective component which coincide with the
equations in a single component situation in Eqs.\
(\ref{E-conservation}),(\ref{Mom-conservation}).

To the background order, from Eqs.\
(\ref{E-conservation-i}),(\ref{Raychaudhury-eq}) we have \bea
   & & \dot \mu_{(i)}
       + 3 H \left( \mu_{(i)} + p_{(i)} \right)
       = - {c \over a} I_{(i)0},
   \label{BG-eq1} \\
   & & \dot \mu
       + 3 H \left( \mu + p \right)
       = 0,
   \label{BG-eq2} \\
   & & 3 \dot H + 3 H^2
       = - {4 \pi G \over c^2} \sum_j \left( \mu_{(j)} + 3 p_{(j)} \right)
       + \Lambda c^2,
   \label{BG-eq3} \\
   & & H^2 = {8 \pi G \over 3 c^2} \sum_j \mu_{(j)}
       - {K c^2 \over a^2} + {\Lambda c^2 \over 3},
   \label{BG-eq4}
\eea where we recovered the speed of light $c$. Dimensions are \bea
   & &
       [G \varrho] = T^{-2}, \quad
       [c] = LT^{-1}, \quad
       [\eta] = 1, \quad
       [p] = [\mu] = [\varrho c^2], \quad
       [a] = L, \quad
       [K] = 1, \quad
       [\Lambda] = L^{-2}, \quad
       [I_{(i)0}] = [\mu].
\eea Equation (\ref{BG-eq2}) follows from the sum of Eq.\
(\ref{BG-eq1}) over components. Equation (\ref{BG-eq4}) follows from
integrating Eq.\ (\ref{BG-eq3}) where $K$-term can be regarded as an
integration constant; in Einstein's gravity $K$-term can be
normalized as the sign of spatial curvature. Compared with the
Newtonian background equations in Eq.\ (\ref{BG-N-eqs}), ignoring
the direct interaction terms in Eq.\ (\ref{BG-eq1}), the presence of
pressure terms in Eqs.\ (\ref{BG-eq1})-(\ref{BG-eq4}) is the pure
general relativistic effect. The cosmological constant $\Lambda$ can
be introduced by hand even in the Newtonian case.

%%%%%%%%%%%%%%%%%%%%%%%%%%%%%%%%%%%%%%%%%%%%%%%%%%%%%%%%%%%%%%
\subsection{Decomposition}

We decompose the metric to three perturbation types \bea
   & & A \equiv \alpha, \quad
       B_\alpha \equiv \beta_{,\alpha} + B^{(v)}_\alpha, \quad
       C_{\alpha\beta} \equiv \varphi g^{(3)}_{\alpha\beta}
       + \gamma_{,\alpha|\beta}
       + C^{(v)}_{(\alpha|\beta)}
       + C^{(t)}_{\alpha\beta},
\eea where superscripts $(v)$ and $(t)$ indicate the transverse
vector-type, and transverse-tracefree tensor-type perturbations,
respectively. We introduce \bea
   & & \chi \equiv a \left( \beta + c^{-1} a \dot \gamma \right), \quad
       \Psi^{(v)}_\alpha \equiv B^{(v)}_\alpha + c^{-1} a \dot
       C^{(v)}_\alpha,
   \label{chi-def}
\eea which are spatially gauge-invariant to the linear order. We set
\bea
   & & K^\alpha_\alpha \equiv - 3 H + \kappa.
\eea We will identify $\kappa$ with Newtonian velocity variable
which will be an important step in our analysis, see Eqs.\
(\ref{kappa-identification}),(\ref{identify-second}),(\ref{identification-K1}).

For the fluid quantities we decompose \bea
   & & \tilde u_{(i)\alpha}
       = a \left( V_{(i)\alpha} - B_\alpha
       + A B_\alpha
       + 2 V_{(i)}^\beta C_{\alpha\beta} \right)
       \equiv a v_{(i)\alpha}
       \equiv a \left( - v_{(i),\alpha} + v_{(i)\alpha}^{(v)}
       \right),
   \label{u-decomp} \\
   & & \Pi_{(i)\alpha\beta}
       \equiv {1 \over a^2} \left( \Pi_{(i),\alpha|\beta}
       - {1 \over 3} g^{(3)}_{\alpha\beta} \Delta \Pi_{(i)} \right)
       + {1 \over a} \Pi^{(v)}_{(i)(\alpha|\beta)}
       + \Pi^{(t)}_{(i)\alpha\beta}, \quad
       \delta I_{(i)\alpha} \equiv \delta I_{(i),\alpha}
       + \delta I_{(i)\alpha}^{(v)}.
\eea The perturbed fluid velocity variables $v_{(i)}$ and
$v_{(i)\alpha}^{(v)}$ subtly differ from the ones introduced in
\cite{NL}; see Sec.\ \ref{sec:CG}. For the collective fluid
component or for a single component case, we simply delete ${(i)}$
subindices. For isotropic pressure we introduce \bea
   & & \delta p \equiv c_s^2 \delta \mu + e, \quad
       c_s^2 \equiv {\dot p \over \dot \mu}.
\eea The perturbation variable $e$ is called an entropic
perturbation. Defined in this way $e$ is gauge-invariant only to the
linear order. To the second order, from Eq.\ (\ref{GT}) we can
derive the following gauge-invariant combination \bea
   & & \delta p_{\delta \mu}
       \equiv e - {\delta \mu \over \dot \mu}
       \left[ \dot e + {1 \over 2} \left( c_s^2 \right)^\cdot \delta
       \mu \right].
\eea In our notation, $\delta p_{\delta \mu}$ is a gauge-invariant
combination which is the same as $\delta p$ in the $\delta \mu = 0$
slicing (temporal gauge) condition to the second order; as the
spatial gauge we take $\gamma \equiv 0$ to the second order; for the
derivation, see the prescription below Eq.\ (266) of \cite{NL}. In
the multi-component case, we similarly have \bea
   & & \delta p_{(i)\delta \mu_{(i)}}
       \equiv e_{(i)} - {\delta \mu_{(i)} \over \dot \mu_{(i)}}
       \left[ \dot e_{(i)} + {1 \over 2} \left( c_{(i)}^2 \right)^\cdot
       \delta \mu_{(i)} \right],
\eea where \bea
   & & \delta p_{(i)} \equiv c_{(i)}^2 \delta \mu_{(i)} + e_{(i)}, \quad
       c_{(i)}^2 \equiv {\dot p_{(i)} \over \dot \mu_{(i)}}.
\eea We have \cite{Kodama-Sasaki-1984} \bea
   & & e = e_{\rm rel} + e_{\rm int}, \quad
       e_{\rm rel} \equiv \sum_j \left( c_{(j)}^2 - c_s^2 \right)
       \delta \mu_{(j)}, \quad
       e_{\rm int} \equiv \sum_j e_{(j)}.
\eea Equation (\ref{sum-1}) gives \bea
   & & \delta \mu
       = \sum_j \left[
       \delta \mu_{(j)}
       + \left( \mu_{(j)} + p_{(j)} \right) v_{(j)}^\alpha
       \left( v_{(j)\alpha} - v_\alpha \right) \right],
   \nonumber \\
   & & \delta p
       = \sum_j \left[
       \delta p_{(j)}
       + {1 \over 3} \left( \mu_{(j)} + p_{(j)} \right) v_{(j)}^\alpha
       \left( v_{(j)\alpha} - v_\alpha \right) \right],
   \nonumber \\
   & & \left( \mu + p \right) v_\alpha
       = \sum_j \left[ \left( \mu_{(j)} + p_{(j)} \right) v_{(j)\alpha}
       + \left( \delta \mu_{(j)} + \delta p_{(j)} \right)
       \left( v_{(j)\alpha} - v_\alpha \right)
       + \Pi^{\;\;\;\;\beta}_{(j)\alpha}
       \left( v_{(j)\beta} - v_\beta \right) \right],
   \nonumber \\
   & & \Pi^\alpha_\beta
       = \sum_j \left\{ \Pi^{\;\;\;\;\alpha}_{(j)\beta}
       + \left( \mu_{(j)} + p_{(j)} \right)  \left[ v_{(j)}^\alpha
       \left( v_{(j)\beta} - v_\beta \right)
       - {1 \over 3} \delta^\alpha_\beta v_{(j)}^\gamma
       \left( v_{(j)\gamma} - v_\gamma \right) \right] \right\},
   \label{sum}
\eea

By recovering $c$, dimensions of the variables are \bea
   & & [\tilde g_{ab}] = [\tilde g^{ab}] = [\tilde u_a] = 1, \quad
       [\tilde T_{ab}] = [\mu], \quad
       [ g^{(3)}_{\alpha\beta} ] = 1, \quad
       [\nabla] = [\Delta] = 1, \quad
       [w] = [c_s^2] = 1,
   \nonumber \\
   & & [A] = [B_\alpha] = [C_{\alpha\beta}] = 1, \quad
       [\alpha] = [\varphi] = [\beta] = [\gamma]
       = [B^{(v)}_\alpha] = [C^{(v)}_\alpha] = [\Psi^{(v)}_\alpha]
       = [C^{(t)}_{\alpha\beta}] = 1, \quad
       [\chi] = L, \quad
       [\kappa] = T^{-1},
   \nonumber \\
   & & [\delta \mu] = [\delta p] = [e] = [\Pi_{\alpha\beta}]
       = [\Pi^{(t)}_{\alpha\beta}] = [\mu], \quad
       [\delta] = [V_\alpha] = 1, \quad
       [v] = [v^{(v)}_\alpha] = 1, \quad
       [\Pi] = L^2 [\mu], \quad
       [\Pi^{(v)}_\alpha] = L [\mu].
\eea

Scalar-type perturbation equations can be derived from Eqs.\
(\ref{def-of-kappa})-(\ref{Mom-conservation-i}) \bea
   & & \kappa - 3 H \alpha + 3 \dot \varphi
       + {\Delta \over a^2} \chi = n_0,
   \label{eq1} \\
   & & 4 \pi G \delta \mu + H \kappa
       + {\Delta + 3 K \over a^2} \varphi
       = {1 \over 4} n_1,
   \label{eq2} \\
   & & \kappa + {\Delta + 3 K \over a^2} \chi
       - 12 \pi G \left( \mu + p \right) a v
       = n_2 \equiv {3 \over 2} \Delta^{-1} \nabla^\alpha
       n_{2\alpha},
   \label{eq3} \\
   & & \dot \kappa + 2 H \kappa
       - 4 \pi G \left( \delta \mu + 3 \delta p \right)
       + \left( 3 \dot H + {\Delta \over a^2} \right) \alpha
       = n_3,
   \label{eq4} \\
   & & \dot \chi + H \chi - \varphi - \alpha
       - 8 \pi G \Pi
       = n_4 \equiv {3 \over 2} a^2 \left( \Delta + 3 K \right)^{-1}
       \Delta^{-1} \nabla^\alpha \nabla_\beta
       n_{4\alpha}^{\;\;\beta},
   \label{eq5} \\
   & & \delta \dot \mu
       + 3 H \left( \delta \mu + \delta p \right)
       - \left( \mu + p \right) \left( \kappa - 3 H \alpha
       + {\Delta \over a} v \right)
       = n_5,
   \label{eq6} \\
   & & {\left[ a^4 \left( \mu + p \right) v \right]^\cdot
       \over a^4 \left( \mu + p \right)}
       - {1 \over a} \alpha
       - {1 \over a \left( \mu + p \right)} \left(
       \delta p + {2 \over 3} {\Delta + 3 K \over a^2} \Pi \right)
       = n_6 \equiv - {1 \over \mu + p} \Delta^{-1} \nabla^\alpha
       n_{6\alpha},
   \label{eq7} \\
   & & \delta \dot \mu_{(i)}
       + 3 H \left( \delta \mu_{(i)} + \delta p_{(i)} \right)
       - \left( \mu_{(i)} + p_{(i)} \right) \left( \kappa - 3 H \alpha
       + {\Delta \over a} v_{(i)} \right)
       + {1 \over a} \delta I_{(i)0}
       = n_{5(i)},
   \label{eq6-i} \\
   & & {\left[ a^4 \left( \mu_{(i)} + p_{(i)} \right) v_{(i)} \right]^\cdot
       \over a^4 \left( \mu_{(i)} + p_{(i)} \right)}
       - {1 \over a} \alpha
       - {1 \over a \left( \mu_{(i)} + p_{(i)} \right)} \left(
       \delta p_{(i)} + {2 \over 3} {\Delta + 3 K \over a^2} \Pi_{(i)}
       - \delta I_{(i)} \right)
   \nonumber \\
   & & \qquad
       = n_{6(i)}
       \equiv - {1 \over \mu_{(i)} + p_{(i)}} \Delta^{-1} \nabla^\alpha
       n_{6\alpha(i)}.
   \label{eq7-i}
\eea Equations for the vector-type perturbation follow from Eqs.\
(\ref{Mom-constraint}),(\ref{tracefree-ADM-propagation}),(\ref{Mom-conservation}),(\ref{Mom-conservation-i})
\bea
   & & {\Delta + 2 K \over 2 a^2} \Psi^{(v)}_\alpha
       + 8 \pi G (\mu + p) v^{(v)}_\alpha
       = {1 \over a} \left( n_{2\alpha}
       - \nabla_\alpha \Delta^{-1} \nabla^{\beta} n_{2\beta} \right)
       \equiv n_{2\alpha}^{({v})},
   \label{vector-2} \\
   & & \dot \Psi^{(v)}_\alpha + 2 H \Psi^{(v)}_\alpha
       - 8 \pi G \Pi^{(v)}_\alpha
       = 2 a \left( \Delta + 2 K \right)^{-1} \left(
       \nabla_\beta n_{4\alpha}^{\;\;\beta}
       - \nabla_\alpha \Delta^{-1}
       \nabla^{\gamma} \nabla_\beta n_{4\gamma}^{\;\;\beta} \right)
       \equiv n_{4\alpha}^{(v)},
   \label{vector-4} \\
   & & {[a^4 (\mu + p) v^{(v)}_\alpha]^\cdot \over a^4 (\mu + p)}
       + {\Delta + 2K \over 2 a^2} {\Pi^{(v)}_\alpha \over \mu + p}
       = {1 \over \mu + p} \left( n_{6\alpha}
       - \nabla_\alpha \Delta^{-1} \nabla^{\beta} n_{6\beta} \right)
       \equiv n_{6\alpha}^{({v})},
   \label{vector-6} \\
   & & {[a^4 (\mu_{(i)} + p_{(i)}) v^{(v)}_{(i)\alpha}]^\cdot
       \over a^4 (\mu_{(i)} + p_{(i)})}
       + {\Delta + 2K \over 2 a^2}
       {\Pi^{(v)}_{(i)\alpha} \over \mu_{(i)} + p_{(i)}}
       - {1 \over a} {\delta I^{(v)}_{(i)\alpha} \over \mu_{(i)} + p_{(i)}}
       = {1 \over \mu_{(i)} + p_{(i)}} \left( n_{6(i)\alpha}
       - \nabla_\alpha \Delta^{-1} \nabla^{\beta} n_{6(i)\beta} \right)
       \equiv n_{6(i)\alpha}^{({v})}.
   \label{vector-6-i}
\eea Equations for the tensor-type perturbation follow from Eq.\
(\ref{tracefree-ADM-propagation}) \bea
   & & \ddot C^{({t})}_{\alpha\beta}
       + 3 H \dot C^{({t})}_{\alpha\beta}
       - {\Delta - 2K \over a^2} C^{({t})}_{\alpha\beta}
       - 8 \pi G \Pi^{({t})}_{\alpha\beta}
   \nonumber \\
   & & \qquad
       = n_{4\alpha\beta}
       - {3 \over 2} \left( \nabla_{\alpha} \nabla_\beta
       - {1\over 3} g^{(3)}_{\alpha\beta} \Delta \right)
       \left( \Delta + 3K \right)^{-1}
       \Delta^{-1} \nabla^{\gamma} \nabla_\delta
       n_{4\gamma}^{\;\;\delta}
   \nonumber \\
   & & \qquad \qquad
       - 2 \nabla_{(\alpha} \left( \Delta + 2K \right)^{-1}
       \left( \nabla^{\gamma} n_{4\beta)\gamma}
       - \nabla_{\beta)} \Delta^{-1} \nabla^{\gamma}
       \nabla_\delta n_{4\gamma}^{\;\;\delta} \right)
       \equiv n_{4\alpha\beta}^{(t)}.
   \label{tensor-4}
\eea
          In order to derive eqs. (\ref{eq5},\ref{vector-4},\ref{tensor-4})
          it is convenient to show
          \bea
             & & {1 \over a^2} \left( \nabla_\alpha \nabla_\beta
                 - {1 \over 3} g^{(3)}_{\alpha\beta} \Delta \right)
                 \left( \dot \chi + H \chi - \varphi - \alpha
                 - 8 \pi G \Pi \right)
                 + {1 \over a^3}
                 \left( a^2 \Psi^{(v)}_{(\alpha|\beta)} \right)^\cdot
                 - 8 \pi G {1 \over a} \Pi^{(v)}_{(\alpha|\beta)}
             \nonumber \\
             & & \qquad
                 + \ddot C^{({t})}_{\alpha\beta}
                 + 3 H \dot C^{({t})}_{\alpha\beta}
                 - {\Delta - 2 K \over a^2} C^{({t})}_{\alpha\beta}
                 - 8 \pi G \Pi^{({t})}_{\alpha\beta}
                = n_{4\alpha\beta},
             \label{chi-eq}
          \eea
          which follows from Eq.\ (\ref{tracefree-ADM-propagation}).
Quadratic combinations of linear-order perturbation variables of all
three types of perturbations contribute to all three types of
perturbation to the second order.

%%%%%%%%%%%%%%%%%%%%%%%%%%%%%%%%%%%%%%%%%%%%%%%%%%%%%%%%%%%%%%
\subsection{Comoving gauge and irrotational condition}
                                       \label{sec:CG}

In Eq.\ (180) of \cite{NL} we introduced \bea
   & & Q_{(i)\alpha} \equiv \left( \mu_{(i)} + p_{(i)} \right)
       \left( - \bar v_{(i),\alpha} + \bar v_{(i)\alpha}^{(v)}
       \right),
   \label{Q-decomp}
\eea where we put overbars to $\bar v_{(i)}$ and $\bar
v_{(i)\alpha}^{(v)}$ in order to distinguish these from our new
notations to be used in this paper; in \cite{NL} we didn't have
overbars. From Eq.\ (\ref{normal-to-energy-frame}) we have \bea
   & & - \bar v_{(i),\alpha} + \bar v_{(i)\alpha}^{(v)}
       \equiv V_{(i)\alpha} - B_\alpha
       + A B_\alpha
       + 2 V_{(i)}^\beta C_{\alpha\beta}
       + {\delta \mu_{(i)} + \delta p_{(i)} \over \mu_{(i)} + p_{(i)}}
       \left( V_{(i)\alpha} - B_\alpha \right)
       + {\Pi_{(i)\alpha\beta} \over \mu_{(i)} + p_{(i)}}
       \left( V^{\beta}_{(i)} - B^\beta \right).
\eea It is more convenient to introduce the decomposition in Eq.\
(\ref{u-decomp}). Thus, we have \bea
   & & - \bar v_{(i),\alpha} + \bar v_{(i)\alpha}^{(v)}
       \equiv - v_{(i),\alpha} + v_{(i)\alpha}^{(v)}
       + {\delta \mu_{(i)} + \delta p_{(i)} \over \mu_{(i)} + p_{(i)}}
       \left( - v_{(i),\alpha} + v_{(i)\alpha}^{(v)} \right)
       + {\Pi_{(i)\alpha\beta} \over \mu_{(i)} + p_{(i)}}
       \left( - v_{(i)}^{,\beta} + v_{(i)}^{(v)\beta}\right).
\eea The variable $\bar v_{(i)\alpha}^{(v)}$ introduced in \cite{NL}
cannot be regarded as a proper vector-type perturbation. However, if
we also take the temporal comoving gauge in \cite{NL} which sets
$\bar v \equiv 0$ together with $\bar v_{(i)\alpha}^{(v)} = 0$, we
have $v = 0 = v_{(i)\alpha}^{(v)}$; these are the same as taking the
proper irrotational condition ($v_{(i)\alpha}^{(v)} = 0$) and the
temporal comoving gauge ($v = 0$). Our analyses in
\cite{second-order,third-order-PRD} are, in fact, based on taking
these two conditions together. In the irrotational fluids, the
temporal comoving gauge $v \equiv 0$ leads to $\tilde u_\alpha = 0$,
thus $\tilde u_a$ coincides with the normal frame four-vector
$\tilde n_a$.

From Eq.\ (\ref{kinematic-quantities-pert}) we have \bea
   & & \tilde \omega_{(i)\alpha\beta} = a v_{(i)[\alpha|\beta]}^{(v)}
       + {a \over \mu_{(i)} + p_{(i)}}
       \left( - v_{(i),[\alpha} + v_{(i)[\alpha}^{(v)} \right)
       \left[ \left( \delta p_{(i)}
       + {2 \over 3} {\Delta + 3 K \over a^2} \Pi_{(i)}
       - \delta I_{(i)} \right)_{,\beta]}
       + {\Delta + 3 K \over 2 a} \Pi_{(i)\beta]}^{(v)}
       - \delta I_{(i)\beta]}^{(v)} \right],
   \nonumber \\
\eea where we used Eq.\ (\ref{Mom-conservation-i}) to the linear
order. For vanishing vector-type perturbation we set
$v_{(i)\alpha}^{(v)} \equiv 0$, etc. In this case, we have \bea
   & & \tilde \omega_{(i)\alpha\beta} = - {a \over \mu_{(i)} + p_{(i)}}
       v_{(i),[\alpha}
       \left( \delta p_{(i)}
       + {2 \over 3} {\Delta + 3 K \over a^2} \Pi_{(i)}
       - \delta I_{(i)} \right)_{,\beta]},
\eea which vanishes for $\delta p_{(i)} = 0 = \Pi_{(i)}$ and $\delta
I_{(i)} = 0$.

%%%%%%%%%%%%%%%%%%%%%%%%%%%%%%%%%%%%%%%%%%%%%%%%%%%%%%%%%%%%%%%
%
%  Effects of pressure
%
%%%%%%%%%%%%%%%%%%%%%%%%%%%%%%%%%%%%%%%%%%%%%%%%%%%%%%%%%%%%%%
\section{Effects of pressure}
                                      \label{sec:pressure}

We consider a single component situation without rotational
perturbations. Equations
(\ref{Raychaudhury-eq}),(\ref{E-conservation}),(\ref{Mom-conservation}),
and Eq.\ (\ref{Mom-constraint}) to the linear order provide a
complete set of equations we need in the following.

%%%%%%%%%%%%%%%%%%%%%%%%%%%%%%%%%%%%%%%%%%%%%%%%%%%%%%%%%%%%%%
\subsection{Irrotational case}

As an irrotational fluid we {\it ignore} all vector-type
perturbations, thus $v^{(v)}_\alpha = B^{(v)}_\alpha =
C^{(v)}_\alpha = \Pi^{(v)}_\alpha = 0$. Quadratic combinations of
linear-order perturbations of all three-types of perturbations
contribute to each type of second-order perturbations. Thus,
concerning the second-order scalar-type and vector-type
perturbations, ignoring the pure vector-type perturbation
corresponds to ignoring the vector-type perturbation only to the
linear order; to the linear order the vector-type perturbation only
has a decaying solution in the expanding phase. Assuming the
background equations are valid Eqs.\
(\ref{E-conservation}),(\ref{Mom-conservation}),(\ref{Raychaudhury-eq}),(\ref{Mom-constraint})
give \bea
   & & \dot \delta
       + 3 H \left( {\delta p \over \mu} - w \delta \right)
       - \left( 1 + w \right) \left( \kappa
       + {\Delta \over a} v
       - 3 H \alpha \right)
       = - {1 \over a} \delta_{,\alpha} \beta^{,\alpha}
       + \left( \delta + {\delta p \over \mu} \right)
       \left( \kappa - 3 H \alpha \right)
       + \left( 1 + w \right) \left( \kappa + {\Delta \over a} v
       \right) \alpha
   \nonumber \\
   & & \qquad
       + {3 \over 2} H \left( 1 + w \right)
       \left( \alpha^2 - \beta^{,\alpha} \beta_{,\alpha} \right)
       - {1 \over \mu} \Pi^{\alpha \beta}
       \left( {1 \over a^2} \chi_{,\alpha|\beta}
       + \dot C^{(t)}_{\alpha\beta} \right)
       - {2 \over a} \left( 1 + w \right) v^{,\alpha|\beta}
       \left( \varphi g^{(3)}_{\alpha\beta}
       + \gamma_{,\alpha|\beta}
       + C^{(t)}_{\alpha\beta} \right)
   \nonumber \\
   & & \qquad
       + {1 \over a} \left( 1 + w \right) v^{,\alpha}
       \left\{ a \left( H + {\dot p \over \mu + p} \right)
       v_{,\alpha}
       - {2 \over \mu + p} \left( \delta p_{,\alpha}
       + \Pi^\beta_{\alpha|\beta} \right)
       + \varphi_{,\alpha}
       - \left[ \left( \Delta + 4 K \right) \gamma \right]_{,\alpha}
       \right\}
   \nonumber \\
   & & \qquad
       + {1 \over a \mu} \left[ \left( \delta \mu + \delta p \right) v^{,\alpha}
       + \Pi^{\alpha\beta} v_{,\beta} \right]_{|\alpha},
   \label{E-conservation-single-irrotational} \\
   & & {1 \over a^4} \left[ a^4 \left( \mu + p \right) v_{,\alpha}
       \right]^\cdot
       - {1 \over a} \left[ \left( \mu + p \right) \alpha_{,\alpha}
       + \delta p_{,\alpha}
       + \Pi^\beta_{\alpha|\beta} \right]
       = {1 \over a} \left( \mu + p \right)
       \left( v - \beta \right)^{,\beta}
       \left( v - \beta \right)_{,\alpha|\beta}
   \nonumber \\
   & & \qquad
       + \left( \mu + p \right) v_{,\alpha}
       \left( \kappa + {\Delta \over a} v
       - 3 H \alpha \right)
       - {1 \over a^4} \left\{ a^4 \left[ \left( \delta \mu +
       \delta p \right) v_{,\alpha}
       + \Pi^\beta_\alpha v_{,\beta} \right] \right\}^\cdot
   \nonumber \\
   & & \qquad
       - {1 \over a} \Bigg\{
       - \left( \delta \mu + \delta p \right) \alpha_{,\alpha}
       + \alpha \left[ \left( \mu + p \right) \alpha_{,\alpha}
       - \delta p_{,\alpha}
       - \Pi^\beta_{\alpha|\beta} \right]
       - \alpha_{,\beta} \Pi^\beta_\alpha
       + 2 \left( C^{\gamma\beta} \Pi_{\alpha\gamma}
       \right)_{|\beta}
       - \Pi^\gamma_\alpha C^\beta_{\beta|\gamma}
       + \Pi^\beta_\gamma C^\gamma_{\beta|\alpha}
       \Bigg\},
   \label{Mom-conservation-single-irrotational} \\
   & & \dot \kappa + 2 H \kappa
       - 4 \pi G \left( \delta \mu + 3 \delta p \right)
       + \left( 3 \dot H + {\Delta \over a^2} \right) \alpha
       = - {1 \over a} \kappa_{,\alpha} \beta^{,\alpha}
       + \alpha \dot \kappa
       + {1 \over 3} \kappa^2
       + {3 \over 2} \dot H \left( 3 \alpha^2
       - \beta^{,\alpha} \beta_{,\alpha} \right)
       + 8 \pi G \left( \mu + p \right) v^{,\alpha} v_{,\alpha}
   \nonumber \\
   & & \qquad
       + {1 \over a^2} \left\{
       2 \Delta \alpha \left( \alpha + \varphi \right)
       + \alpha^{,\alpha} \left[ \alpha - \varphi
       + \left( \Delta + 4 K \right) \gamma
       \right]_{,\alpha}
       + 2 \alpha^{,\alpha|\beta} \left( \gamma_{,\alpha|\beta}
       + C^{(t)}_{\alpha\beta} \right)
       - {1 \over 2} \Delta \left( \beta^{,\alpha} \beta_{,\alpha}
       \right)
       \right\}
   \nonumber \\
   & & \qquad
       + \left( {1 \over a^2} \chi^{,\alpha|\beta}
       + \dot C^{(t)\alpha\beta} \right)
       \left( {1 \over a^2} \chi_{,\alpha|\beta}
       + \dot C^{(t)}_{\alpha\beta} \right)
       - {1 \over 3} \left( {\Delta \over a^2} \chi \right)^2,
   \label{Raychaudhury-eq-single-irrotational} \\
   & & \kappa = - {\Delta + 3 K \over a^2} \chi
       + 12 \pi G a \left( \mu + p \right) v.
   \label{Mom-constraint-single-irrotational}
\eea Equation (\ref{Mom-constraint-single-irrotational}) is valid to
the linear order.

%%%%%%%%%%%%%%%%%%%%%%%%%%%%%%%%%%%%%%%%%%%%%%%%%%%%%%%%%%%%%%
\subsection{Comoving gauge}

As the temporal and spatial gauge conditions we set \bea
   & & v \equiv 0 \equiv \gamma,
\eea thus, $\beta = \chi/a$. Under these gauge conditions Eq.\
(\ref{Mom-conservation-single-irrotational}) gives \bea
   & & \alpha
       = - {1 \over 2 a^2} \chi^{,\alpha} \chi_{,\alpha}
       - {c_s^2 \over 1 + w} \delta \left[
       1 - {1 + 3 c_s^2 \over 2(1 + w)} \delta \right]
       + \alpha_\Pi,
   \label{Mom-conservation-single-irotational-CG}
\eea where the imperfect fluid contributions (stresses) are denoted
by $\alpha_\Pi$ with \bea
   & & \alpha_\Pi
       \equiv {1 \over \mu + p} \left( e + {2 \over 3}
       {\Delta + 3 K \over a^2} \Pi \right)
       \left[ - 1 + {2 c_s^2 \over 1 + w} \delta
       + {1 \over \mu + p} \left( e + {2 \over 3}
       {\Delta + 3 K \over a^2} \Pi \right) \right]
   \nonumber \\
   & & \qquad
       + {1 \over \mu + p} \Delta^{-1} \nabla^\alpha \Bigg[
       {1 + c_s^2 \over 1 + w} \delta \left( e_{,\alpha}
       + \Pi^\beta_{\alpha|\beta} \right)
       - \left( {c_s^2 \over 1 + w}
       \delta_{,\beta}
       + {e_{,\beta} + \Pi^\gamma_{\beta|\gamma} \over \mu + p}
       \right) \Pi^\beta_\alpha
       + 2 \left( C^{\beta\gamma} \Pi_{\alpha\gamma}
       \right)_{|\beta}
       - \Pi^\gamma_\alpha C^\beta_{\beta|\gamma}
       + \Pi^\beta_\gamma C^\gamma_{\beta|\alpha} \Bigg].
   \nonumber \\
\eea Using this, Eqs.\
(\ref{E-conservation-single-irrotational})-(\ref{Mom-constraint-single-irrotational})
give \bea
   & & \dot \delta - 3 H w \delta - \left( 1 + w \right) \kappa
       = - { 1\over a^2} \delta_{,\alpha} \chi^{,\alpha}
       + \kappa \delta
       + {3 \over 2} H {c_s^2 \over 1 + w} \delta^2
       + \delta_\Pi,
   \label{E-conservation-single-irrotational-CG} \\
   & & \dot \kappa + 2 H \kappa
       - 4 \pi G \mu \delta
       - {c_s^2 \over 1 + w} {\Delta + 3 K \over a^2} \delta
       = - {1 \over a^2} \kappa_{,\alpha} \chi^{,\alpha}
       + {1 \over 3} \kappa^2
       + {c_s^2 \over 2 (1 + w)}
       \left( 4 \pi G \mu
       - {1 + 2 c_s^2 \over 1 + w} {\Delta + 3 K \over a^2} \right)
       \delta^2
   \nonumber \\
   & & \qquad
       + {c_s^2 \over 1 + w} \left[ 2 H \delta \kappa
       + {1 \over a^2} \left( \varphi^{,\alpha} \delta_{,\alpha}
       - 2 \varphi \Delta \delta
       - 2 \delta^{,\alpha|\beta} C^{(t)}_{\alpha\beta} \right)
       \right]
       + \left( {1 \over a^2} \chi^{,\alpha|\beta}
       + \dot C^{(t)\alpha\beta} \right)
       \left( {1 \over a^2} \chi_{,\alpha|\beta}
       + \dot C^{(t)}_{\alpha\beta} \right)
       - {1 \over 3} \left( {\Delta \over a^2} \chi \right)^2
   \nonumber \\
   & & \qquad
       + \kappa_\Pi,
   \label{Raychaudhury-eq-single-irrotational-CG} \\
   & & \kappa = - {\Delta + 3 K \over a^2} \chi,
   \label{Mom-constraint-single-irrotational-CG}
\eea where Eq.\ (\ref{Mom-constraint-single-irrotational-CG}) is
valid only to the linear order, and \bea
   & & \delta_\Pi
       \equiv 2 H {\Delta + 3 K \over a^2} {\Pi \over \mu}
       + {e \over \mu} \left( \kappa + 3 H {c_s^2 \over 1 + w}
       \delta \right)
       - {\Pi^{\alpha\beta} \over \mu} \left( {1 \over a^2}
       \chi_{,\alpha|\beta} + \dot C^{(t)}_{\alpha\beta} \right)
   \nonumber \\
   & & \qquad
       + {1 \over \mu + p} \left( e + {2 \over 3}
       {\Delta + 3 K \over a^2} \Pi \right)
       \left[ 3 H \delta - \left( 1 + w \right) \kappa
       + {3 \over 2} H {1 \over \mu} \left( e - {2 \over 3}
       {\Delta + 3 K \over a^2} \Pi \right) \right]
   \nonumber \\
   & & \qquad
       - 3 H {1 \over \mu} \Delta^{-1} \nabla^\alpha \Bigg[
       {1 + c_s^2 \over 1 + w} \delta \left( e_{,\alpha}
       + \Pi^\beta_{\alpha|\beta} \right)
       - \left( {c_s^2 \over 1 + w}
       \delta_{,\beta}
       + {e_{,\beta} + \Pi^\gamma_{\beta|\gamma} \over \mu + p}
       \right) \Pi^\beta_\alpha
       + 2 \left( C^{\beta\gamma} \Pi_{\alpha\gamma}
       \right)_{|\beta}
       - \Pi^\gamma_\alpha C^\beta_{\beta|\gamma}
       + \Pi^\beta_\gamma C^\gamma_{\beta|\alpha} \Bigg],
   \nonumber \\
   & & \kappa_\Pi
       \equiv 12 \pi G e \left( 1 - {c_s^2 \over 1 + w} \delta
       \right)
   \nonumber \\
   & & \qquad
       - \left( 3 \dot H + {\Delta \over a^2} \right)
       \Bigg\{
       {1 \over \mu + p} \left( e + {2 \over 3}
       {\Delta + 3 K \over a^2} \Pi \right)
       \left[ -1 + {2 c_s^2 \over 1 + w} \delta
       + {1 \over \mu + p} \left( e + {2 \over 3}
       {\Delta + 3 K \over a^2} \Pi \right) \right]
   \nonumber \\
   & & \qquad
       + {1 \over \mu + p} \Delta^{-1} \nabla^\alpha \Bigg[
       {1 + c_s^2 \over 1 + w} \delta \left( e_{,\alpha}
       + \Pi^\beta_{\alpha|\beta} \right)
       - \left( {c_s^2 \over 1 + w}
       \delta_{,\beta}
       + {e_{,\beta} + \Pi^\gamma_{\beta|\gamma} \over \mu + p}
       \right) \Pi^\beta_\alpha
   \nonumber \\
   & & \qquad
       + 2 \left( C^{\beta\gamma} \Pi_{\alpha\gamma}
       \right)_{|\beta}
       - \Pi^\gamma_\alpha C^\beta_{\beta|\gamma}
       + \Pi^\beta_\gamma C^\gamma_{\beta|\alpha} \Bigg]
       \Bigg\}
   \nonumber \\
   & & \qquad
       - \left[
       - 2 H \kappa
       + 4 \pi G \left( 1 + 6 c_s^2 \right) \mu \delta
       + 12 \pi G e
       - {c_s^2 \over 1 + w} {3 K \over a^2} \delta
       - {3 \over 2} \dot H {1 \over \mu + p} \left( e + {2 \over 3}
       {\Delta + 3 K \over a^2} \Pi \right)
       + 2 \varphi {\Delta \over a^2} \right]
   \nonumber \\
   & & \qquad
       \times
       {1 \over \mu + p} \left( e + {2 \over 3}
       {\Delta + 3 K \over a^2} \Pi \right)
       + {1 \over \mu + p} {\Delta \over a^2} \left\{
       \left( e + {2 \over 3} {\Delta + 3 K \over a^2} \Pi \right)
       \left[ {1 \over 2} {1 \over \mu + p} \left( e + {2 \over 3}
       {\Delta + 3 K \over a^2} \Pi \right)
       + {c_s^2 \over 1 + w} \delta \right] \right\}
   \nonumber \\
   & & \qquad
       + {1 \over \mu + p} {1 \over a^2} \left[
       \left( e + {2 \over 3}
       {\Delta + 3 K \over a^2} \Pi \right)^{,\alpha}
       \varphi_{,\alpha}
       - 2 \left( e + {2 \over 3}
       {\Delta + 3 K \over a^2} \Pi \right)^{,\alpha|\beta}
       C^{(t)}_{\alpha\beta} \right].
\eea Combining Eqs.\
(\ref{E-conservation-single-irrotational-CG}),(\ref{Raychaudhury-eq-single-irrotational-CG})
we can derive \bea
   & & {1 + w \over a^2 H}
       \left[ {H^2 \over (\mu + p) a} \left( {a^3 \mu \over H}
       \delta \right)^\cdot \right]^\cdot
       - c_s^2 {\Delta \over a^2} \delta
       = \left( 1 + w \right) \left\{
       {1 \over a^4} \chi^{,\alpha|\beta} \chi_{,\alpha|\beta}
       - {1 \over a^2} \kappa_{,\alpha} \chi^{,\alpha}
       + {1 \over 3} \left[ \kappa^2
       - \left( {\Delta \over a} \chi \right)^2 \right] \right\}
   \nonumber \\
   & & \qquad
       + {1 + w \over a^2} \left[ {1 \over 1 + w}
       \left( a^2 \kappa \delta
       - \delta_{,\alpha} \chi^{,\alpha}
       + {3 \over 2} a^2 H {c_s^2 \over 1 + w} \delta^2 \right)
       \right]^\cdot
       + {1 \over 2} c_s^2 \left( 4 \pi G \mu
       - {1 + 2 c_s^2 \over 1 + w} {\Delta + 3 K \over a^2} \right)
       \delta^2
   \nonumber \\
   & & \qquad
       + c_s^2 {1 \over a^2} \left( 2 a^2 H \delta \kappa
       + \varphi^{,\alpha} \delta_{,\alpha}
       - 2 \varphi \Delta \delta
       - 2 \delta^{,\alpha|\beta} C^{(t)}_{\alpha\beta} \right)
       + \left( 1 + w \right) \dot C^{(t)\alpha\beta}
       \left( {2 \over a^2} \chi_{,\alpha|\beta}
       + \dot C^{(t)}_{\alpha\beta} \right)
   \nonumber \\
   & & \qquad
       + \left( 1 + w \right) \kappa_\Pi
       + {1 + w \over a^2} \left( {a^2 \over 1 + w} \delta_\Pi \right)^\cdot
       - {1 \over a^2} \delta_{\Pi,\alpha} \chi^{,\alpha}.
   \label{ddot-delta-single-irrotational-CG}
\eea Notice that the equations above are valid in the presence of
general $K$ and $\Lambda$.

%%%%%%%%%%%%%%%%%%%%%%%%%%%%%%%%%%%%%%%%%%%%%%%%%%%%%%%%%%%%%%
\subsection{Newtonian correspondence}

Guided by our success in the zero-pressure case, we continue to {\it
identify} \bea
   & & \kappa \equiv - {1 \over a} \nabla \cdot {\bf u},
   \label{kappa-identification}
\eea to the second order. Using Eq.\
(\ref{Mom-constraint-single-irrotational-CG}), {\it assuming} $K =
0$, we can identify \bea
   & & \kappa = - c {\Delta \over a^2} \chi
       = - {1 \over a} \nabla \cdot {\bf u}, \quad
       {\bf u} \equiv {c \over a} \nabla \chi
       = - c \nabla v_\chi,
\eea to the linear order. We have recovered the speed of light $c$.
We have $[{\bf u}] = LT^{-1}$. Equation
(\ref{Mom-conservation-single-irotational-CG}) becomes \bea
   & & \alpha
       = - {1 \over 2 c^2} {\bf u}^2
       - {c_s^2 \over 1 + w} \delta \left[ 1 - {1 + 3 c_s^2 \over
       2(1 +w)} \delta \right]
       + \alpha_\Pi.
   \label{Mom-conserv-eq4}
\eea Equations (\ref{E-conservation-single-irrotational-CG}),
(\ref{Raychaudhury-eq-single-irrotational-CG}) give \bea
   & & \dot \delta - 3 w H \delta
       + \left( 1 + w \right) {1 \over a} \nabla \cdot {\bf u}
       = - {1 \over a} \nabla \cdot \left( \delta {\bf u} \right)
       + {3 \over 2} {c_s^2 \over 1 + w} H \delta^2
       + \delta_\Pi,
   \label{E-conserv-eq4} \\
   & & {1 \over a} \nabla \cdot \left( \dot {\bf u} + H {\bf u}
       \right)
       + {4 \pi G \mu \over c^2} \delta
       + {c_s^2 \over 1 + w} c^2 {\Delta \over a^2} \delta
       = - {1 \over a^2} \nabla \cdot \left( {\bf u} \cdot \nabla
       {\bf u} \right)
       - \dot C^{(t)\alpha\beta}
       \left( {2 \over a} u_{\alpha|\beta}
       + \dot C^{(t)}_{\alpha\beta} \right)
   \nonumber \\
   & & \qquad
       + {c_s^2 \over 1 + w} \left\{
       {1 \over 2} \left( - {4 \pi G \mu \over c^2}
       + {1 + 2 c_s^2 \over 1 + w} c^2 {\Delta \over a^2} \right) \delta^2
       + 2 H \delta {1 \over a} \nabla \cdot {\bf u}
       + {c^2 \over a^2} \left[ 2 \varphi \Delta \delta
       - \left( \nabla \varphi \right) \cdot \nabla \delta
       + 2 \delta^{,\alpha|\beta} C^{(t)}_{\alpha \beta} \right]
       \right\}
       - \kappa_\Pi.
   \label{Raychaudhury-eq4}
\eea We have $[\alpha_\Pi] = 1$, $[\delta_\Pi] = T^{-1}$, and
$[\kappa_\Pi] = T^{-2}$. Combining these equations or Eq.\
(\ref{ddot-delta-single-irrotational-CG}) gives \bea
   & & {1 + w \over a^2 H}
       \left[ {H^2 \over (\mu + p) a} \left( {a^3 \mu \over H}
       \delta \right)^\cdot \right]^\cdot
       - c_s^2 c^2 {\Delta \over a^2} \delta
   \nonumber \\
   & & \qquad
       = {1 + w \over a^2}
       \nabla \cdot \left( {\bf u} \cdot \nabla {\bf u} \right)
       - {1 + w \over a^2} \left\{ {a \over 1 + w}
       \left[ \nabla \cdot \left( \delta {\bf u} \right)
       - {3 \over 2} a H {c_s^2 \over 1 + w} \delta^2 \right] \right\}^\cdot
       + \left( 1 + w \right) \dot C^{(t)\alpha\beta}
       \left( {2 \over a} u_{\alpha|\beta}
       + \dot C^{(t)}_{\alpha\beta} \right)
   \nonumber \\
   & & \qquad
       + {1 \over 2} c_s^2 \left( {4 \pi G \mu \over c^2}
       - {1 + 2 c_s^2 \over 1 + w} c^2 {\Delta \over a^2} \right)
       \delta^2
       - c_s^2 {1 \over a^2} \left[ 2 a H \delta \nabla \cdot {\bf u}
       + 2 \varphi c^2 \Delta \delta
       - c^2 \left( \nabla \varphi \right) \cdot \nabla \delta
       + 2 c^2 \delta^{,\alpha|\beta} C^{(t)}_{\alpha\beta} \right]
   \nonumber \\
   & & \qquad
       + \left( 1 + w \right) \kappa_\Pi
       + {1 + w \over a^2} \left( {a^2 \over 1 + w} \delta_\Pi \right)^\cdot
       - {1 \over a} {\bf u} \cdot \nabla \delta_\Pi.
   \label{ddot-delta-eq4}
\eea We note that, to the linear order, Eq.\ (\ref{ddot-delta-eq4})
is valid in the presence of general $K$ and $\Lambda$, see Eq.\
(\ref{ddot-delta-single-irrotational-CG}).

%%%%%%%%%%%%%%%%%%%%%%%%%%%%%%%%%%%%%%%%%%%%%%%%%%%%%%%%%%%%%%
\subsection{Linear-order relativistic pressure corrections}

To the linear order, {\it ignoring} the entropic perturbation $e$
and the anisotropic stress $\Pi$, Eqs.\
(\ref{E-conserv-eq4})-(\ref{ddot-delta-eq4}) become \bea
   & & \dot \delta - 3 w H \delta
       + \left( 1 + w \right) {1 \over a} \nabla \cdot {\bf u}
       = 0,
   \label{P-lin-eq1} \\
   & & {1 \over a} \nabla \cdot \left( \dot {\bf u} + H {\bf u}
       \right)
       + 4 \pi G \varrho \delta
       = - {1 \over 1 + w} {\Delta \over a^2} {\delta p \over \varrho},
   \label{P-lin-eq2} \\
   & & {1 + w \over a^2 H}
       \left[ {H^2 \over a (1 + w) \varrho} \left( {a^3 \varrho \over H}
       \delta \right)^\cdot \right]^\cdot
       = {\Delta \over a^2} {\delta p \over \varrho},
   \label{P-lin-eq3}
\eea where {\it ignoring} the specific internal energy density
$\epsilon$ we used $\mu = \varrho c^2$, thus $\delta = \delta
\varrho/\varrho$; in general we have $\mu = \varrho \left( c^2 +
\epsilon \right)$ \cite{covariant}. Equation (\ref{P-lin-eq3}) can
be expanded to give \bea
   & & {1 + w \over a^2 H}
       \left[ {H^2 \over a (1 + w) \varrho} \left( {a^3 \varrho \over H}
       \delta \right)^\cdot \right]^\cdot
       = {\Delta \over a^2} {\delta p \over \varrho}
   \nonumber \\
   & & \qquad
       = \ddot \delta
       + \left( 2 - 6 w + 3 c_s^2 \right) H \dot \delta
       - \left[ \left( 1 + 8 w - 6 c_s^2 - 3 w^2 \right) 4 \pi G
       \varrho
       - 12 \left( w - c_s^2 \right) {K c^2 \over a^2}
       + \left( 5 w - 3 c_s^2 \right) \Lambda c^2 \right] \delta,
\eea which is valid in the presence of $K$ and $\Lambda$. Equation
of density perturbation in the comoving gauge was first derived by
Nariai in \cite{Nariai-1969}.

In a single component case the Newtonian equations in Eqs.\
(\ref{dot-delta-eq-N-i}),(\ref{dot-velocity-eq-N-i}),(\ref{ddot-delta-eq-N-i})
give
 \bea
   & & \dot \delta
       + {1 \over a} \nabla \cdot {\bf u} = 0,
   \label{N-lin-eq1} \\
   & & {1 \over a} \nabla \cdot \left( \dot {\bf u} + H {\bf u}
       \right)
       + 4 \pi G \varrho \delta
       =
       - {\Delta \over a^2} {\delta p \over \varrho},
   \label{N-lin-eq2} \\
   & & \ddot \delta + 2 H \dot \delta
       - 4 \pi G \varrho \delta
       = {1 \over a^2 H}
       \left[ a^2 H^2 \left( {1 \over H}
       \delta \right)^\cdot \right]^\cdot
       = {\Delta \over a^2} {\delta p \over \varrho}.
   \label{N-lin-eq3}
\eea Comparing Eqs.\ (\ref{P-lin-eq1}),(\ref{P-lin-eq2}) with Eqs.\
(\ref{N-lin-eq1}),(\ref{N-lin-eq2}), we notice the presence of $w
\equiv p/(\varrho c^2)$ term in three places in the relativistic
equations. Even to the linear order the presence of these terms
should be regarded as pure general relativistic effect of the
isotropic pressure. The effects of pressures to the second order
compared with the Newtonian equations can be found in Eqs.\
(\ref{E-conserv-eq4})-(\ref{ddot-delta-eq4}) which should be
compared with Newtonian equations in Eqs.\
(\ref{dot-delta-eq-N-i}),(\ref{dot-velocity-eq-N-i}),(\ref{ddot-delta-eq-N-i}).
Pressure has the genuine relativistic role in cosmology even in the
background level.

%%%%%%%%%%%%%%%%%%%%%%%%%%%%%%%%%%%%%%%%%%%%%%%%%%%%%%%%%%%%%%%
%
%  Effects of multi-component
%
%%%%%%%%%%%%%%%%%%%%%%%%%%%%%%%%%%%%%%%%%%%%%%%%%%%%%%%%%%%%%%
\section{Effects of multi-component}
                                      \label{sec:multi-component}

We {\it assume} zero-pressure medium, thus set \bea
   & & p_{(i)} \equiv 0, \quad
       \delta p_{(i)} \equiv 0 \equiv \Pi_{(i)\alpha\beta}.
\eea From Eq.\ (\ref{sum}) we notice that to the second order the
collective fluid quantities differ from simple sum of individual
fluid quantities. Even in the zero-pressure mediums, we have \bea
   & & \delta \mu
       = \sum_j \left[
       \delta \mu_{(j)}
       + \mu_{(j)} v_{(j)}^\alpha
       \left( v_{(j)\alpha} - v_\alpha \right) \right],
   \nonumber \\
   & & \delta p
       = {1 \over 3} \sum_j
       \mu_{(j)} v_{(j)}^\alpha
       \left( v_{(j)\alpha} - v_\alpha \right),
   \nonumber \\
   & & \mu  v_\alpha
       = \sum_j \left[ \mu_{(j)} v_{(j)\alpha}
       + \delta \mu_{(j)}
       \left( v_{(j)\alpha} - v_\alpha \right) \right],
   \nonumber \\
   & & \Pi^\alpha_\beta
       = \sum_j
       \mu_{(j)} \left[ v_{(j)}^\alpha
       \left( v_{(j)\beta} - v_\beta \right)
       - {1 \over 3} \delta^\alpha_\beta v_{(j)}^\gamma
       \left( v_{(j)\gamma} - v_\gamma \right) \right],
   \label{sum-zero-pressure}
\eea thus $\delta p \neq 0 \neq \Pi^\alpha_\beta$ to the second
order.

Equations
(\ref{E-conservation-i}),(\ref{Mom-conservation-i}),(\ref{Raychaudhury-eq}),(\ref{E-conservation}),(\ref{Mom-conservation}),(\ref{Mom-constraint})
give \bea
   & & \dot \delta_{(i)}
       - \delta K + 3 H A
       + {1 \over a} \left[ \left( 1 + \delta_{(i)} \right)
       v_{(i)}^\alpha \right]_{|\alpha}
       =
       - {1 \over a} \delta_{(i),\alpha} B^\alpha
       + \delta_{(i)} \left( \delta K - 3 H A \right)
       + A \delta K
       + {3 \over 2} H \left( A^2 - B^\alpha B_\alpha \right)
   \nonumber \\
   & & \qquad
       - v_{(i)\alpha}
       \left( 2 \dot v_{(i)}^\alpha
       + H v_{(i)}^\alpha \right)
       - {1 \over a} A v_{(i)|\alpha}^\alpha
       - {2 \over a} A_{,\alpha} v_{(i)}^\alpha
       + {2 \over a} \left( C^{\alpha\beta}
       v_{(i)\beta} \right)_{|\alpha}
       - { 1\over a} C^{\alpha|\beta}_\alpha
       v_{(i)\beta},
   \label{E-conservation-zero-pressure-i} \\
   & & {1 \over a} \left[ a \left( 1 + \delta_{(i)} \right)
       v_{(i)\alpha} \right]^\cdot
       + {1 \over a} A_{,\alpha}
       =
       \left( \delta K - 3 H A \right)
       v_{(i)\alpha}
   \nonumber \\
   & & \qquad
       + {1 \over a} \left[
       A A_{,\alpha}
       - \delta_{(i)} A_{,\alpha}
       - B^\beta B_{\beta|\alpha}
       - \left( v_{(i)\alpha} v_{(i)}^\beta \right)_{|\beta}
       - v_{(i)\alpha|\beta} B^\beta
       - v_{(i)\beta} B^\beta_{\;\;|\alpha}
       \right],
   \label{Mom-conservation-zero-pressure-i} \\
   & & \delta \dot K + 2 H \delta K
       - 4 \pi G \left( \delta \mu + 3 \delta p \right)
       + \left( 3 \dot H + {\Delta \over a^2} \right) A
       = A \delta \dot K
       - {1 \over a} \delta K_{,\alpha} B^\alpha
       + {1 \over 3} \delta K^2
       + {3 \over 2} \dot H \left( 3 A^2 - B^\alpha B_\alpha \right)
   \nonumber \\
   & & \qquad
       + {1 \over a^2} \left[ 2 A \Delta A
       + A^{,\alpha} A_{,\alpha}
       - {1 \over 2} \Delta \left( B^\alpha  B_\alpha \right)
       + A^{,\alpha} \left( 2 C^\beta_{\alpha|\beta}
       - C^\beta_{\beta|\alpha} \right)
       + 2 C^{\alpha\beta} A_{,\alpha|\beta} \right]
   \nonumber \\
   & & \qquad
       + \left( \dot C^{\alpha\beta}
       + {1 \over a} B^{(\alpha|\beta)} \right)
       \left( \dot C_{\alpha\beta} + {1 \over a} B_{\alpha|\beta}
       \right)
       - {1 \over 3} \left( \dot C^\alpha_\alpha
       + {1 \over a} B^\alpha_{\;\;|\alpha} \right)^2
       + 8 \pi G \mu v^\alpha v_\alpha,
   \label{Raychaudhury-eq-zero-pressure} \\
   & & \dot \delta
       - \delta K + 3 H A
       + {1 \over a} \left[ \left( 1 + \delta \right)
       v^\alpha \right]_{|\alpha}
       =
       - {1 \over a} \delta_{,\alpha} B^\alpha
       + \delta \left( \delta K - 3 H A \right)
       + A \delta K
       + {3 \over 2} H \left( A^2 - B^\alpha B_\alpha \right)
   \nonumber \\
   & & \qquad
       - v_{\alpha}
       \left( 2 \dot v^\alpha
       + H v^\alpha \right)
       - {1 \over a} A v_{\;\;|\alpha}^\alpha
       - {2 \over a} A_{,\alpha} v^\alpha
       + {2 \over a} \left( C^{\alpha\beta}
       v_{\beta} \right)_{|\alpha}
       - {1 \over a} C^{\alpha|\beta}_\alpha
       v_{\beta}
       - 3 H {\delta p \over \mu},
   \label{E-conservation-zero-pressure} \\
   & & {1 \over a} \left[ a \left( 1 + \delta \right)
       v_{\alpha} \right]^\cdot
       + {1 \over a} A_{,\alpha}
       =
       \left( \delta K - 3 H A \right)
       v_{\alpha}
   \nonumber \\
   & & \qquad
       + {1 \over a} \left[
       A A_{,\alpha}
       - \delta_{(i)} A_{,\alpha}
       - B^\beta B_{\beta|\alpha}
       - \left( v_{\alpha} v^\beta \right)_{|\beta}
       - v_{\alpha|\beta} B^\beta
       - v_{\beta} B^\beta_{\;\;|\alpha}
       - {1 \over \mu} \left( \delta p_{,\alpha}
       + \Pi^\beta_{\alpha|\beta} \right)
       \right],
   \label{Mom-conservation-zero-pressure} \\
   & & \left[ \dot C^\beta_\alpha
       + {1 \over 2a} \left( B^\beta_{\;\;|\alpha}
       + B_\alpha^{\;\;|\beta} \right) \right]_{|\beta}
       - {1 \over 3} \left( \dot C^\gamma_\gamma
       + {1 \over a} B^\gamma_{\;\;|\gamma} \right)_{,\alpha}
       + {2 \over 3} \delta K_{,\alpha}
       = - 8 \pi G a \sum_j \mu_{(j)}
       v_{(j)\alpha}.
   \label{Mom-constraint-zero-pressure}
\eea where Eq.\ (\ref{Mom-constraint-zero-pressure}) is valid to the
linear order.

%%%%%%%%%%%%%%%%%%%%%%%%%%%%%%%%%%%%%%%%%%%%%%%%%%%%%%%%%%%%%%
\subsection{Irrotational case}

Assuming an irrotational condition we ignore all vector-type
perturbations. As the spatial gauge condition we take \bea
   & & \gamma \equiv 0,
   \label{gamma=0}
\eea thus, $\beta \equiv \chi/a$. Equations
(\ref{E-conservation-zero-pressure-i})-(\ref{Mom-constraint-zero-pressure})
become \bea
   & & \dot \delta_{(i)}
       - \kappa
       + 3 H \alpha
       - {1 \over a} \left[ \left( 1 + \delta_{(i)} \right)
       v_{(i)}^{,\alpha} \right]_{|\alpha}
       =
       - {1 \over a^2} \delta_{(i),\alpha} \chi^{,\alpha}
       + \delta_{(i)} \left( \kappa - 3 H \alpha \right)
       + \alpha \kappa
       + {3 \over 2} H \left( \alpha^2
       - {1 \over a^2} \chi^{,\alpha} \chi_{,\alpha} \right)
   \nonumber \\
   & & \qquad
       + H v_{(i),\alpha} v_{(i)}^{,\alpha}
       + {1 \over a} \alpha \Delta v_{(i)}
       + {1 \over a} \varphi^{,\alpha} v_{(i),\alpha}
       - {2 \over a} \varphi \Delta v_{(i)}
       - {2 \over a} C^{(t)\alpha\beta} v_{(i),\alpha|\beta},
   \label{E-conservation-zero-pressure-irrotation} \\
   & & {1 \over a} \left[ a \left( 1 + \delta_{(i)} \right)
       v_{(i)} \right]^\cdot_{,\alpha}
       - {1 \over a} \alpha_{,\alpha}
   \nonumber \\
   & & \qquad
       =
       \left( \kappa - 3 H \alpha \right)
       v_{(i),\alpha}
       + {1 \over a} \left[
       - \alpha \alpha_{,\alpha}
       + {1 \over a^2} \chi^{,\beta} \chi_{,\beta|\alpha}
       + \delta_{(i)} \alpha_{,\alpha}
       + \left( v_{(i),\alpha} v_{(i)}^{,\beta} \right)_{|\beta}
       - {1 \over a} \left( v_{(i),\beta} \chi^{,\beta} \right)_{,\alpha}
       \right],
   \label{Mom-conservation-zero-pressure-irrotation} \\
   & & \dot \kappa + 2 H \kappa
       - 4 \pi G \left( \delta \mu +  3\delta p \right)
       + \left( 3 \dot H + {\Delta \over a^2} \right) \alpha
       = \alpha \dot \kappa
       - {1 \over a^2} \kappa_{,\alpha} \chi^{,\alpha}
       + {1 \over 3} \kappa^2
       + {3 \over 2} \dot H \left( 3 \alpha^2
       - {1 \over a^2} \chi^{,\alpha} \chi_{,\alpha} \right)
   \nonumber \\
   & & \qquad
       + {1 \over a^2} \left[ 2 \alpha \Delta \alpha
       + \alpha^{,\alpha} \alpha_{,\alpha}
       - {\Delta \over 2 a^2} \left( \chi^{,\alpha} \chi_{,\alpha} \right)
       - \alpha^{,\alpha} \varphi_{,\alpha}
       + 2 \varphi \Delta \alpha
       + 2 C^{(t)\alpha\beta} \alpha_{,\alpha|\beta} \right]
   \nonumber \\
   & & \qquad
       + \left( \dot C^{(t)\alpha\beta}
       + {1 \over a^2} \chi^{,\alpha|\beta} \right)
       \left( \dot C^{(t)}_{\alpha\beta}
       + {1 \over a^2} \chi_{,\alpha|\beta} \right)
       - {1 \over 3} \left( {\Delta \over a^2} \chi \right)^2
       + 8 \pi G \mu v^{,\alpha} v_{,\alpha},
   \label{Raychaudhury-eq-zero-pressure-irrotation} \\
   & & \dot \delta
       - \kappa
       + 3 H \alpha
       - {1 \over a} \left[ \left( 1 + \delta \right)
       v^{,\alpha} \right]_{|\alpha}
       =
       - {1 \over a^2} \delta_{,\alpha} \chi^{,\alpha}
       + \delta \left( \kappa - 3 H \alpha \right)
       + \alpha \kappa
       + {3 \over 2} H \left( \alpha^2
       - {1 \over a^2} \chi^{,\alpha} \chi_{,\alpha} \right)
   \nonumber \\
   & & \qquad
       + H v_{,\alpha} v^{,\alpha}\
       + {1 \over a} \alpha \Delta v
       + {1 \over a} \varphi^{,\alpha} v_{,\alpha}
       - {2 \over a} \varphi \Delta v
       - {2 \over a} C^{(t)\alpha\beta} v_{,\alpha|\beta}
       - 3 H {\delta p \over \mu},
   \label{E-conservation-zero-pressure-irrotation-total} \\
   & & {1 \over a} \left[ a \left( 1 + \delta \right)
       v \right]^\cdot_{,\alpha}
       - {1 \over a} \alpha_{,\alpha}
       =
       \left( \kappa - 3 H \alpha \right)
       v_{,\alpha}
   \nonumber \\
   & & \qquad
       + {1 \over a} \left[
       - \alpha \alpha_{,\alpha}
       + {1 \over a^2} \chi^{,\beta} \chi_{,\beta|\alpha}
       + \delta \alpha_{,\alpha}
       + \left( v_{,\alpha} v^{,\beta} \right)_{|\beta}
       - {1 \over a} \left( v_{,\beta} \chi^{,\beta} \right)_{,\alpha}
       - {1 \over \mu} \left( \delta p_{,\alpha}
       + \Pi^\beta_{\alpha|\beta} \right)
       \right],
   \label{Mom-conservation-zero-pressure-irrotation-total} \\
   & & {\Delta + 3 K \over a^2} \chi + \kappa
       = 12 \pi G a \sum_j \mu_{(j)} v_{(j)}.
   \label{Mom-constraint-zero-pressure-irrotation}
\eea Equation (\ref{Mom-constraint-zero-pressure-irrotation}) is
valid to the linear order.

%%%%%%%%%%%%%%%%%%%%%%%%%%%%%%%%%%%%%%%%%%%%%%%%%%%%%%%%%%%%%%
\subsection{Linear perturbations}

To the linear order Eqs.\
(\ref{E-conservation-zero-pressure-irrotation})-(\ref{Mom-conservation-zero-pressure-irrotation-total})
give \bea
   & & \dot \delta_{(i)}
       - \kappa
       + 3 H \alpha
       - c {\Delta \over a} v_{(i)}
       = 0,
   \label{E-conservation-i-lin} \\
   & & \dot v_{(i)} + H v_{(i)}
       - {c \over a} \alpha
       = 0,
   \label{Mom-conservation-i-lin} \\
   & & \dot \kappa + 2 H \kappa
       - 4 \pi G \sum_j \varrho_{(j)} \delta_{(j)}
       + \left( 3 \dot H + c^2 {\Delta \over a^2} \right) \alpha
       = 0,
   \label{Raychaudhury-eq-lin} \\
   & & c {\Delta + 3 K \over a^2} \chi + \kappa
       = {12 \pi G \over c} a \sum_j \varrho_{(j)} v_{(j)},
   \label{Mom-constraint-lin} \\
   & & \dot \delta
       - \kappa
       + 3 H \alpha
       - c {\Delta \over a} v
       = 0,
   \label{E-conservation-total-lin} \\
   & & \dot v + H v
       - {c \over a} \alpha
       = 0,
   \label{Mom-conservation-total-lin}
\eea where we have recovered the speed of light $c$. The following
additional equations can be found in Eqs.\ (195),(196),(199) of
\cite{NL} \bea
   & & \kappa - 3 H \alpha + 3 \dot \varphi
       + c {\Delta \over a^2} \chi = 0,
   \label{kappa-definition-lin} \\
   & & 4 \pi G \varrho \delta + H \kappa
       + c^2 {\Delta + 3 K \over a^2} \varphi = 0,
   \label{E-constraint-lin} \\
   & & {1 \over c} \left( \dot \chi + H \chi \right)
       - \varphi - \alpha = 0.
   \label{ADM-tracefree-propagation-lin}
\eea Equation (\ref{ADM-tracefree-propagation-lin}) gives \bea
   & & \alpha_\chi = - \varphi_\chi.
\eea Equation (\ref{Mom-conservation-total-lin}) gives \bea
   & & \alpha_v = 0.
\eea Equations (\ref{Mom-constraint-lin}),
(\ref{kappa-definition-lin}) give \bea
   & & \dot \varphi_v = {K c \over a^2} \chi_v.
\eea Thus, for $K = 0$ we have \bea
   & & \dot \varphi_v = 0,
   \label{dot-varphi_v=0}
\eea which is valid even in the presence of multi-components and the
cosmological constant. Equations (\ref{Mom-conservation-i-lin}),
(\ref{Mom-conservation-total-lin}) give \bea
   & & \dot v_\chi + H v_\chi
       + {c \over a} \varphi_\chi
       = 0,
   \label{dot-velocity-eq-lin} \\
   & & \dot v_{(i)\chi} + H v_{(i)\chi}
       + {c \over a} \varphi_\chi
       = 0.
   \label{dot-velocity-i-eq-lin}
\eea Equations (\ref{Mom-constraint-lin}), (\ref{E-constraint-lin})
give \bea
   & & c^2 {\Delta + 3 K \over a^2} \varphi_\chi
       = - 4 \pi G \varrho \delta_v
       = - 4 \pi G \sum_j \varrho_{(j)} \delta_{(j)v}
       = - 4 \pi G \sum_j \varrho_{(j)} \delta_{(j)v_{(j)}}.
   \label{Poisson-eq-lin}
\eea We can derive density perturbation equation in many different
temporal gauge (hypersurface) conditions all of which naturally
correspond to gauge-invariant variables. In a single component case,
density perturbation in the comoving gauge ($v = 0$) is known to
give Newtonian result. In the multi-component situation we have many
different comoving gauge conditions.  Here, we consider two such
gauges for $\delta_{(i)}$ variable: one based on $v = 0$ gauge, and
the other based on $v_{(\ell)} = 0$ gauge for a specific $\ell$.

(I) Equation (\ref{E-conservation-i-lin}) evaluated in the $v = 0$
gauge, and using Eq.\ (\ref{Mom-constraint-lin}) we can derive \bea
   & & \dot \delta_{(i)v}
       - c {\Delta \over a} v_{(i)\chi}
       - c {3 K \over a} v_{\chi}
       = 0,
       \label{dot-density-i-eq-lin}
\eea where we used $\chi_v \equiv \chi - a v \equiv - a v_{\chi}$,
$v_{v_{(i)}} \equiv v - v_{(i)}$, and $\alpha_v \equiv \alpha -
c^{-1} \left( a v \right)^\cdot = 0$. Using Eqs.\
(\ref{dot-velocity-eq-lin}),(\ref{dot-velocity-i-eq-lin}),(\ref{Poisson-eq-lin}),
we have \bea
   & & \ddot \delta_{(i)v}
       + 2 H \dot \delta_{(i)v}
       - 4 \pi G \sum_j \varrho_{(j)} \delta_{(j)v}
       = 0.
   \label{ddot-density-i-eq-lin}
\eea This coincides exactly with the Newtonian result in Eq.\
(\ref{ddot-delta-eq-N-i}) to the linear order, even in the presence
of $K$. Thus, we may identify $\delta_{(i)v}$ as the Newtonian
density perturbation $\delta_i$ to the linear order even in the
presence of $K$. For $K = 0$ we may also identify $- c \nabla
v_{(i)\chi}$ as the Newtonian velocity perturbation ${\bf u}_i$.
However, in the presence of $K$, we cannot identify the relativistic
variables which correspond to the Newtonian velocity perturbation of
individual component. Therefore, to the linear order we have the
following Newtonian correspondences \bea
   & & \delta_i = \delta_{(i)v}, \quad
       {\bf u}_i = - c \nabla v_{(i)\chi},
\eea where the latter one is valid only for $K = 0$.

(II) Evaluating Eq.\ (\ref{E-conservation-i-lin}) in the $v_{(\ell)}
= 0$ gauge for a specific $\ell$, and using Eq.\
(\ref{Mom-constraint-lin}) we can derive \bea
   & & \dot \delta_{(i)v_{(\ell)}}
       - c {\Delta + 3 K \over a} v_{(\ell)\chi}
       = {12 \pi G \over c} a \sum_j \varrho_{(j)} \left( v_{(j)} - v_{(\ell)} \right)
       + c {\Delta \over a} \left( v_{(i)} - v_{(\ell)} \right),
       \label{dot-density-ii-eq-lin}
\eea where we used $\chi_{v_{(\ell)}} \equiv \chi - a v_{(\ell)}
\equiv - a v_{(\ell)\chi}$, $v_{(i)v_{(\ell)}} \equiv v_{(i)} -
v_{(\ell)}$, and $\alpha_{v_{(\ell)}} \equiv \alpha - c^{-1} \left(
a v_{(\ell)} \right)^\cdot = 0$. Using Eqs.\
(\ref{dot-velocity-eq-lin}),(\ref{dot-velocity-i-eq-lin}),(\ref{Poisson-eq-lin}),
we have \bea
   & & \ddot \delta_{(i)v_{(\ell)}}
       + 2 H \dot \delta_{(i)v_{(\ell)}}
       - 4 \pi G \sum_j \varrho_{(j)} \delta_{(j)v_{(j)}}
       = {12 \pi G \over c} a H \sum_j \varrho_{(j)}
       \left( v_{(\ell)} - v_{(j)} \right).
   \label{ddot-density-ii-eq-lin}
\eea The terms in right-hand-sides of Eqs.\
(\ref{dot-density-ii-eq-lin}),(\ref{ddot-density-ii-eq-lin}) look
like relativistic correction terms present even to the linear order
based on the variable $\delta_{(i)v_{(\ell)}}$. Since no such
correction terms appear in Eq.\ (\ref{ddot-density-i-eq-lin}) based
on the variable $\delta_{(i)v}$, the relativistic correction terms
in Eq.\ (\ref{ddot-density-ii-eq-lin}) can be regarded as being
caused by a complicated hypersurface (gauge) choice.

%%%%%%%%%%%%%%%%%%%%%%%%%%%%%%%%%%%%%%%%%%%%%%%%%%%%%%%%%%%%%%
\subsubsection{Exact solutions}
                                      \label{sec:exact-solutions}

{\it Assuming} $K = 0$, we can identify Newtonian perturbation
variables as \bea
   & & \delta \equiv \delta_v, \quad
       \kappa_v \equiv - {1 \over a} \nabla \cdot {\bf u}, \quad
       {\bf u} \equiv - c \nabla v_\chi, \quad
       \delta \Phi \equiv - c^2 \varphi_\chi, \quad
       \delta_i \equiv \delta_{(i)v}, \quad
       {\bf u}_i \equiv - c \nabla v_{(i)\chi}.
   \label{identification-linear}
\eea Equations
(\ref{E-conservation-total-lin}),(\ref{Raychaudhury-eq-lin}),(\ref{E-conservation-i-lin}),(\ref{Mom-conservation-i-lin}),(\ref{Poisson-eq-lin})
become \bea
   & & \dot \delta
       = - {1 \over a} \nabla \cdot {\bf u},
   \label{N-delta-eq} \\
   & & \dot {\bf u} + H {\bf u}
       = - {1 \over a} \nabla \delta \Phi,
   \label{N-v-eq} \\
   & & \dot \delta_i
       = - {1 \over a} \nabla \cdot {\bf u}_i,
   \label{N-delta-i-eq} \\
   & & \dot {\bf u}_i + H {\bf u}_i
       = - {1 \over a} \nabla \delta \Phi,
   \label{N-v-i-eq} \\
   & & {\Delta \over a^2} \delta \Phi
       = 4 \pi G \varrho \delta
       = 4 \pi G \sum_j \varrho_j \delta_j.
   \label{N-Poisson-eq}
\eea Under the identification in Eq.\ (\ref{identification-linear})
these equations are valid in both Newton's and Einstein's gravity
theories. Equations (\ref{N-delta-eq}), (\ref{N-v-eq}),
(\ref{N-Poisson-eq}), and Eqs.\
(\ref{N-delta-i-eq}),(\ref{N-v-i-eq}),(\ref{N-Poisson-eq}),
respectively, give \bea
   & & \ddot \delta
       + 2 H \dot \delta
       - 4 \pi G \varrho \delta
       = {1 \over a^2 H} \left[ a^2 H^2 \left( {\delta \over H}
       \right)^\cdot \right]^\cdot
       = 0,
   \label{N-ddot-delta-eq} \\
   & & \ddot \delta_i
       + 2 H \dot \delta_i
       - 4 \pi G \sum_j \varrho_j \delta_j
       = 0.
   \label{N-ddot-delta-i-eq}
\eea

Equation (\ref{N-ddot-delta-eq}) has an exact solution \bea
   & & \delta ({\bf x}, t)
       = H \left[ c_g ({\bf x}) \int^t {dt \over a^2 H^2}
       + c_d ({\bf x}) \right],
   \label{N-delta-exact-solution}
\eea where $c_g$ and $c_d$ are integration constants which indicate
the relatively growing and decaying solutions in expanding phase; we
do not consider the lower bound of integration which is absorbed to
the $c_d$ mode. Equations (\ref{N-ddot-delta-eq}),
(\ref{N-ddot-delta-i-eq}), and the solution in Eq.\
(\ref{N-delta-exact-solution}) are valid considering general $K$ and
$\Lambda$ in the background world model. Equation
(\ref{N-Poisson-eq}) can be solved to give \bea
   & & \delta \Phi = - G \varrho a^2 \int
       { \delta ({\bf x}^\prime, t) \over
       \left| {\bf x}^\prime - {\bf x} \right| }
       d^3 x^\prime.
\eea From Eqs.\
(\ref{N-delta-eq}),(\ref{N-v-eq}),(\ref{N-Poisson-eq}) we can show
\cite{Peebles-1980} \bea
   & & {\bf u}
       = - a \left( {\nabla \delta \Phi \over 4 \pi G \varrho a^2}
       \right)^\cdot
       + {1 \over a} {\bf D} ({\bf x}), \quad
       \nabla \cdot {\bf D} \equiv 0,
   \label{u-lin-sol}
\eea where the ${\bf D}$ term is the solution of the homogeneous
part of Eq.\ (\ref{N-v-eq}); it decouples from the density
inhomogeneity and corresponds to the peculiar velocity in the
background world model. Since the ${\bf D}$ term is not connected to
the density inhomogeneity and simply decays, we may ignore it to the
linear order.

Now, for the individual component, from Eqs.\
(\ref{N-delta-i-eq}),(\ref{N-v-i-eq}) we have \bea
   & & {\bf u}_i
       = - a \left( {\nabla \delta \Phi \over 4 \pi G \varrho a^2}
       \right)^\cdot
       + {1 \over a} \nabla d_i ({\bf x})
       + {1 \over a} {\bf D}_i ({\bf x}), \quad
       \nabla \cdot {\bf D}_i \equiv 0,
   \label{u-i-lin-sol} \\
   & & \delta_i = \delta + c_i ({\bf x}) - \Delta d_i ({\bf x}) \int^t
       {dt^\prime \over a^2 (t^\prime)},
\eea with \bea
   & & \sum_j \varrho_j d_j \equiv 0 \equiv \sum_j \varrho_j c_j.
\eea The $c_i$ and $d_i$ are the two isocurvature-type ($\delta =
0$, thus $\delta \Phi = 0$) solutions. It happens that the
relatively decaying isocurvature-type solution, i.e., $d_i$-mode,
temporally behaves the same as the peculiar velocity in the
background, i.e., ${\bf D}_i$-mode. The relatively growing
isocurvature-type solution $c_i$ does not contribute to the ${\bf
u}_i$, see Eq.\ (\ref{N-delta-i-eq}). From Eqs.\
(\ref{u-lin-sol}),(\ref{u-i-lin-sol}) we have \bea
   & & {\bf u}_i - {\bf u} =
       {1 \over a} \left[ \nabla d_i ({\bf x})
       + {\bf D}_i ({\bf x})
       - {\bf D} ({\bf x}) \right],
   \label{u_i-u-solution}
\eea which simply decays; the ${\bf D}_i$ and ${\bf D}$ solutions
are divergence-free and decoupled from the density perturbation, and
are the peculiar velocity perturbation present in the background
world model.

%%%%%%%%%%%%%%%%%%%%%%%%%%%%%%%%%%%%%%%%%%%%%%%%%%%%%%%%%%%%%%
\subsection{Comoving gauge}

To the linear order, only the hypersurface condition $v=0$ (the
comoving temporal gauge) allows the density perturbation equation
presented in the Newtonian form. Thus, we take \bea
   & & v \equiv 0,
\eea even to the second order. We take $\gamma \equiv 0$ in Eq.\
(\ref{gamma=0}) as the spatial gauge condition. Equation
(\ref{Mom-conservation-zero-pressure-irrotation-total}) gives \bea
   & & \alpha
       = - {1 \over 2 a^2} \chi^{,\alpha} \chi_{,\alpha}
       - \sum_j {\mu_{(j)} \over \mu} \left[ {1 \over 2}
       v_{(j)}^\alpha v_{(j)\alpha}
       + \Delta^{-1} \nabla_\alpha \left(
       v_{(j)}^\alpha v^\beta_{(j)|\beta} \right) \right].
\eea Using this, Eqs.\
(\ref{E-conservation-zero-pressure-irrotation})-(\ref{E-conservation-zero-pressure-irrotation-total})
give \bea
   & & \dot \delta
       - \kappa
       =
       - {c \over a^2} \delta_{,\alpha} \chi^{,\alpha}
       + \delta \kappa
       + {1 \over 2} H \sum_j
       \mu_{(j)} v_{(j)}^\alpha v_{(j)\alpha}
       + 3 H \sum_j {\mu_{(j)} \over \mu} \Delta^{-1} \nabla_\alpha \left(
       v_{(j)}^\alpha v^\beta_{(j)|\beta} \right),
   \label{E-conservation-zero-pressure-irrotation-total-CG} \\
   & & \dot \kappa + 2 H \kappa
       - 4 \pi G \varrho \delta
       =
       - {c \over a^2} \kappa_{,\alpha} \chi^{,\alpha}
       + {1 \over 3} \kappa^2
       + \left( \dot C^{(t)\alpha\beta}
       + {c \over a^2} \chi^{,\alpha|\beta} \right)
       \left( \dot C^{(t)}_{\alpha\beta}
       + {c \over a^2} \chi_{,\alpha|\beta} \right)
       - {1 \over 3} \left( c {\Delta \over a^2} \chi \right)^2
   \nonumber \\
   & & \qquad
       + {1 \over 2} \left( 3 \dot H + 8 \pi G \varrho + c^2 {\Delta \over a^2} \right)
       \sum_j \mu_{(j)} v_{(j)}^\alpha v_{(j)\alpha}
       + \left( 3 \dot H + c^2 {\Delta \over a^2} \right)
       \sum_j {\mu_{(j)} \over \mu} \Delta^{-1} \nabla_\alpha \left(
       v_{(j)}^\alpha v^\beta_{(j)|\beta} \right),
   \label{Raychaudhury-eq-zero-pressure-irrotation-CG} \\
   & & \dot \delta_{(i)}
       - \kappa
       + c {1 \over a} \left[ \left( 1 + \delta_{(i)} \right)
       v_{(i)}^{\alpha} \right]_{|\alpha}
       =
       - {c \over a^2} \delta_{(i),\alpha} \chi^{,\alpha}
       + \delta_{(i)} \kappa
       + H v_{(i)\alpha} v_{(i)}^{\alpha}
   \nonumber \\
   & & \qquad
       - {c \over a} \left( \varphi^{,\alpha} v_{(i)\alpha}
       - 2 \varphi v^\alpha_{(i)|\alpha}
       - 2 v_{(i)}^{\alpha|\beta} C^{(t)}_{\alpha\beta} \right)
       + {3 \over 2} H \sum_j \mu_{(j)} v_{(j)}^\alpha v_{(j)\alpha}
       + 3 H \sum_j {\mu_{(j)} \over \mu} \Delta^{-1} \nabla_\alpha \left(
       v_{(j)}^\alpha v^\beta_{(j)|\beta} \right),
   \label{E-conservation-zero-pressure-irrotation-CG} \\
   & & {1 \over a} \left[ a \left( 1 + \delta_{(i)} \right)
       v_{(i)\alpha} \right]^\cdot
       =
       - {c \over a^2} \left( v_{(i)\beta} \chi^{,\beta} \right)_{,\alpha}
       + \kappa v_{(i)\alpha}
       - {c \over a} \left( v_{(i)\alpha} v_{(i)}^{\beta} \right)_{|\beta}
   \nonumber \\
   & & \qquad
       + {c \over 2 a} \sum_j \mu_{(j)}
       \nabla_\alpha \left( v_{(j)}^\beta v_{(j)\beta} \right)
       + {c \over a} \sum_j {\mu_{(j)} \over \mu} \nabla_\alpha
       \Delta^{-1} \nabla_\gamma \left(
       v_{(j)}^\gamma v^\beta_{(j)|\beta} \right).
   \label{Mom-conservation-zero-pressure-irrotation-CG}
\eea From Eqs.\
((\ref{E-conservation-zero-pressure-irrotation-total-CG}),\ref{Raychaudhury-eq-zero-pressure-irrotation-CG}),
and Eqs.\
(\ref{Raychaudhury-eq-zero-pressure-irrotation-CG})-(\ref{Mom-conservation-zero-pressure-irrotation-CG})
we can derive, respectively, \bea
   & & {1 \over a^2} \left[ a^2 \left( \dot \delta
       + {c \over a^2} \delta_{,\alpha} \chi^{,\alpha} \right)
       \right]^\cdot
       - 4 \pi G \varrho \delta \left( 1 + \delta \right)
       = - {c \over a^2} \kappa_{,\alpha} \chi^{,\alpha}
       + {4 \over 3} \kappa^2
   \nonumber \\
   & & \qquad
       + \left( \dot C^{(t)\alpha\beta}
       + {c \over a^2} \chi^{,\alpha|\beta} \right)
       \left( \dot C^{(t)}_{\alpha\beta}
       + {c \over a^2} \chi_{,\alpha|\beta} \right)
       - {1 \over 3} \left( c {\Delta \over a^2} \chi \right)^2
   \nonumber \\
   & & \qquad
       + \left( 2 \dot H + 4 \pi G \varrho
       + {1 \over 2} c^2 {\Delta \over a^2}\right)
       \sum_j \mu_{(j)} v_{(j)}^\alpha v_{(j)\alpha}
       + \left( 6 \dot H + c^2 {\Delta \over a^2} \right)
       \sum_j {\mu_{(j)} \over \mu} \Delta^{-1} \nabla_\alpha \left(
       v_{(j)}^\alpha v^\beta_{(j)|\beta} \right),
   \label{ddot-eq-CG} \\
   & & {1 \over a^2} \left[ a^2 \left( \dot \delta_{(i)}
       + {c \over a^2} \delta_{(i),\alpha} \chi^{,\alpha} \right)
       \right]^\cdot
       - 4 \pi G \varrho \delta \left( 1 + \delta_{(i)} \right)
       = - {c \over a^2} \kappa_{,\alpha} \chi^{,\alpha}
       + {4 \over 3} \kappa^2
   \nonumber \\
   & & \qquad
       + \left( \dot C^{(t)\alpha\beta}
       + {c \over a^2} \chi^{,\alpha|\beta} \right)
       \left( \dot C^{(t)}_{\alpha\beta}
       + {c \over a^2} \chi_{,\alpha|\beta} \right)
       - {1 \over 3} \left( c {\Delta \over a^2} \chi \right)^2
       - {c \over a} \left[ \left( \kappa + \dot \varphi \right)^{,\alpha}
       v_{(i)\alpha}
       + 2 \left( \kappa - \dot \varphi \right) v^\alpha_{(i)|\alpha}
       \right]
   \nonumber \\
   & & \qquad
       + c^2 {\Delta \over a^3} \left( v_{(i)\alpha} \chi^{,\alpha} \right)
       + {c^2 \over a^2} \left( v_{(i)}^\alpha v_{(i)}^\beta \right)_{|\alpha\beta}
       + \dot H v_{(i)\alpha} v_{(i)}^{\alpha}
       + {2 c \over a} v_{(i)}^{\alpha|\beta} \dot C^{(t)}_{\alpha\beta}
   \nonumber \\
   & & \qquad
       + \left( 3 \dot H + 4 \pi G \varrho \right)
       \sum_j \mu_{(j)} v_{(j)}^\alpha v_{(j)\alpha}
       + 6 \dot H \sum_j {\mu_{(j)} \over \mu} \Delta^{-1} \nabla_\alpha \left(
       v_{(j)}^\alpha v^\beta_{(j)|\beta} \right).
   \label{ddot-eq-CG-i}
\eea Notice the ${\cal O} (v_{(j)}^\alpha v_{(j)\alpha})$ correction
terms are present in Eqs.\
(\ref{E-conservation-zero-pressure-irrotation-total-CG}),(\ref{Raychaudhury-eq-zero-pressure-irrotation-CG}),(\ref{ddot-eq-CG})
even in the single component situation. Except for these ${\cal O}
(v_{(j)}^\alpha v_{(j)\alpha})$ terms, the remaining parts of these
equations coincide with the ones in the single component situation.

%%%%%%%%%%%%%%%%%%%%%%%%%%%%%%%%%%%%%%%%%%%%%%%%%%%%%%%%%%%%%%
\subsection{Newtonian correspondence}
                                                   \label{sec:multi-Newtonian}

For $K = 0$, to the linear order, we have \bea
   & & \chi_v \equiv \chi - a v \equiv - a v_\chi, \quad
       \kappa_v = - c {\Delta \over a^2} \chi_v
       = c {\Delta \over a} v_\chi, \quad
       v_{(i)v} \equiv v_{(i)} - v = v_{(i)\chi} - v_\chi, \quad
       \dot \varphi_v = 0.
\eea To the linear order we identify \bea
   & & {\bf u} \equiv - c \nabla v_\chi, \quad
       {\bf u}_i \equiv - c \nabla v_{(i)\chi}.
\eea Now, to the second order, we attempt identifying the Newtonian
perturbation variables $\delta$, $\delta_i$, ${\bf u}$, and ${\bf
u}_i$ as \bea
   & & \kappa_v \equiv - {1 \over a} \nabla \cdot {\bf u}, \quad
       \chi_v \equiv {a \over c} u, \quad
       {\bf u} \equiv \nabla u, \quad
       {\bf v}_{(i)v} \equiv {1 \over c} \left( {\bf u}_i - {\bf u}
       \right), \quad
       \delta \equiv \delta_v, \quad
       \delta_i \equiv \delta_{(i)v}.
   \label{identify-second}
\eea Using these identifications Eqs.\
(\ref{E-conservation-zero-pressure-irrotation-total-CG})-(\ref{ddot-eq-CG-i})
can be written as \bea
   & & \dot \delta + {1 \over a} \nabla \cdot {\bf u}
       = - {1 \over a} \nabla \cdot \left( \delta {\bf u} \right)
       + H \sum_j {\varrho_j \over \varrho} {1 \over c^2}
       \left\{
       {1 \over 2} |{\bf u}_j - {\bf u}|^2
       + 3 \Delta^{-1} \nabla \cdot \left[
       \left( {\bf u}_j - {\bf u} \right)
       \nabla \cdot \left( {\bf u}_j - {\bf u} \right) \right]
       \right\},
   \label{ddot-delta-eq-N} \\
   & & {1 \over a} \nabla \cdot \left( \dot {\bf u} + H {\bf u}
       \right)
       + 4 \pi G \varrho \delta
       = - {1 \over a^2} \nabla \cdot \left( {\bf u} \cdot \nabla
       {\bf u} \right)
       - \dot C^{(t)\alpha\beta} \left(
       {2 \over a} u_{\alpha|\beta} + \dot C^{(t)}_{\alpha\beta}
       \right)
   \nonumber \\
   & & \qquad
       + \sum_j {\varrho_j \over \varrho} {1 \over c^2}
       \left\{ {1 \over 2} \left( 4 \pi G \varrho - c^2 {\Delta
       \over a^2} \right)
       |{\bf u}_j - {\bf u}|^2
       + \left( 12 \pi G \varrho - c^2 {\Delta
       \over a^2} \right) \Delta^{-1} \nabla \cdot \left[
       \left( {\bf u}_j - {\bf u} \right)
       \nabla \cdot \left( {\bf u}_j - {\bf u} \right) \right]
       \right\},
   \\
   & & \dot \delta_i
       + {1 \over a} \nabla \cdot {\bf u}_i
       = - {1 \over a} \nabla \cdot \left( \delta_i {\bf u}_i
       \right)
       + {1 \over a} \left[ 2 \varphi \nabla \cdot
       \left( {\bf u}_i - {\bf u} \right)
       - \left( {\bf u}_i - {\bf u} \right) \cdot \nabla \varphi
       + 2 \left( u^\alpha_i - u^\alpha \right)^{|\beta}
       C^{(t)}_{\alpha\beta} \right]
   \nonumber \\
   & & \qquad
       + H {1 \over c^2} |{\bf u}_i - {\bf u}|^2
       + 3 H \sum_j {\varrho_j \over \varrho} {1 \over c^2}
       \left\{
       {1 \over 2} |{\bf u}_j - {\bf u}|^2
       + \Delta^{-1} \nabla \cdot \left[
       \left( {\bf u}_j - {\bf u} \right)
       \nabla \cdot \left( {\bf u}_j - {\bf u} \right) \right]
       \right\},
   \\
   & & {1 \over a} \nabla \cdot \left( \dot {\bf u}_i
       + H {\bf u}_i \right)
       + 4 \pi G \varrho \delta
       = - {1 \over a^2} \nabla \cdot \left( {\bf u}_i \cdot
       \nabla {\bf u}_i \right)
       - \dot C^{(t)\alpha\beta} \left(
       {2 \over a} u_{\alpha|\beta} + \dot C^{(t)}_{\alpha\beta}
       \right)
   \nonumber \\
   & & \qquad
       + 4 \pi G \sum_j \varrho_j {1 \over c^2}
       \left\{
       {1 \over 2} |{\bf u}_j - {\bf u}|^2
       + 3 \Delta^{-1} \nabla \cdot \left[
       \left( {\bf u}_j - {\bf u} \right)
       \nabla \cdot \left( {\bf u}_j - {\bf u} \right) \right]
       \right\},
   \\
   & & {1 \over a^2} \left( a^2 \dot \delta \right)^\cdot
       - 4 \pi G \varrho \delta
       = - {1 \over a^2} \left[ a \nabla \cdot \left( \delta {\bf u}
       \right) \right]^\cdot
       + {1 \over a^2} \nabla \cdot \left( {\bf u} \cdot \nabla
       {\bf u} \right)
       + \dot C^{(t)\alpha\beta} \left(
       {2 \over a} u_{\alpha|\beta} + \dot C^{(t)}_{\alpha\beta}
       \right)
   \nonumber \\
   & & \qquad
       - \sum_j {\varrho_j \over \varrho} {1 \over c^2}
       \left\{ \left( 4 \pi G \varrho
       - {c^2 \over 2} {\Delta \over a^2} \right) |{\bf u}_j - {\bf u}|^2
       + \left( 24 \pi G \varrho
       - c^2 {\Delta \over a^2} \right)
       \Delta^{-1} \nabla \cdot \left[
       \left( {\bf u}_j - {\bf u} \right)
       \nabla \cdot \left( {\bf u}_j - {\bf u} \right) \right]
       \right\},
   \label{ddot-delta-eq} \\
   & & {1 \over a^2} \left( a^2 \dot \delta_i \right)^\cdot
       - 4 \pi G \varrho \delta
       = - {1 \over a^2} \left[ a \nabla \cdot \left( \delta_i {\bf
       u}_i \right) \right]^\cdot
       + {1 \over a^2} \nabla \cdot \left( {\bf u}_i \cdot \nabla
       {\bf u}_i \right)
       + \dot C^{(t)}_{\alpha\beta} \left(
       {2 \over a} u_i^{\alpha|\beta} + \dot C^{(t)\alpha\beta}
       \right)
   \nonumber \\
   & & \qquad
       + {1 \over a^2} \left\{ \Delta \left[ {\bf u} \cdot \left(
       {\bf u}_i - {\bf u} \right) \right]
       - \nabla \cdot \left[ \left( {\bf u}_i - {\bf u} \right)
       \cdot \nabla {\bf u}
       + {\bf u} \cdot \nabla \left( {\bf u}_i - {\bf u} \right)
       \right] \right\}
   \nonumber \\
   & & \qquad
       - {4 \pi G \varrho \over c^2} |{\bf u}_i - {\bf u}|^2
       - 8 \pi G \sum_j \varrho_j {1 \over c^2} \left\{
       |{\bf u}_j - {\bf u}|^2
       + 3 \Delta^{-1} \nabla \cdot \left[
       \left( {\bf u}_j - {\bf u} \right)
       \nabla \cdot \left( {\bf u}_j - {\bf u} \right) \right]
       \right\}.
   \label{ddot-delta-eq-i}
\eea

In Sec.\ \ref{sec:exact-solutions} we have shown that, to the linear
order, $({\bf u}_i - {\bf u})$ simply decays ($\propto a^{-1}$) in
an expanding phase. From Eq.\ (\ref{u_i-u-solution}) we have \bea
   & & {\bf u}_i - {\bf u} =
       {1 \over a} \left[ \nabla d_i ({\bf x})
       + {\bf D}_i ({\bf x})
       - {\bf D} ({\bf x}) \right],
   \nonumber \\
   & & \sum_j \varrho_j d_j \equiv 0, \quad
       \nabla \cdot {\bf D} \equiv 0 \equiv \nabla \cdot {\bf D}_i.
   \label{u_i-u-solution2}
\eea Thus, ${\bf u}_i - {\bf u}$ simply decays in an expanding
background. If we {\it ignore} these contributions from the velocity
differences, except for the presence of the tensor-type
perturbation, Eq.\ (\ref{ddot-delta-eq-i}) coincides exactly with
the zero-pressure limit of Newtonian result in Eq.\
(\ref{ddot-delta-eq-N-i}). Terms in the last two lines of Eq.\
(\ref{ddot-delta-eq-i}) are relativistic correction terms which
vanish for a single component case leading to Eq.\
(\ref{ddot-delta-eq}); the second line in Eq.\ (\ref{ddot-delta-eq})
also vanishes in the single component case.

Ignoring quadratic combination of $({\bf u}_i - {\bf u})$ terms, we
have \bea
   & & \dot \delta + {1 \over a} \nabla \cdot {\bf u}
       = - {1 \over a} \nabla \cdot \left( \delta {\bf u} \right),
   \label{dot-delta-second-2} \\
   & & {1 \over a} \nabla \cdot \left( \dot {\bf u} + H {\bf u}
       \right)
       + 4 \pi G \varrho \delta
       = - {1 \over a^2} \nabla \cdot \left( {\bf u} \cdot \nabla
       {\bf u} \right)
       - \dot C^{(t)\alpha\beta} \left(
       {2 \over a} u_{\alpha|\beta} + \dot C^{(t)}_{\alpha\beta}
       \right),
   \label{dot-u-second-2} \\
   & & {1 \over a^2} \left( a^2 \dot \delta \right)^\cdot
       - 4 \pi G \varrho \delta
       = - {1 \over a^2} \left[ a \nabla \cdot \left( \delta {\bf u}
       \right) \right]^\cdot
       + {1 \over a^2} \nabla \cdot \left( {\bf u} \cdot \nabla
       {\bf u} \right)
       + \dot C^{(t)\alpha\beta} \left(
       {2 \over a} u_{\alpha|\beta} + \dot C^{(t)}_{\alpha\beta}
       \right),
   \label{ddot-delta-second-2}
\eea and \bea
   & & \dot \delta_i
       + {1 \over a} \nabla \cdot {\bf u}_i
       = - {1 \over a} \nabla \cdot \left( \delta_i {\bf u}_i
       \right)
       + {1 \over a} \left[ 2 \varphi \nabla \cdot
       \left( {\bf u}_i - {\bf u} \right)
       - \left( {\bf u}_i - {\bf u} \right) \cdot \nabla \varphi
       + 2 \left( u^\alpha_i - u^\alpha \right)^{|\beta}
       C^{(t)}_{\alpha\beta} \right],
   \label{dot-delta-i-second-2} \\
   & & {1 \over a} \nabla \cdot \left( \dot {\bf u}_i
       + H {\bf u}_i \right)
       + 4 \pi G \varrho \delta
       = - {1 \over a^2} \nabla \cdot \left( {\bf u}_i \cdot
       \nabla {\bf u}_i \right)
       - \dot C^{(t)\alpha\beta} \left(
       {2 \over a} u_{\alpha|\beta} + \dot C^{(t)}_{\alpha\beta}
       \right),
   \label{dot-u-i-second-2} \\
   & & {1 \over a^2} \left( a^2 \dot \delta_i \right)^\cdot
       - 4 \pi G \varrho \delta
       = - {1 \over a^2} \left[ a \nabla \cdot \left( \delta_i {\bf
       u}_i \right) \right]^\cdot
       + {1 \over a^2} \nabla \cdot \left( {\bf u}_i \cdot \nabla
       {\bf u}_i \right)
       + \dot C^{(t)}_{\alpha\beta} \left(
       {2 \over a} u_i^{\alpha|\beta} + \dot C^{(t)\alpha\beta}
       \right)
   \nonumber \\
   & & \qquad
       + {1 \over a^2} \left\{ \Delta \left[ {\bf u} \cdot \left(
       {\bf u}_i - {\bf u} \right) \right]
       - \nabla \cdot \left[ \left( {\bf u}_i - {\bf u} \right)
       \cdot \nabla {\bf u}
       + {\bf u} \cdot \nabla \left( {\bf u}_i - {\bf u} \right)
       \right] \right\}.
   \label{ddot-delta-i-second-2}
\eea Equations
(\ref{dot-delta-second-2})-(\ref{ddot-delta-second-2}) coincide with
the density and velocity perturbation equations of a single
component medium \cite{second-order}; thus, except for the
contribution from gravitational waves, these equations coincide with
ones in the Newtonian context.

If we further ignore $({\bf u}_i - {\bf u})$ terms appearing in the
pure second-order combinations, Eqs.\
(\ref{dot-delta-i-second-2})-(\ref{ddot-delta-i-second-2}) become
\bea
   & & \dot \delta_i
       + {1 \over a} \nabla \cdot {\bf u}_i
       = - {1 \over a} \nabla \cdot \left( \delta_i {\bf u}_i
       \right),
   \label{dot-delta-i-second-3} \\
   & & {1 \over a} \nabla \cdot \left( \dot {\bf u}_i
       + H {\bf u}_i \right)
       + 4 \pi G \varrho \delta
       = - {1 \over a^2} \nabla \cdot \left( {\bf u}_i \cdot
       \nabla {\bf u}_i \right)
       - \dot C^{(t)\alpha\beta} \left(
       {2 \over a} u_{\alpha|\beta} + \dot C^{(t)}_{\alpha\beta}
       \right),
   \label{dot-u-i-second-3} \\
   & & {1 \over a^2} \left( a^2 \dot \delta_i \right)^\cdot
       - 4 \pi G \varrho \delta
       = - {1 \over a^2} \left[ a \nabla \cdot \left( \delta_i {\bf
       u}_i \right) \right]^\cdot
       + {1 \over a^2} \nabla \cdot \left( {\bf u}_i \cdot \nabla
       {\bf u}_i \right)
       + \dot C^{(t)}_{\alpha\beta} \left(
       {2 \over a} u_i^{\alpha|\beta} + \dot C^{(t)\alpha\beta}
       \right).
   \label{ddot-delta-i-second-3}
\eea Notice that, by ignoring $i$-indices, Eqs.
(\ref{dot-delta-i-second-3})-(\ref{ddot-delta-i-second-3}) coincide
with Eqs. (\ref{dot-delta-second-2})-(\ref{ddot-delta-second-2}). In
this context, except for the contribution from gravitational waves,
the above equations coincide exactly with ones in the Newtonian
context even in the multi-component case; compare with Eqs.\
(\ref{dot-delta-eq-N-i}),(\ref{dot-v-eq-N-i}),(\ref{ddot-delta-eq-N-i})
without pressure. In the single component situation such a
relativistic/Newtonian correspondence to the second order was shown
in \cite{NL,second-order}. In the present case, the same equation
valid in the single component is now valid in the multi-component
case for the collective fluid variables. This justifies our
identifications of Newtonian perturbation variables in Eq.\
(\ref{identify-second}).

%%%%%%%%%%%%%%%%%%%%%%%%%%%%%%%%%%%%%%%%%%%%%%%%%%%%%%%%%%%%%%%
%
%  Effects of curvature
%
%%%%%%%%%%%%%%%%%%%%%%%%%%%%%%%%%%%%%%%%%%%%%%%%%%%%%%%%%%%%%%%
\section{Effects of curvature}
                                          \label{sec:curvature}

We {\it consider} a single zero-pressure, irrotational fluid. We
{\it take} the temporal comoving gauge ($v \equiv 0$) and the
spatial $\gamma \equiv 0$ gauge. In the presence of background
curvature the basic equations are presented in Eqs.\
(\ref{E-conservation-single-irrotational-CG})-(\ref{Mom-constraint-single-irrotational-CG})
for nonvanishing pressure, or Eqs.\
(\ref{E-conservation-zero-pressure-irrotation-CG})-(\ref{ddot-eq-CG-i})
for zero-pressure multiple component fluids. By setting pressures
equal to zero in Eqs.\
(\ref{E-conservation-single-irrotational-CG})-(\ref{Mom-constraint-single-irrotational-CG}),
or from Eqs.\
(\ref{Raychaudhury-eq-zero-pressure-irrotation-CG})-(\ref{ddot-eq-CG}),
we have \bea
   & & \dot \delta
       - \kappa
       =
       - {c \over a^2} \delta_{,\alpha} \chi^{,\alpha}
       + \delta \kappa,
   \label{E-conservation-zK} \\
   & & \dot \kappa + 2 H \kappa
       - 4 \pi G \varrho \delta
       =
       - {c \over a^2} \kappa_{,\alpha} \chi^{,\alpha}
       + {1 \over 3} \kappa^2
       - {1 \over 3} \left( c {\Delta \over a^2} \chi \right)^2
       + \left( \dot C^{(t)\alpha\beta}
       + {c \over a^2} \chi^{,\alpha|\beta} \right)
       \left( \dot C^{(t)}_{\alpha\beta}
       + {c \over a^2} \chi_{,\alpha|\beta} \right),
   \label{kappa-eq-K} \\
   & & \ddot \delta + 2 H \dot \delta
       - 4 \pi G \varrho \delta
       = 4 \pi G \varrho \delta^2
       - {c \over a^2} \left(
       \delta_{,\alpha} \chi^{,\alpha} \right)^\cdot
       - {c \over a^2} \kappa_{,\alpha} \chi^{,\alpha}
       + {4 \over 3} \kappa^2
       - {1 \over 3} \left( c {\Delta \over a^2} \chi \right)^2
   \nonumber \\
   & & \qquad
       + \left( \dot C^{(t)\alpha\beta}
       + {c \over a^2} \chi^{,\alpha|\beta} \right)
       \left( \dot C^{(t)}_{\alpha\beta}
       + {c \over a^2} \chi_{,\alpha|\beta} \right),
   \label{ddot-eq-K} \\
   & & c {\Delta + 3 K \over a^2} \chi + \kappa
       = 0,
   \label{Mom-constraint-K}
\eea where Eq.\ (\ref{Mom-constraint-K}) is valid to the linear
order. Compared with the situation with vanishing curvature, the
effects of curvature in the above perturbed set of equations appear
only in the linear-order relation between $\kappa$ and $\chi$ in
Eq.\ (\ref{Mom-constraint-K}).

%%%%%%%%%%%%%%%%%%%%%%%%%%%%%%%%%%%%%%%%%%%%%%%%%%%%%%%%%%%%%%
\subsection{Newtonian correspondence}

Considering the successful Newtonian correspondence to the linear
order even in the presence of the background curvature, we {\it
assume} the identification in Eq.\ (\ref{kappa-identification}) is
valid to the second order. Then, to the linear order, from Eq.\
(\ref{Mom-constraint-K}) we have \bea
   & & \kappa \equiv - {1 \over a} \nabla \cdot {\bf u}
       \equiv - {\Delta \over a} u
       = - c {\Delta + 3 K \over a^2} \chi,
   \label{identification-K1}
\eea where ${\bf u} \equiv \nabla u$ and $\chi = \chi_v = - a
v_\chi$. Thus, to the linear order, formally we have \bea
   & & {c \over a} \chi
       = \left( 1 - {3 K \over \Delta + 3 K} \right) u.
   \label{identification-K2}
\eea

In the presence of curvature, the scalar-type perturbation can be
handled by solving Eqs.\
(\ref{E-conservation-zK}),(\ref{kappa-eq-K}),(\ref{Mom-constraint-K})
together with the identifications made above. We can formally
separate the effects of pure curvature contribution. Using Eqs.\
(\ref{identification-K1}),(\ref{identification-K2}), Eqs.\
(\ref{E-conservation-zK}),(\ref{kappa-eq-K}),(\ref{Mom-constraint-K})
become \bea
   & & \dot \delta + {1 \over a} \nabla \cdot {\bf u}
       = - {1 \over a} \nabla \cdot \left( \delta {\bf u} \right)
       + {1 \over a} \left( \nabla \delta \right) \cdot \nabla
       \left( {3K \over \Delta + 3K} u \right),
   \label{density-pert-K-N} \\
   & & {1 \over a} \nabla \cdot \left( \dot {\bf u}
       + H {\bf u} \right)
       + 4 \pi G \varrho \delta
       = - {1 \over a^2} \nabla \cdot \left(
       {\bf u} \cdot \nabla {\bf u} \right)
       - \dot C^{(t)\alpha\beta} \left( \dot C^{(t)}_{\alpha\beta}
       + {2 \over a} u_{,\alpha|\beta} \right)
       + {1 \over a^2} \nabla \cdot \left[ \left( {3 K \over \Delta
       + 3 K} {\bf u} \right) \cdot \nabla {\bf u} \right]
   \nonumber \\
   & & \qquad
       - {1 \over 3} {1 \over a^2} \left( {3K \over \Delta + 3K}
       \nabla \cdot {\bf u} \right)
       \left( {2 \Delta + 3 K \over \Delta + 3K} \nabla \cdot {\bf u}
       \right)
       + {1 \over a} \left( {3 K \over \Delta + 3 K} u
       \right)^{,\alpha|\beta}
       \left[ 2 \dot C^{(t)}_{\alpha\beta}
       + {1 \over a} \left( {\Delta \over \Delta + 3K} u
       \right)_{,\alpha|\beta} \right].
   \label{velocity-pert-K-N}
\eea From Eq.\ (\ref{BG-eq4}), the $K$ term can be written as \bea
   & & K = \left( {aH \over c} \right)^2 \left( \Omega_t - 1 \right),
       \quad
       \Omega_t \equiv \Omega + \Omega_\Lambda, \quad
       \Omega \equiv {8 \pi G \varrho \over 3 H^2}, \quad
       \Omega_\Lambda \equiv {\Lambda c^2 \over 3 H^2}.
\eea

%%%%%%%%%%%%%%%%%%%%%%%%%%%%%%%%%%%%%%%%%%%%%%%%%%%%%%%%%%%%%%%
%
%  Effects of rotation
%
%%%%%%%%%%%%%%%%%%%%%%%%%%%%%%%%%%%%%%%%%%%%%%%%%%%%%%%%%%%%%%
\section{Effects of vector-type perturbation}
                                        \label{sec:rotation}

The spatial $C$-gauge sets \bea
   & & \gamma \equiv 0 \equiv C^{(v)}_\alpha.
\eea The remaining variables under this gauge condition are
completely free of the spatial gauge modes and have unique spatially
gauge-invariant counterparts. If we simultaneously take {\it any}
temporal gauge which also removes the temporal gauge mode
completely, all the remaining variables have corresponding unique
gauge-invariant counterparts. The above statements are true to all
orders in perturbations, see Sec. VI of \cite{NL}. From Eq.\
(\ref{chi-def}) we have \bea
   & & \beta \equiv {1 \over a} \chi, \quad
       B^{(v)}_\alpha \equiv \Psi^{(v)}_\alpha.
\eea Thus, we have \bea
   & & B_\alpha = {1 \over a} \chi_{,\alpha}
       + \Psi^{(v)}_\alpha, \quad
       C_{\alpha\beta}
       \equiv \varphi g^{(3)}_{\alpha\beta}
       + C^{(t)}_{\alpha\beta}.
\eea As the temporal comoving gauge we set \bea
   & & v \equiv 0.
\eea As we mentioned, the remaining variables under these gauge
conditions are completely free of the gauge modes and have unique
gauge-invariant counterparts to all orders in perturbation, see
\cite{NL}.

%%%%%%%%%%%%%%%%%%%%%%%%%%%%%%%%%%%%%%%%%%%%%%%%%%%%%%%%%%%%%%
\subsection{Linear perturbations}

To the linear order, the three types of perturbations decouple, and
evolve independently. The rotational perturbation is described by
Eqs.\ (206)-(209) in \cite{NL}: \bea
   & & {\Delta + 2 K \over 2 a^2} \Psi_\alpha^{(v)}
       + {8 \pi G \over c^4} \left( \mu + p \right) v^{(v)}_\alpha = 0,
   \label{rot-linear-eq1} \\
   & & {\left[ a^4 \left( \mu + p \right) v^{(v)}_\alpha
       \right]^\cdot \over a^4 \left( \mu + p \right)}
       = - c {\Delta + 2 K \over 2 a^2} {\Pi^{(v)}_\alpha \over \mu + p},
   \\
   & & \dot \Psi^{(v)}_\alpha
       + 2 H \Psi^{(v)}_\alpha
       = {8 \pi G \over c^3} \Pi^{(v)}_\alpha,
\eea where the last equation follows from the first two. Compared
with Bardeen's notation in \cite{Bardeen-1980}, we have \bea
   & & B^{(v)}_\alpha = B^{(1)} Q^{(1)}_\alpha, \quad
       C^{(v)}_\alpha = - {1 \over k} H^{(1)}_T Q^{(1)}_\alpha,
       \quad
       \Psi^{(v)}_\alpha = \Psi Q^{(1)}_\alpha
       \equiv \left( B^{(1)} - {a \over k} \dot H^{(1)}_T \right)
       Q^{(1)}_\alpha,
   \nonumber \\
   & & v^{(v)}_\alpha
       = v_c Q^{(1)}_\alpha
       \equiv \left( v^{(1)} - B^{(1)} \right)
       Q^{(1)}_\alpha, \quad
       v^{(v)}_\alpha + \Psi^{(v)}_\alpha
       = v^{(1)}_s Q^{(1)}_\alpha
       \equiv \left( v^{(1)} - {a \over k} \dot H^{(1)}_T \right)
       Q^{(1)}_\alpha,
   \label{Bardeen-rotation}
\eea where Bardeen's $v^{(1)}_c$ and $v^{(1)}_s$, thus our
$v^{(v)}_\alpha$ and $v^{(v)}_\alpha + \Psi^{(v)}_\alpha$, are
related to the vorticity and the shear, respectively. From Eq.\
(\ref{kinematic-quantities-pert}), to the linear order, we have \bea
   & & \tilde \omega_{\alpha\beta}
       = a v^{(v)}_{[\alpha|\beta]},
   \nonumber \\
   & & \tilde \sigma_{\alpha\beta}
       = a \left[ - \left( \nabla_\alpha \nabla_\beta
       - {1 \over 3} g^{(3)}_{\alpha\beta} \Delta \right)
       \left( v - {1 \over a} \chi \right)
       + \left( v^{(v)}_{(\alpha} + \Psi^{(v)}_{(\alpha} \right)_{|\beta)}
       + C^{(t)\prime}_{\alpha\beta} \right].
\eea Bardeen called $\Psi^{(v)}_\alpha$ a `frame-dragging
potential'. The difference between $v_c^{(1)}$ and $v_s^{(1)}$ is
crucially important to show Mach's principle including the linear
order rotational perturbation in \cite{Schmid}.

In the absence of anisotropic stress which can act as a sink or
source of the angular momentum, we have \bea
   & & {\rm Angular \;\; momentum}
       \propto a^4 \left( \mu + p \right) v^{(v)}_\alpha
       \propto a^2 \Psi^{(v)}_\alpha
       \propto {\rm constant \; in \; time}.
\eea Thus, for vanishing anisotropic stress, we have \bea
   & & v^{(v)}_\alpha \propto {1 \over a^4 (\mu + p)}, \quad
       \Psi^{(v)}_\alpha \propto {1 \over a^2}.
\eea In the zero-pressure limit we have \bea
   & & v^{(v)}_\alpha \propto {1 \over a}, \quad
       \Psi^{(v)}_\alpha \propto {1 \over a^2}.
\eea

%%%%%%%%%%%%%%%%%%%%%%%%%%%%%%%%%%%%%%%%%%%%%%%%%%%%%%%%%%%%%%
\subsection{Second-order perturbations}

We consider a single component situation with general pressure. We
set $K \equiv 0$. To the linear-order we use \bea
   & & \chi \equiv \chi_v \equiv - a v_\chi, \quad
       \kappa \equiv \kappa_v = c {\Delta \over a} v_\chi.
\eea The scalar-type perturbation is described by Eqs.\
(\ref{E-conservation}),(\ref{Raychaudhury-eq}) which give \bea
   & & \dot \delta
       + 3 H \left( c_s^2 - w \right) \delta
       + 3 H {e \over \mu}
       - \left( 1 + w \right) \kappa
   \nonumber \\
   & & \qquad
       = - 3 H \left( 1 + w \right) \left[
       \alpha - {1 \over 2} \alpha^2
       + {1 \over 2} \left( v_\chi^{\;\;,\alpha} -
       \Psi^{(v)\alpha} \right) \left( v_{\chi,\alpha} -
       \Psi^{(v)}_\alpha \right) \right]
       + {c \over a} \delta_{,\alpha} \left( v_\chi^{\;\;,\alpha}
       - \Psi^{(v)\alpha} \right)
   \nonumber \\
   & & \qquad
       + \left( \delta + {\delta p \over \mu} \right)
       \left( c {\Delta \over a} v_\chi
       - 3 H \alpha \right)
       + \left( 1 + w \right) \alpha c {\Delta \over a} v_\chi
       + \left( 1 + w \right) v^{(v)}_\alpha
       \left[ \left( 1 - 3 c_s^2 \right) H v^{(v)\alpha}
       + {c \over \mu + p} {\Delta \over a^2} \Pi^{(v)\alpha}
       \right]
   \nonumber \\
   & & \qquad
       + {c \over a} \left( 1 + w \right) \left[ - \left( 2 \alpha +
       \varphi \right)_{,\alpha} v^{(v)\alpha}
       + 2 C^{(t)\alpha\beta} v^{(v)}_{\beta|\alpha} \right]
       - \dot \varphi {1 \over \mu} \Pi^\alpha_\alpha
       - \left( \dot C^{(t)}_{\alpha\beta}
       - {c \over a} v_{\chi,\alpha|\beta}
       + {c \over a} \Psi^{(v)}_{\alpha|\beta} \right)
       {1 \over \mu} \Pi^{\alpha\beta}
   \nonumber \\
   & & \qquad
       - {1 \over a \mu} \left[ \left( \delta \mu + \delta p \right)_{,\alpha}
       v^{(v)\alpha}
       + \left( \Pi^{\alpha\beta} v^{(v)}_\beta \right)_{|\alpha} \right],
   \label{E-conservation-single-rot} \\
   & & \dot \kappa
       + 2 H \kappa
       - {4 \pi G \over c^2} \left( \delta \mu + 3 \delta p \right)
       = - \left( 3 \dot H + c^2 {\Delta \over a^2} \right)
       \left[ \alpha
       + {1 \over 2} \left( v_\chi^{\;\;,\alpha} -
       \Psi^{(v)\alpha} \right) \left( v_{\chi,\alpha} -
       \Psi^{(v)}_\alpha \right) \right]
       - 2 H \alpha c {\Delta \over a} v_\chi
   \nonumber \\
   & & \qquad
       + {4 \pi G \over c^2} \left( \delta \mu + 3 \delta p \right) \alpha
       + {3 \over 2} \dot H \alpha^2
       + \left( \alpha + 2 \varphi \right) c^2 {\Delta \over a^2} \alpha
       + {c^2 \over a^2} \alpha^{,\alpha}
       \left( \alpha - \varphi \right)_{,\alpha}
       + {2 c^2 \over a^2} C^{(t)\alpha\beta} \alpha_{,\alpha|\beta}
       + {8 \pi G \over c^2} \left( \mu + p \right) v^{(v)\alpha} v^{(v)}_\alpha
   \nonumber \\
   & & \qquad
       + \left( v_\chi^{\;\;,\alpha} -
       \Psi^{(v)\alpha} \right) \left( c^2 {\Delta \over a^2} v_\chi
       \right)_{,\alpha}
       + \left( \dot C^{(t)\alpha\beta}
       - {c \over a} v_\chi^{\;\;,\alpha|\beta}
       + {c \over a} \Psi^{(v)(\alpha|\beta)} \right)
       \left( \dot C^{(t)}_{\alpha\beta}
       - {c \over a} v_{\chi,\alpha|\beta}
       + {c \over a} \Psi^{(v)}_{\alpha|\beta} \right).
    \label{Raychaudhury-eq-single-rot}
\eea The vector-type perturbation is described by Eq.\
(\ref{Mom-conservation}) which gives \bea
   & & {1 \over a^4} \left[ a^4 \left( \mu + p \right)
       v^{(v)}_\alpha \right]^\cdot
       + {c \over a} \left( \mu + p \right) \alpha_{,\alpha}
       + {c \over a} \left( \delta p_{,\alpha}
       + \Pi^\beta_{\alpha|\beta} \right)
   \nonumber \\
   & & \qquad
       = {c \over a} \left\{
       - \alpha \delta p
       + {1 \over 2} \left( \mu + p \right)
       \left[ \alpha^2 - \left( v_\chi^{\;\;,\beta} -
       \Psi^{(v)\beta} \right) \left( v_{\chi,\beta} -
       \Psi^{(v)}_\beta \right) \right] \right\}_{,\alpha}
       - {c \over a} \alpha_{,\alpha} \delta \mu
   \nonumber \\
   & & \qquad \qquad
       + \left( \mu + p \right) \left( c {\Delta \over a} v_\chi
       - 3 H \alpha \right) v^{(v)}_\alpha
       - {c \over a} \left( \mu + p \right) \left[
       v^{(v)}_\beta \left( - v^{\;\;,\beta}_\chi + \Psi^{(v)\beta}
       \right)_{|\alpha}
       + v^{(v)}_{\alpha|\beta} \left( v^{(v)\beta}
       - v_\chi^{\;\;,\beta} + \Psi^{(v)\beta} \right) \right]
   \nonumber \\
   & & \qquad \qquad
       + {c \over a} \left[
       2 \varphi \Pi^\beta_{\alpha|\beta}
       - \varphi_{,\beta} \Pi^\beta_\alpha
       + \varphi_{,\alpha} \Pi^\beta_\beta
       - \left( \alpha \Pi^\beta_\alpha \right)_{|\beta}
       + 2 C^{(t)\beta\gamma} \Pi_{\alpha\beta|\gamma}
       + C^{(t)\gamma}_{\;\;\;\;\;\beta|\alpha} \Pi^\beta_\gamma
       \right]
   \nonumber \\
   & & \qquad \qquad
       - {1 \over a^4} \left\{ a^4 \left[
       \left( \delta \mu + \delta p \right)
       v^{(v)}_\alpha
       + \Pi^\beta_\alpha v^{(v)}_\beta \right] \right\}^\cdot
   \nonumber \\
   & & \qquad
       \equiv \mu c A_\alpha.
   \label{Mom-conservation-single-rot}
\eea We have $[A_\alpha] = L^{-1}$. From this we have \bea
   & & \left( \mu + p \right) \alpha
       + \delta p + { 2\over 3} {\Delta \over a^2} \Pi
       = a \mu \Delta^{-1} \nabla \cdot {\bf A},
   \label{Mom-conservation-single-rot-scalar}
   \\
   & & {1 \over a^4} \left[ a^4 \left( \mu + p \right)
       v^{(v)}_\alpha \right]^\cdot
       + c {\Delta \over 2 a^2} \Pi^{(v)}_\alpha
       = \mu c \left[ A_\alpha
       - \left( \Delta^{-1} \nabla \cdot {\bf A}
       \right)_{,\alpha} \right].
   \label{Mom-conservation-single-rot-vector}
\eea Equation for the tensor-type perturbation (gravitational waves)
follows from Eqs.\ (\ref{tracefree-ADM-propagation}),(\ref{chi-eq}).

%%%%%%%%%%%%%%%%%%%%%%%%%%%%%%%%%%%%%%%%%%%%%%%%%%%%%%%%%%%%%%
\subsection{Zero-pressure case}

In the zero-pressure limit, Eqs.\
(\ref{Mom-conservation-single-rot-scalar}),(\ref{Mom-conservation-single-rot})
give \bea
   & & \alpha = a \Delta^{-1} \nabla \cdot {\bf A},
   \\
   & & a A_\alpha
       \equiv - {1 \over 2} \left[ \left( v_\chi^{\;\;,\beta} -
       \Psi^{(v)\beta} \right) \left( v_{\chi,\beta} -
       \Psi^{(v)}_\beta \right) \right]_{,\alpha}
       + v^{(v)}_{\alpha|\beta} \left( v_\chi^{\;\;,\beta}
       - \Psi^{(v)\beta} \right)
       + v^{(v)}_\beta \left( v_\chi^{\;\;,\beta}
       - \Psi^{(v)\beta} \right)_{|\alpha}
       - v^{(v)}_{\alpha|\beta} v^{(v)\beta},
\eea thus, $\alpha$ is purely second-order, and \bea
   & & \alpha
       + {1 \over 2} \left( v_\chi^{\;\;,\beta} -
       \Psi^{(v)\beta} \right) \left( v_{\chi,\beta} -
       \Psi^{(v)}_\beta \right)
       = \Delta^{-1} \nabla^\alpha \left[
       v^{(v)}_{\alpha|\beta} \left( v_\chi^{\;\;,\beta}
       - \Psi^{(v)\beta} \right)
       + v^{(v)}_\beta \left( v_\chi^{\;\;,\beta}
       - \Psi^{(v)\beta} \right)_{|\alpha}
       - v^{(v)}_{\alpha|\beta} v^{(v)\beta}
       \right].
\eea Equations (\ref{E-conservation-single-rot}),
(\ref{Raychaudhury-eq-single-rot}),
(\ref{Mom-conservation-single-rot-vector}) give \bea
   & & \dot \delta
       - \kappa
       =
       {c \over a} \left[ \delta \left( v_\chi^{\;\;,\alpha} -
       \Psi^{(v)\alpha} - v^{(v)\alpha} \right) \right]_{,\alpha}
       + H v^{(v)}_\alpha v^{(v)\alpha}
       + {c \over a} \left( - \varphi^{,\alpha} v^{(v)}_{\alpha}
       + 2 C^{(t)\alpha\beta} v^{(v)}_{\beta|\alpha} \right)
   \nonumber \\
   & & \qquad
       - 3 H \Delta^{-1} \nabla^\alpha \left[
       v^{(v)}_{\alpha|\beta} \left( v_\chi^{\;\;,\beta}
       - \Psi^{(v)\beta} \right)
       + v^{(v)}_\beta \left( v_\chi^{\;\;,\beta}
       - \Psi^{(v)\beta} \right)_{|\alpha}
       - v^{(v)}_{\alpha|\beta} v^{(v)\beta}
       \right],
   \label{E-conservation-single-rot-zero-p} \\
   & & \dot \kappa
       + 2 H \kappa
       - 4 \pi G \varrho \delta
       =
       {c^2 \over a^2} \left[
       \left( v_{\chi}^{\;\;,\alpha} - \Psi^{(v)\alpha} \right)
       \left( v_{\chi}^{\;\;,\beta} - \Psi^{(v)\beta} \right)_{|\alpha}
       \right]_{|\beta}
       + \dot C^{(t)\alpha\beta}
       \left[ \dot C^{(t)}_{\alpha\beta}
       - {2 c \over a} \left( v_{\chi,\alpha|\beta}
       - \Psi^{(v)}_{\alpha|\beta} \right) \right]
   \nonumber \\
   & & \qquad
       + 8 \pi G \varrho v^{(v)\alpha} v^{(v)}_\alpha
       - \left( 3 \dot H + c^2 {\Delta \over a^2} \right)
       \Delta^{-1} \nabla^\alpha \left[
       v^{(v)}_{\alpha|\beta} \left( v_\chi^{\;\;,\beta}
       - \Psi^{(v)\beta} \right)
       + v^{(v)}_\beta \left( v_\chi^{\;\;,\beta}
       - \Psi^{(v)\beta} \right)_{|\alpha}
       - v^{(v)}_{\alpha|\beta} v^{(v)\beta}
       \right],
    \label{Raychaudhury-eq-single-rot-zero-p} \\
   & & \dot v^{(v)}_\alpha + H v^{(v)}_\alpha
       = {c \over a} \left[
       v^{(v)}_{\alpha|\beta} \left( v_\chi^{\;\;,\beta}
       - \Psi^{(v)\beta} \right)
       + v^{(v)}_\beta \left( v_\chi^{\;\;,\beta}
       - \Psi^{(v)\beta} \right)_{|\alpha}
       - v^{(v)}_{\alpha|\beta} v^{(v)\beta} \right]
   \nonumber \\
   & & \qquad
       - {c \over a} \nabla_\alpha \Delta^{-1} \nabla^\gamma
       \left[
       v^{(v)}_{\gamma|\beta} \left( v_\chi^{\;\;,\beta}
       - \Psi^{(v)\beta} \right)
       + v^{(v)}_\beta \left( v_\chi^{\;\;,\beta}
       - \Psi^{(v)\beta} \right)_{|\gamma}
       - v^{(v)}_{\gamma|\beta} v^{(v)\beta} \right].
   \label{Mom-conservation-single-rot-vector-zero-p}
\eea

%%%%%%%%%%%%%%%%%%%%%%%%%%%%%%%%%%%%%%%%%%%%%%%%%%%%%%%%%%%%%%
\subsection{Newtonian correspondence}

In order to compare with Newtonian equations we continue identifying
$\kappa$ as in Eq.\ (\ref{kappa-identification}) to the second
order. To the linear order we identify \bea
   & & {\bf u} \equiv - c \left( \nabla v_\chi - {\bf v}^{(v)}
       \right) \equiv \nabla u + {\bf u}^{(v)},
\eea thus \bea
   & & u \equiv - c v_\chi, \quad
       {\bf u}^{(v)} \equiv c {\bf v}^{(v)}.
\eea We introduce the following notations \bea
   & & U_\alpha \equiv u_\alpha + c \Psi^{(v)}_\alpha, \quad
       \tilde U_{\alpha} \equiv u_{,\alpha} + c \Psi^{(v)}_\alpha.
\eea As mentioned below Eq.\ (\ref{Bardeen-rotation}), to the linear
order, $u_\alpha$ and $U_\alpha$ are related to the vorticity and
the shear, respectively. Equations
(\ref{E-conservation-single-rot-zero-p})-(\ref{Mom-conservation-single-rot-vector-zero-p})
become \bea
   & & \dot \delta + {1 \over a} \nabla \cdot {\bf u}
       = - {1 \over a} \nabla \cdot \left( \delta {\bf U}
       \right)
       + {2 \over a} C^{(t)\alpha\beta} u^{(v)}_{\alpha|\beta}
       - {1 \over a} {\bf u}^{(v)} \cdot \nabla \varphi
       + {1 \over c^2} H \left[
       {\bf u}^{(v)} \cdot {\bf u}^{(v)}
       + 3 \Delta^{-1} \nabla^\alpha \left(
       u^{(v)}_{\alpha|\beta} U^\beta
       + u^{(v)\beta} \tilde U_{\beta|\alpha}
       \right) \right],
   \label{rot-second-order-eq1} \\
   & & {1 \over a} \nabla \cdot \left( \dot {\bf u} + H {\bf u}
       \right)
       + 4 \pi G \varrho \delta
       = - {1 \over a^2} \nabla \cdot \left( {\bf U} \cdot
       \nabla {\bf U} \right)
       - \dot C^{(t)\alpha\beta} \left( \dot C^{(t)}_{\alpha\beta}
       + {2 \over a} \tilde U_{\alpha|\beta} \right)
   \nonumber \\
   & & \qquad
       + {8 \pi G \varrho \over c^2} \left[
       - {\bf u}^{(v)} \cdot {\bf u}^{(v)}
       + {3 \over 2} \Delta^{-1} \nabla^\alpha
       \left(
       u^{(v)}_{\alpha|\beta} U^\beta
       + u^{(v)\beta} \tilde U_{\beta|\alpha}
       \right) \right],
   \label{rot-second-order-eq2} \\
   & & \dot {\bf u}^{(v)}
       + H {\bf u}^{(v)}
       = - {1 \over a} \left[
       {\bf U} \cdot \nabla {\bf u}
       - \nabla \Delta^{-1} \nabla \cdot \left(
       {\bf U} \cdot \nabla {\bf u} \right) \right]
       - {1 \over a} \left[
       U^\beta c \Psi^{(v)}_{\beta|\alpha}
       - \nabla_\alpha \Delta^{-1} \nabla^\gamma \left(
       U^\beta c \Psi^{(v)}_{\beta|\gamma} \right) \right].
   \label{rot-second-order-eq3}
\eea From Eq.\ (\ref{rot-second-order-eq3}) we notice that the
tensor-type perturbation does not affect the vector-type
perturbation to the second order. The pure scalar-type perturbation
also cannot generate the vector-type perturbation to the
second-order; the same is true in the Newtonian case, see below Eq.\
(\ref{dot-rotation-eq-N-i2}).

From Eq.\ (\ref{rot-linear-eq1}), to the linear order, we have \bea
   & & v^{(v)}_\alpha = {1 \over 6} \left( {c k \over a H} \right)^2
       \Psi^{(v)}_\alpha,
\eea where $k$ is a comoving wavenumber with $\Delta \equiv - k^2$,
thus $[k] = 1$. Since $(ck)/(aH) \sim$ (visual-horizon)/(scale), we
have \bea
   & & {\rm far \; inside \; horizon:} \quad \;\;
       v_\alpha^{(v)} \gg \Psi^{(v)}_\alpha, \quad
       {\bf U} \simeq {\bf u}
       = \nabla u + {\bf u}^{(v)}, \quad
       \tilde {\bf U} \simeq \nabla u,
   \nonumber \\
   & & {\rm far \; outside \; horizon:} \quad
       v_\alpha^{(v)} \ll \Psi^{(v)}_\alpha, \quad
       {\bf U} \simeq \tilde {\bf U}
       = u_{,\alpha} + c \Psi^{(v)}_\alpha.
\eea Apparently, contributions of vector-type perturbation to the
second order depend on the visual-horizon scale.

Far inside the horizon, we can ignore $c\Psi^{(v)}_\alpha$ compared
with $u^{(v)}_\alpha$. In the matter dominated era we have
\bea
   & & \varphi_v = {5 \over 3} \varphi_\chi
       = {5 \over 3} {a^2 \over k^2} {4 \pi G \varrho \over c^2} \delta_v
       = {5 \over 2} \left( {aH \over k c} \right)^2 \delta_v,
\eea where we used Eqs.\ (293),(329),(330) of \cite{NL}. Thus, the
third term in the right-hand-side of Eq.\
(\ref{rot-second-order-eq1}) is $(aH/kc)^2$-order smaller than the
first term. The fourth (and last) term in the right-hand-side of
Eq.\ (\ref{rot-second-order-eq1}) is $(aH/kc) [{\bf u}^{(v)}/(c
\delta)]$-order smaller than the first term. The third (and last)
term in the right-hand-side of Eq.\ (\ref{rot-second-order-eq2}) is
also $(aH/kc)^2$-order smaller than the first term. Thus, Eqs.\
(\ref{rot-second-order-eq1})-(\ref{rot-second-order-eq3}) give \bea
   & & \dot \delta + {1 \over a} \nabla \cdot {\bf u}
       = - {1 \over a} \nabla \cdot \left( \delta {\bf u}
       \right)
       + {2 \over a} C^{(t)\alpha\beta} u^{(v)}_{\alpha|\beta},
   \label{rot-second-order-N-eq1} \\
   & & {1 \over a} \nabla \cdot \left( \dot {\bf u} + H {\bf u}
       \right)
       + 4 \pi G \varrho \delta
       = - {1 \over a^2} \nabla \cdot \left( {\bf u} \cdot
       \nabla {\bf u} \right)
       - \dot C^{(t)\alpha\beta} \left[ \dot C^{(t)}_{\alpha\beta}
       + {2 \over a} \left( u_{,\alpha|\beta}
       + c \Psi^{(v)}_{\alpha|\beta} \right) \right],
   \label{rot-second-order-N-eq2} \\
   & & \dot {\bf u}^{(v)}
       + H {\bf u}^{(v)}
       = - {1 \over a} \left[
       {\bf u} \cdot \nabla {\bf u}
       - \nabla \Delta^{-1} \nabla \cdot \left(
       {\bf u} \cdot \nabla {\bf u} \right) \right].
   \label{rot-second-order-N-eq3}
\eea Thus, if we could ignore the tensor-type combination in Eqs.\
(\ref{rot-second-order-N-eq1}),(\ref{rot-second-order-N-eq2}), Eqs.\
(\ref{rot-second-order-N-eq1})-(\ref{rot-second-order-N-eq3})
coincide exactly with the Newtonian equations: see Eqs.
(\ref{dot-delta-eq-N-i}),(\ref{dot-velocity-eq-N-i}),(\ref{dot-rotation-eq-N-i2})
ignoring the pressure terms and the subindices $i$. The
vector-tensor combinations in Eq.\
(\ref{rot-second-order-N-eq1}),(\ref{rot-second-order-N-eq2}) are
new relativistic contributions of the vector-type perturbations;
compare these two equations with Eqs.\
(\ref{E-conserv-eq5}),(\ref{Raychaudhury-eq5}) which are valid in
the absence of the vector-type perturbations. Notice the form of
last term $u_{,\alpha|\beta} + c \Psi^{(v)}_{\alpha|\beta}$ in Eq.\
(\ref{rot-second-order-N-eq2}) which subtly differs from the
expression $u_{\alpha|\beta}$ in Eq.\ (\ref{Raychaudhury-eq5}).

Contributions from the vector-type perturbation become more
complicated near and outside the horizon scale. The presence of
vector-type metric perturbation $\Psi^{(v)}_\alpha$, the scalar-type
curvature perturbation $\varphi$, and the tensor-type perturbation
$C^{(t)}_{\alpha\beta}$ coupled with the vector-type perturbation
give additional effects.

%%%%%%%%%%%%%%%%%%%%%%%%%%%%%%%%%%%%%%%%%%%%%%%%%%%%%%%%%%%%%%
\subsection{Pure vector-type perturbations}

In Sec.\ VII-E of \cite{NL} we have considered a situation with pure
vector-type perturbation. As the analysis was made based on the
fluid quantities in the normal frame, in the following we present
the case based on the fluid quantities in the energy frame.
Considering only the vector-type perturbation of a fluid, Eq.\
(\ref{Mom-conservation}) gives \bea
   & & {1 \over a^4 (\mu + p)} \left\{ a^4 \left[ \left( \mu + p
       \right) v^{(v)}_\alpha
       + {1 \over a} \Pi^{(v)}_{(\alpha|\beta)} v^{(v)\beta} \right]
       \right\}^\cdot
   \nonumber \\
   & & \qquad
       = - {c (\Delta + 2 K) \Pi^{(v)}_\alpha \over 2 a^2 (\mu + p)}
       - {c \over a} \left[ \left( v^{(v)\beta} + \Psi^{(v)|\beta}
       \right) \left( v^{(v)}_{\alpha|\beta}
       + \Psi^{(v)}_{\beta|\alpha} \right)
       - {2 \over 3} v^{(v)\beta} v^{(v)}_{\beta|\alpha} \right].
\eea Thus, for $\Pi^{(v)}_\alpha = 0$ we have \bea
   & & {1 \over a^4 (\mu + p)} \left[ a^4 \left( \mu + p
       \right) v^{(v)}_\alpha \right]^\cdot
       =
       - {c \over a} \left[ \left( v^{(v)\beta} + \Psi^{(v)|\beta}
       \right) \left( v^{(v)}_{\alpha|\beta}
       + \Psi^{(v)}_{\beta|\alpha} \right)
       - {2 \over 3} v^{(v)\beta} v^{(v)}_{\beta|\alpha} \right].
\eea This differs from Eq.\ (365) of \cite{NL} which is due to the
difference in the frame choice. The momentum constraint equation in
Eq.\ (101) of \cite{NL} becomes \bea
   & & {\Delta + 2 K \over 2 a^2} \Psi_\alpha^{(v)}
       + {8 \pi G \over c^4} \left( \mu + p \right) v^{(v)}_\alpha
       = - {8 \pi G \over c^4} {1 \over a}
       \Pi^{(v)}_{(\alpha|\beta)} v^{(v)\beta}.
\eea

Under the gauge transformation, from Eq.\ (\ref{GT}) we have \bea
   & & \hat v^{(v)}_\alpha
       = v^{(v)}_\alpha
       - v^{(v)}_\beta \xi^{(v)\beta}_{\;\;\;\;\;\;\;,\alpha}
       - v^{(v)}_{\alpha,\beta} \xi^{(v)\beta}.
   \label{GT-vector}
\eea To the linear order, from Eq.\ (230) of \cite{NL} we have $\hat
B^{(v)}_\alpha = B^{(v)}_\alpha + \xi^{(v)\prime}_\alpha$. We
consider a gauge transformation from the $C$-gauge ($C^{(v)}_\alpha
\equiv 0$, without hat) to the $B$-gauge ($B^{(v)}_\alpha \equiv 0$,
with hat). We have $\hat B^{(v)}_\alpha \equiv 0$, and
$B^{(v)}_\alpha |_{C-{\rm gauge}} = - \xi^{(v)\prime}_\alpha$. Thus,
\bea
   & & \xi^{(v)}_\alpha = - \int^\eta B^{(v)}_\alpha |_{C-{\rm gauge}} d \eta
       = - a^2 \Psi^{(v)}_\alpha \int^\eta {d \eta \over a^2},
\eea where we used $B^{(v)}_\alpha |_{C-{\rm gauge}} =
\Psi^{(v)}_\alpha \propto a^{-2}$. Thus, Eq.\ (\ref{GT-vector})
gives \bea
   & & v^{(v)}_\alpha |_{B-{\rm gauge}}
       = \left[ v^{(v)}_\alpha
       - \left( v^{(v)}_\beta \Psi^{(v)\beta}_{\;\;\;\;\;\;\;,\alpha}
       - v^{(v)}_{\alpha,\beta} \Psi^{(v)\beta} \right)
       a^2 \int^\eta {d \eta \over a^2}
       \right]_{C-{\rm gauge}}.
\eea

%%%%%%%%%%%%%%%%%%%%%%%%%%%%%%%%%%%%%%%%%%%%%%%%%%%%%%%%%%%%%%%
%
% MSFs
%
%%%%%%%%%%%%%%%%%%%%%%%%%%%%%%%%%%%%%%%%%%%%%%%%%%%%%%%%%%%%%%%
\section{Equations with fields}
                                      \label{sec:fields}

%%%%%%%%%%%%%%%%%%%%%%%%%%%%%%%%%%%%%%%%%%%%%%%%%%%%%%%%%%%%%%%
\subsection{A minimally coupled scalar field}

Equations in the case of a minimally coupled scalar field are
presented in Eqs.\ (112)-(114) of \cite{NL}. The equation of motion
in Eq.\ (112) and the full Einstein's equations in Eqs.\ (99)-(105)
expressed using the normal-frame fluid quantities together with the
normal-frame fluid quantities for the scalar field in Eqs.\ (114)
all in \cite{NL} provide a complete set of equations we need to the
second-order. The fluid quantities in Eq.\ (114) of \cite{NL} are
presented in the normal-frame four-vector and it is convenient to
know the conventionally used fluid quantities which are based on the
energy-frame four-vector. These latter quantities can be read from
Eqs.\ (88),(114) of \cite{NL} and Eq.\ (\ref{u-decomp}) as \bea
   & & Q_\alpha^{N(\phi)}
       = - {1 \over a} \left[ \dot \phi \delta \phi_{,\alpha}
       + \delta \phi_{,\alpha} \left( \delta \dot \phi
       - \dot \phi A \right) \right]
   \nonumber \\
   & & \qquad
       = \left( \mu^{(\phi)} + p^{(\phi)} \right)
       \left( V^{(\phi)}_\alpha - B_\alpha
       + A B_\alpha
       + 2 V^{(\phi)\beta} C_{\alpha\beta} \right)
       + \left( \delta \mu^{(\phi)} + \delta p^{(\phi)} \right)
       \left( V^{(\phi)}_\alpha - B_\alpha \right)
   \nonumber \\
   & & \qquad
       = \left( \mu^{(\phi)} + p^{(\phi)}
       + \delta \mu^{(\phi)} + \delta p^{(\phi)} \right)
       \left( - v^{(\phi)}_{,\alpha} + v^{(\phi,v)}_\alpha \right),
   \nonumber \\
   & &
       \delta \mu^{(\phi)}
       = \delta \mu^{N(\phi)}
       - {1 \over a^2} \delta \phi^{,\alpha} \delta \phi_{,\alpha},
       \quad
       \delta p^{(\phi)}
       = \delta p^{N(\phi)}
       - {1 \over 3 a^2} \delta \phi^{,\alpha} \delta
       \phi_{,\alpha},
   \nonumber \\
   & &
       \Pi^{(\phi)}_{\alpha\beta}
       = \Pi^{N(\phi)}_{\alpha\beta}
       - {1 \over a^2} \left( \delta \phi_{,\alpha} \delta
       \phi_{,\beta}
       - {1 \over 3} g^{(3)}_{\alpha\beta}
       \delta \phi^{,\gamma} \delta \phi_{,\gamma} \right).
\eea Thus, using Eq.\ (114) of \cite{NL} we have \bea
   & & v^{(\phi)}
       = {1 \over a \dot \phi} \delta \phi
       - {1 \over a \dot \phi^2} \Delta^{-1} \nabla^\alpha
       \left[
       \left( \delta \dot \phi - A \dot \phi \right) \delta \phi_{,\alpha} \right],
   \nonumber \\
   & &
       v^{(\phi,v)}_\alpha
       = {1 \over a \dot \phi^2} \left\{
       \left( \delta \dot \phi - \dot \phi A
       \right) \delta \phi_{,\alpha}
       - \nabla_\alpha \Delta^{-1} \nabla^\beta \left[
       \left( \delta \dot \phi - \dot \phi A
       \right) \delta \phi_{,\beta} \right] \right\},
   \nonumber \\
   & &
       \delta \mu^{(\phi)}
       = \dot \phi \delta \dot \phi
       - \dot \phi^2 A
       + V_{,\phi} \delta \phi
       + {1 \over 2} \delta \dot \phi^2
       - {1 \over 2 a^2} \delta \phi^{,\alpha} \delta \phi_{,\alpha}
       + {1 \over 2} V_{,\phi\phi} \delta \phi^2
       - 2 \dot \phi \delta \dot \phi A
       + {1 \over a} \dot \phi \delta \phi_{,\alpha} B^\alpha
       + 2 \dot \phi^2 A^2
       - {1 \over 2} \dot \phi^2 B^\alpha B_\alpha,
   \nonumber \\
   & &
       \delta p^{(\phi)}
       = \dot \phi \delta \dot \phi
       - \dot \phi^2 A
       - V_{,\phi} \delta \phi
       + {1 \over 2} \delta \dot \phi^2
       - {1 \over 2 a^2} \delta \phi^{,\alpha} \delta \phi_{,\alpha}
       - {1 \over 2} V_{,\phi\phi} \delta \phi^2
       - 2 \dot \phi \delta \dot \phi A
       + {1 \over a} \dot \phi \delta \phi_{,\alpha} B^\alpha
       + 2 \dot \phi^2 A^2
       - {1 \over 2} \dot \phi^2 B^\alpha B_\alpha,
   \nonumber \\
   & &
       \Pi^{(\phi)}_{\alpha\beta}
       = 0.
   \label{fluids-MSF}
\eea Notice that no anisotropic stress is caused by a minimally
coupled scalar field even to the second order in perturbations.  The
uniform-field gauge takes $\delta \phi \equiv 0$ as a temporal gauge
(slicing) condition to the second-order in perturbation. The
uniform-field gauge gives $v^{(\phi)} = 0$ which is the comoving
gauge, and vice versa. Thus, \bea
   & & \delta \phi = 0 \quad \leftrightarrow \quad v^{(\phi)} = 0.
\eea We also have \bea
   & & \delta \mu^{(\phi)} - \delta p^{(\phi)}
       = 2 V_{,\phi} \delta \phi + V_{,\phi\phi} \delta \phi^2,
\eea and under the uniform-field gauge we have \bea
   & & \delta \mu^{(\phi)}_{\delta \phi}
       = \delta p^{(\phi)}_{\delta \phi}
       = - \dot \phi^2 \left[ A \left( 1 - 2 A \right)
       + {1 \over 2} B^\alpha B_\alpha \right]_{\delta \phi}.
    \label{delta-p-MSF-CG}
\eea

Equation (\ref{fluids-MSF}) apparently shows that the vector-type
perturbation $v^{(\phi,v)}_\alpha$ does not vanish to the
second-order. However, the second-order quantities in
right-hand-side depend on the temporal gauge condition for the
scalar-type perturbations, and trivially vanish for the
uniform-field gauge. We can also show that it vanishes for the
uniform-density gauge or the uniform-pressure gauge where we have
$\delta \dot \phi - \dot \phi A \propto \delta \phi$ to the linear
order. In fact, we can show that the vector-type perturbation is not
sourced by the scalar field to the second order by evaluating the
rotational tensor $\tilde \omega_{\alpha\beta}$ in Eq.\
(\ref{kinematic-quantities-pert}) for the scalar field: i.e., for a
minimally coupled scalar field we have \bea
   & & \tilde \omega_{\alpha\beta}^{(\phi)} = 0,
\eea to the second order. Thus, we conclude that a minimally coupled
scalar field does {\it not} contribute to the rotational
perturbation to the second order in perturbations. In fact, we can
show that a single scalar field do not support vector-type
perturbations to all orders in perturbations. As we take the energy
frame, thus $\tilde q_a \equiv 0$, from Eq.\ (23) of \cite{NL} we
can show \bea
   & & \tilde u_a = {\tilde \phi_{,a}
       \over \sqrt{ - \tilde \phi^{,c} \tilde \phi_{,c}} }.
\eea Using the definition of the vorticity tensor in Eq.\
(\ref{kinematic_quantities-i}) we can show that $\tilde \omega_{ab}
= 0$.

Using the fluid quantities in Eq.\ (\ref{fluids-MSF}) we can handle
the scalar field using our non-ideal fluid formulation. The fluid
equations in the energy frame, like Eqs.\
(\ref{def-of-kappa})-(\ref{Mom-conservation}), remain valid in the
case of scalar field with the fluid quantities expressed as in Eq.\
(\ref{fluids-MSF}). The anisotropic stress vanishes and the entropic
perturbation $e$ is given as \bea
   & & e^{(\phi)} \equiv \delta p^{(\phi)}
       - c_{(\phi)}^2 \delta \mu^{(\phi)}, \quad
       c_{(\phi)}^2 \equiv {\dot p^{(\phi)} \over \dot \mu^{(\phi)}}
       = {\ddot \phi - V_{,\phi} \over \ddot \phi + V_{,\phi}}.
\eea Under the comoving gauge $v \equiv 0$, using Eq.\
(\ref{delta-p-MSF-CG}) we have \bea
   & & \delta p^{(\phi)}_v = \delta \mu^{(\phi)}_v, \quad
       e^{(\phi)}_v = \left( 1
       - c_{(\phi)}^2 \right) \delta \mu^{(\phi)}_v,
   \label{EOS-MSF}
\eea which is a well known relation, now valid to the second order
in perturbations. A fluid formulation of the scalar field to the
linear order is presented in \cite{H-PRW-1991,MSF}. Using Eq.\
(\ref{EOS-MSF}) together with vanishing anisotropic stress, Eqs.\
(\ref{def-of-kappa})-(\ref{Mom-conservation}) or
(\ref{eq1})-(\ref{eq7}) provide the fluid formulation for a
minimally coupled scalar field to the second order in perturbations.
The perturbed equation of motion of the scalar field is presented in
Eq.\ (112) of \cite{NL}.

%%%%%%%%%%%%%%%%%%%%%%%%%%%%%%%%%%%%%%%%%%%%%%%%%%%%%%%%%%%%%%%
\subsection{Minimally coupled scalar fields}

In the case of multiple minimally coupled scalar fields, the
equation of motions in Eq.\ (119) and the full Einstein's equations
in Eqs.\ (99)-(105) expressed using the normal-frame fluid
quantities together with the normal-frame fluid quantities for the
scalar field in Eqs.\ (121) all in \cite{NL} provide a complete set
of equations we need to the second-order. The fluid quantities in
Eq.\ (121) of \cite{NL} are presented in the normal-frame
four-vector and it is convenient to know the conventionally used
fluid quantities which are based on the energy-frame four-vector.
These latter quantities can be read from Eqs.\ (121) of \cite{NL}
and Eq.\ (\ref{normal-to-energy-frame}) as \bea
   & & v^{(\phi)}
       = {1 \over a \sum_n \dot \phi^2_{(n)}}
       \sum_k \left\{
       \dot \phi_{(k)} \delta \phi_{(k)}
       + \Delta^{-1} \nabla^\alpha \left[
       \left( \delta \dot \phi_{(k)} - \dot \phi_{(k)} A \right)
       \left( \delta \phi_{(k)}
       - 2 \dot \phi_{(k)} { \sum_l \dot \phi_{(l)} \delta \phi_{(l)}
       \over \sum_m \dot \phi^2_{(m)} } \right)_{,\alpha} \right] \right\},
   \nonumber \\
   & & v^{(\phi,v)}_\alpha
       = - {1 \over a \sum_n \dot \phi^2_{(n)}}
       \sum_k \Bigg\{
       \left( \delta \dot \phi_{(k)} - \dot \phi_{(k)} A \right)
       \left( \delta \phi_{(k)}
       - 2 \dot \phi_{(k)} { \sum_l \dot \phi_{(l)} \delta \phi_{(l)}
       \over \sum_m \dot \phi^2_{(m)} } \right)_{,\alpha}
   \nonumber \\
   & & \qquad
       - \nabla_\alpha \Delta^{-1} \nabla^\beta \left[
       \left( \delta \dot \phi_{(k)} - \dot \phi_{(k)} A \right)
       \left( \delta \phi_{(k)}
       - 2 \dot \phi_{(k)} { \sum_l \dot \phi_{(l)} \delta \phi_{(l)}
       \over \sum_m \dot \phi^2_{(m)} } \right)_{,\beta} \right] \Bigg\},
   \nonumber \\
   & & \delta \mu^{(\phi)}
       = \delta \mu^{N(\phi)}
       - {1 \over a^2 \sum_m \dot \phi^2_{(m)}}
       \left( \sum_k \dot \phi_{(k)} \delta
       \phi_{(k)}^{\;\;\;\;,\alpha} \right)
       \left( \sum_l \dot \phi_{(l)} \delta
       \phi_{(l),\alpha} \right),
   \nonumber \\
   & & \delta p^{(\phi)}
       = \delta p^{N(\phi)}
       - {1 \over 3} {1 \over a^2 \sum_m \dot \phi^2_{(m)}}
       \left( \sum_k \dot \phi_{(k)} \delta
       \phi_{(k)}^{\;\;\;\;,\alpha} \right)
       \left( \sum_l \dot \phi_{(l)} \delta
       \phi_{(l),\alpha} \right),
   \nonumber \\
   & & \Pi^{(\phi)}_{\alpha\beta}
       = \Pi^{N(\phi)}_{\alpha\beta}
       - {1 \over a^2 \sum_m \dot \phi^2_{(m)}}
       \left[ \left( \sum_k \dot \phi_{(k)} \delta
       \phi_{(k),\alpha} \right)
       \left( \sum_l \dot \phi_{(l)} \delta
       \phi_{(l),\beta} \right)
       - {1 \over 3} g^{(3)}_{\alpha\beta}
       \left( \sum_k \dot \phi_{(k)} \delta
       \phi_{(k)}^{\;\;\;\;,\gamma} \right)
       \left( \sum_l \dot \phi_{(l)} \delta
       \phi_{(l),\gamma} \right) \right],
   \nonumber \\
\eea where the normal-frame fluid quantities are presented in Eq.\
(121) of \cite{NL}. Thus, by moving into the energy-frame fluid
quantities from the normal-frame ones we have rather complicated
terms which do not cancel out nicely any term in the normal-frame
quantities. In single field case we had such cancelations, for
example we have $\Pi^{(\phi)}_{\alpha\beta} = 0$ in Eq.\
(\ref{fluids-MSF}). But, we do not have such a luxury in the
multi-component situation. Apparently $\Pi^{(\phi)}_{\alpha\beta}$
in the energy-frame does not vanish and looks more complicated.
Thus, in the multi-field situation we had better use both the fluid
quantities and Einstein's equations all expressed in the normal
frame: these are Eqs.\ (99)-(105),(121) in \cite{NL}.

As the temporal gauge condition we can set any one field
perturbation, say the specific $\ell$-th one $\delta \phi_{(\ell)}$,
equal to zero which might be called the uniform-$\phi_{(\ell)}$
gauge to the second order. This apparently differs from the comoving
gauge which sets $v^{(\phi)} \equiv 0$. In the multi-component
situation we cannot take a gauge condition which makes
$v^{(\phi,v)}_\alpha = 0$. Thus, the multiple scalar fields source
the vector-type perturbation to the second order in perturbations.
The scalar-, vector-, and tensor-type decomposition of the
anisotropic stress $\Pi^{(\phi)}_{\alpha\beta}$ can be read by using
decomposition formulae in Eq.\ (177) of \cite{NL}.

In multiple-field situation, it is ad hoc and cumbersome (if not
impossible) to introduce individual fluid quantity for each field
variable even to the background and linear order perturbations, see
\cite{HN-multi-CQG}.

%%%%%%%%%%%%%%%%%%%%%%%%%%%%%%%%%%%%%%%%%%%%%%%%%%%%%%%%%%%%%%%
\subsection{Generalized gravity case}

In Sec.\ IV.D of \cite{NL}we presented the equation of motion and
effective fluid quantities in a class of generalized gravity
theories together with additional presence of fluids and fields to
the second order. The equation of motion is in Eq.\ (128) of
\cite{NL}, and the full Einstein's equations in Eqs.\ (99)-(105)
expressed using the normal-frame fluid quantities together with the
normal-frame effective fluid quantities in Eq.\ (130) of \cite{NL}
provide a complete set of equations we need to the second-order. The
effective fluid quantities in Eq.\ (130) of \cite{NL} are presented
in the normal-frame four-vector and using Eq.\ (88) in \cite{NL} we
can easily derive the effective fluid quantities based on the
energy-frame four-vector. As in the multiple field case in a
previous subsection, by moving into the energy-frame, the effective
fluid quantities become more complicated compared with the ones in
the normal-frame. Thus, in these class of generalized gravity
theories we had better use both the fluid quantities and Einstein's
equations all expressed in the normal frame: these are Eqs.\
(99)-(105),(130) in \cite{NL}.

%%%%%%%%%%%%%%%%%%%%%%%%%%%%%%%%%%%%%%%%%%%%%%%%%%%%%%%%%%%%%%%
%
% Conservations
%
%%%%%%%%%%%%%%%%%%%%%%%%%%%%%%%%%%%%%%%%%%%%%%%%%%%%%%%%%%%%%%%
\section{Curvature perturbations and large-scale conservations}
                                               \label{sec:conservations}

In the large-scale limit the spatial curvature perturbation
$\varphi$ in several different gauge conditions is known to remain
constant in expanding phase. Often the conservation properties are
shown based on the first time derivative of the curvature
perturbation. In order to show the conservation properties properly
we have to construct the closed form second-order differential
equations for the curvature perturbation. In the following we will
derive such first-order and second-order differential equations for
$\dot \varphi_v$, $\dot \varphi_\chi$, $\dot \varphi_\kappa$, and
$\dot \varphi_\delta$. First we will derive equations for the linear
perturbation including the background curvature and non-ideal fluid
properties. Then we will derive equations for the second order
perturbation assuming a flat background; including the background
curvature is trivial, though. We consider a single-component fluid.

%%%%%%%%%%%%%%%%%%%%%%%%%%%%%%%%%%%%%%%%%%%%%%%%%%%%%%%%%%%%%%%
\subsection{Linear-order equations}

We introduce a combination \bea
   & & \Phi \equiv \varphi_v
       - {K/a^2 \over 4 \pi G \left( \mu + p \right)} \varphi_\chi.
\eea This combination was first introduced by Field and Shepley in
\cite{Field-Shepley-1968}. From Eqs.\
(\ref{eq1}),(\ref{eq3}),(\ref{eq5}), Eqs.\
(\ref{eq1})-(\ref{eq3}),(\ref{eq5}),(\ref{eq7}), Eqs.\
(\ref{eq1}),(\ref{eq6}), and Eqs.\
(\ref{eq1}),(\ref{eq2}),(\ref{eq4}), respectively, we can derive
\bea
   & & {H \over a} \left( {a \over H} \varphi_\chi \right)^\cdot
       = {4 \pi G \left( \mu + p \right) \over H} \Phi
       - 8 \pi G H \Pi,
   \label{dot-varphi_chi-eq} \\
   & & \dot \Phi
       = {H c_s^2 \over 4 \pi G \left( \mu + p \right)}
       {\Delta \over a^2} \varphi_\chi
       - {H \over \mu + p} \left( e
       + {2 \over 3} {\Delta \over a^2} \Pi \right),
   \label{dot-Phi-eq} \\
   & & \dot \varphi_\delta
       = {\Delta \over 3 a} v_\chi
       - {H e \over \mu + p},
   \label{dot-varphi_delta-eq} \\
   & & \dot \varphi_\kappa
       = - {H \over 3 \dot H + \Delta/a^2} \left[
       \left( 1 + 3 c_s^2 \right) {\Delta + 3 K \over a^2}
       \varphi_\kappa
       - 12 \pi G e \right]
       - {\Delta \over 3 a^2} \chi_\kappa.
   \label{dot-varphi_kappa-eq}
\eea These equations were presented in \cite{H-PRW-1991,IF}

We can derive closed form second-order differential equations for
$\Phi$, $\varphi_\chi$, $\varphi_\delta$, and $\varphi_\kappa$ \bea
   & & {\mu + p \over H} \left[ {H^2 \over \left( \mu + p \right) a}
       \left( {a \over H} \varphi_\chi \right)^\cdot
       + {8 \pi G H^2 \over \mu + p} \Pi \right]^\cdot
       = c_s^2 {\Delta \over a^2} \varphi_\chi
       - 4 \pi G \left( e
       + {2 \over 3} {\Delta \over a^2} \Pi \right),
   \label{ddot-varphi_chi-eq} \\
   & & {H^2 c_s^2 \over \left( \mu + p \right) a^3}
       \left[ {\left( \mu + p \right) a^3 \over H^2 c_s^2}
       \dot \Phi
       + {a^3 \over H c_s^2} \left( e
       + {2 \over 3} {\Delta \over a^2} \Pi \right) \right]^\cdot
       = c_s^2 {\Delta \over a^2} \left( \Phi
       - {2 H^2 \over \mu + p} \Pi \right),
   \label{ddot-Phi-eq} \\
   & & {3 \dot H + \Delta/a^2 \over \left( \mu + p \right) a^3}
       \left[ {\left( \mu + p \right) a^3 \over 3 \dot H +
       \Delta/a^2} \left( \dot \varphi_\delta
       + {H \over \mu + p} e \right) \right]^\cdot
   \nonumber \\
   & & \qquad
       = - { 4 \pi G \left( \mu + p \right) + c_s^2 \left( \Delta + 3 K \right)/ a^2 \over
       12 \pi G \left( \mu + p \right) - \left( \Delta + 3 K \right) / a^2}
       {\Delta \over a^2} \varphi_\delta
       + {1 \over \mu + p} {\Delta \over a^2} \left[ e
       + {2 \over 3} \left(3 \dot H + {\Delta \over a^2} \right) \Pi \right],
   \label{ddot-varphi_delta-eq} \\
   & & {1 \over a^3} \left\{ a^3 \left[ \dot \varphi_\kappa
       + {H \over 3 \dot H + \Delta/a^2} \left(
       \left( 1 + 3 c_s^2 \right) {\Delta + 3 K \over a^2}
       \varphi_\kappa
       - 12 \pi G e \right) \right] \right\}^\cdot
   \nonumber \\
   & & \qquad
       = - {\Delta \over 3 a^2} \left[
       \varphi_\kappa - {1 \over 3 \dot H + \Delta/a^2} \left(
       \left( 1 + 3 c_s^2 \right) {\Delta + 3 K \over a^2}
       \varphi_\kappa
       - 12 \pi G e \right) + 8 \pi G \Pi \right].
   \label{ddot-varphi_kappa-eq}
\eea Equations (\ref{ddot-varphi_chi-eq}),(\ref{ddot-Phi-eq}) follow
by combining Eqs.\ (\ref{dot-varphi_chi-eq}),(\ref{dot-Phi-eq}).
Equations (\ref{ddot-varphi_delta-eq}) and
(\ref{ddot-varphi_kappa-eq}) follow from Eqs.\
(\ref{eq1}),(\ref{eq2}),(\ref{eq4}),(\ref{eq5}), and Eqs.\
(\ref{eq1}),(\ref{eq2}),(\ref{eq4}),(\ref{eq5}), respectively.

From Eqs.\ (\ref{eq2}),(\ref{eq3}) we can derive \bea
   & & \varphi_\delta
       = \left[ 1 - {\Delta + 3 K \over 12 \pi G \left( \mu + p
       \right) a^2} \right] \varphi_\kappa
       = \varphi_v - {\Delta + 3 K \over 12 \pi G \left( \mu + p
       \right) a^2} \varphi_\chi
       = \Phi - {\Delta \over 12 \pi G \left( \mu + p
       \right) a^2} \varphi_\chi.
   \label{varphi-relations}
\eea These relations were presented in \cite{Hwang-Noh-Newtonian}.
In the large-scale limit and in near flat background, thus ignoring
$\Delta$ and $K$ terms, we have \bea
   & & \Phi \simeq \varphi_v \simeq \varphi_\delta
       \simeq \varphi_\kappa,
   \label{varphi-relations-2}
\eea to the leading order in the large-scale expansion.

In the large-scale limit, ignoring $\Delta$ terms, in near flat
background and for an ideal fluid case, thus setting $K = 0$ and $e
= 0 = \Pi$, Eqs.\
(\ref{ddot-varphi_chi-eq})-(\ref{ddot-varphi_kappa-eq}) give \bea
   & & \left( {a \over H} \varphi_\chi \right)^\cdot
       \propto {\left( \mu + p \right) a \over H^2},
   \label{dot-varphi_chi-eq-LS} \\
   & & \dot \varphi_v \propto {H^2 c_s^2 \over \left( \mu + p
       \right) a^3},
   \label{dot-varphi_v-eq-LS} \\
   & & \dot \varphi_\delta
       \propto \dot \varphi_\kappa
       \propto {1 \over a^3}.
   \label{dot-varphi_delta-eq-LS}
\eea Notice that if we simply ignore the $\Delta$ terms in Eqs.\
(\ref{dot-Phi-eq})-(\ref{dot-varphi_kappa-eq}) we simply have $\dot
\varphi_v = \dot \varphi_\delta = \dot \varphi_\kappa = 0$. In such
a way we cannot recover the terms in the right-hand-side of Eqs.\
(\ref{dot-varphi_v-eq-LS}),(\ref{dot-varphi_delta-eq-LS}); these
terms lead to decaying solutions (in an expanding era) in the
large-scale limit and are higher order in the large-scale expansion
compared with the decaying solution of $\varphi_\chi$, see below.
From Eqs.\
(\ref{dot-varphi_chi-eq-LS})-(\ref{dot-varphi_delta-eq-LS}) we have
general large-scale asymptotic solutions \bea
   & & \varphi_\chi = 4 \pi G C ({\bf x}) {H \over a}
       \int^t {a \left( \mu + p \right) \over H^2} dt
       + d ({\bf x}) {H \over a},
   \label{varphi_chi-sol} \\
   & & \varphi_v = C ({\bf x}) + {\Delta \over 4 \pi G} d ({\bf x})
       \int^t {c_s^2 H^2 \over \left( \mu + p \right) a^3} dt,
   \label{varphi_v-sol} \\
   & & \varphi_\delta
       = \varphi_\kappa
       = C ({\bf x})
       + {\Delta \over 3} d ({\bf x}) \int^t {dt \over a^3},
   \label{varphi_delta-sol}
\eea where $C ({\bf x})$ and $d ({\bf x})$ are integration constants
which correspond to the relatively growing and decaying modes,
respectively, in an expanding phase; in a collapsing phase the roles
are reversed. The coefficients are fixed using the relations in
Eqs.\
(\ref{dot-varphi_chi-eq}),(\ref{dot-Phi-eq}),(\ref{varphi-relations}).
Notice that for the $C$-mode the relation in Eq.\
(\ref{varphi-relations-2}) is satisfied, and simply remain constant.
For the $d$-mode, $\varphi_v$, $\varphi_\delta$, and
$\varphi_\kappa$ are $\Delta/(aH)^2$-order higher compared with the
$d$-mode of $\varphi_\chi$. The $\varphi_v$ is one of the well known
conserved quantity in the large-scale even in the context of
generalized gravity theories \cite{H-PRW-1991,conserved}.

In order to evaluate the solutions to the second order in the next
section, we need complete sets of linear order solutions for
different gauge conditions. For an ideal fluid, and for a minimally
coupled scalar field such complete sets of solutions are presented
in tabular forms in \cite{IF,MSF}. In the following we summarize
such sets of solutions in an ideal fluid case for four different
gauge conditions. From Table 8 of \cite{IF} we have \bea
   & & \varphi_\chi
       = - \alpha_\chi
       = C \left( 1 - {H \over a} \int^t a dt \right), \quad
       \delta_\chi
       = - {2 \over 3} {\kappa_\chi \over H}
       = - 2 C {\dot H \over aH} \int^t a dt, \quad
       v_\chi = - C {1 \over a^2} \int^t a dt;
   \nonumber \\
   & & \varphi_v = C, \quad
       H \chi_v = C {H \over a} \int^t a dt, \quad
       \delta_v = - {1 + w \over c_s^2} \alpha_v
       = - {2 \over 3} {\Delta \over a^2 H^2} C
       \left( 1 - {H \over a} \int^t a dt \right), \quad
       {\kappa_v \over H} = - {\Delta \over a^2 H^2} C
       {H \over a} \int^t a dt;
   \nonumber \\
   & & \varphi_\delta = C, \quad
       H \chi_\delta = C {H \over a} \int^t a dt, \quad
       {\kappa_\delta \over H} = 3 \alpha_\delta
       = - {\Delta \over a^2 H^2} C, \quad
       v_\delta = {1 \over 3} {\Delta \over a^2 H^2} C
       {H \over a \dot H}
       \left( 1 - {H \over a} \int^t a dt \right);
   \nonumber \\
   & & \varphi_\kappa = C, \quad
       H \chi_\kappa = C {H \over a} \int^t a dt, \quad
       \delta_\kappa = - 3 {1 + w \over 1 + 3 c_s^2} \alpha_\kappa
       = - {2 \over 3} {\Delta \over a^2 H^2} C, \quad
       v_\kappa = - {1 \over 3} {\Delta \over a^2 \dot H} C
       {1 \over a^2} \int^t a dt.
   \label{linear-solutions-C}
\eea For corresponding sets of solutions for a minimally coupled
scalar field, see Table 1 of \cite{MSF}. Compared with the notation
used in \cite{IF} we have $\gamma \Psi = - 8 \pi G (\mu + p) a v$.
The lower bounds of integration of solutions in Eq.\
(\ref{linear-solutions-C}) give behaviors of $d$-modes. For
solutions without integration, the $d$-modes are
$\Delta/(aH)^2$-order higher than the non-vanishing $d$-mode, for
example, see the solutions in Table 2 of \cite{IF}. Thus, the
$d$-modes are \bea
   & & \varphi_\chi
       = - \alpha_\chi
       = d {H \over a}, \quad
       \delta_\chi
       = - {2 \over 3} {\kappa_\chi \over H}
       = 2 d {\dot H \over aH}, \quad
       v_\chi = d {1 \over a^2};
   \nonumber \\
   & & \varphi_v = {\Delta \over 4 \pi G} d ({\bf x})
       \int^t {c_s^2 H^2 \over \left( \mu + p \right) a^3} dt, \quad
       H \chi_v = - {H \over a} d, \quad
       \delta_v = - {1 + w \over c_s^2} \alpha_v
       = - {2 \over 3} {\Delta \over a^2 H^2} d {H \over a}, \quad
       {\kappa_v \over H} = {\Delta \over a^3 H} d;
   \nonumber \\
   & & \varphi_\delta = {\Delta \over 3} d ({\bf x}) \int^t {dt \over a^3}, \quad
       H \chi_\delta = - {H \over a} d, \quad
       {\kappa_\delta \over H}
       = - {1 \over 3} {\Delta^2 \over a^2 H^2} d \int^t {dt \over
       a^3}, \quad
       \alpha_\delta
       = {1 \over 9} {\Delta^2 \over a^2 H^2} d \left(
       {H \over a^3 \dot H} - \int^t {dt \over a^3} \right), \;\;
       v_\delta = {1 \over 3} {\Delta \over a^4 \dot H} d;
   \nonumber \\
   & & \varphi_\kappa = {\Delta \over 3} d ({\bf x}) \int^t {dt \over a^3}, \quad
       H \chi_\kappa = - {H \over a} d, \quad
       \delta_\kappa = - 3 {1 + w \over 1 + 3 c_s^2} \alpha_\kappa
       = - {2 \over 9} {\Delta^2 \over a^2 H^2} d \int^t {dt \over a^3}, \quad
       v_\kappa = {1 \over 3} {\Delta \over a^4 \dot H} d.
   \label{linear-solutions-d}
\eea

%%%%%%%%%%%%%%%%%%%%%%%%%%%%%%%%%%%%%%%%%%%%%%%%%%%%%%%%%%%%%%%
\subsection{Second-order equations}

We assume $K = 0$. From Eqs.\ (\ref{eq1}),(\ref{eq3}),(\ref{eq5}),
Eqs.\ (\ref{eq1})-(\ref{eq3}),(\ref{eq7}), Eqs.\
(\ref{eq1}),(\ref{eq6}), and Eqs.\
(\ref{eq1}),(\ref{eq2}),(\ref{eq4}), respectively, we can derive
\bea
   & & {H \over a} \left[ {a \over H} \left( \varphi - H \chi \right) \right]^\cdot
       = {4 \pi G \left( \mu + p \right) \over H} \left( \varphi - a
       H v \right)
       - 8 \pi G H \Pi
       + {1 \over 3} \left( n_0 - n_2 \right) - H n_4,
   \label{dot-varphi_chi-eq-NL} \\
   & & \left( \varphi - a H v \right)^\cdot
       = {H c_s^2 \over 4 \pi G \left( \mu + p \right)}
       \left[ {\Delta \over a^2} \left( \varphi - H \chi \right)
       - {1 \over 4} n_1 + H n_2 \right]
       - {H \over \mu + p} \left( e
       + {2 \over 3} {\Delta \over a^2} \Pi \right)
       + {1 \over 3} \left( n_0 - n_2 \right) - a H n_6,
   \label{dot-varphi_v-eq-NL} \\
   & & \dot \varphi_\delta
       = {\Delta \over 3 a} \left( v - {1 \over a} \chi \right)
       - {H e \over \mu + p}
       + {1 \over 3} \left( n_0 + {n_5 \over \mu + p} \right),
   \label{dot-varphi_delta-eq-NL} \\
   & & \dot \varphi_\kappa
       = - {H \over 3 \dot H + \Delta/a^2} \left[
       \left( 1 + 3 c_s^2 \right) {\Delta \over a^2}
       \varphi_\kappa
       - 12 \pi G e
       - {1 + 3 c_s^2 \over 4} n_1 - n_3 \right]
       - {\Delta \over 3 a^2} \chi + {1 \over 3} n_0.
   \label{dot-varphi_kappa-eq-NL}
\eea The perturbed order variables in Eqs.\
(\ref{dot-varphi_delta-eq-NL}),(\ref{dot-varphi_kappa-eq-NL}) are
evaluated in the uniform-density gauge ($\delta \equiv 0$), and the
uniform-expansion gauge ($\kappa \equiv 0$), respectively. Equation
(\ref{dot-varphi_delta-eq-NL}) also follows from Eq.\
(\ref{LV-relation1}) evaluated to the second order.

We can derive closed form second-order differential equations for
$\varphi_v$, $\varphi_\chi$, $\varphi_\delta$, and $\varphi_\kappa$
\bea
   & & {\mu + p \over H} \left\{ {H \over \mu + p} \left[
       {H \over a}
       \left( {a \over H} \left( \varphi - H \chi \right) \right)^\cdot
       + 8 \pi G H \Pi
       - { 1\over 3} \left( n_0 - n_2 \right) + H n_4 \right] \right\}^\cdot
   \nonumber \\
   & & \qquad
       = c_s^2 {\Delta \over a^2} \left( \varphi - H \chi \right)
       - 4 \pi G \left( e
       + {2 \over 3} {\Delta \over a^2} \Pi \right)
       + c_s^2 \left( - {1 \over 4} n_1 + H n_2 \right)
       + {4 \pi G \left( \mu + p \right) \over H}
       \left[ {1 \over 3} \left( n_0 - n_2 \right) - a H n_6 \right],
   \label{ddot-varphi_chi-eq-NL} \\
   & & {H^2 c_s^2 \over 4 \pi G \left( \mu + p \right) a^3}
       \Bigg\{ {4 \pi G \left( \mu + p \right) a^3 \over H^2 c_s^2}
       \left[ \left( \varphi - a H v \right)^\cdot
       + {H \over \mu + p} \left( e
       + {2 \over 3} {\Delta \over a^2} \Pi \right)
       - {1 \over 3} \left( n_0 - n_2 \right) + a H n_6 \right]
   \nonumber \\
   & & \qquad
       + {a^3 \over H} \left( {1 \over 4} n_1 - H n_2 \right)
       \Bigg\}^\cdot
       = c_s^2 {\Delta \over a^2} \left( \varphi - a H v
       - {2 H^2 \over \mu + p} \Pi \right)
       + {H c_s^2 \Delta \over 4 \pi G \left( \mu + p \right) a^2}
       \left[ {1 \over 3} \left( n_0 - n_2 \right) - H n_4 \right],
   \label{ddot-varphi_v-eq-NL} \\
   & & {1 + \Delta/(3 a^2 \dot H) \over a^3}
       \left\{ {a^3 \over 1 + \Delta/(3 a^2 \dot H)} \left( \dot \varphi_\delta
       + {H \over \mu + p} e \right)
       + {a^3 \over 3 \dot H + \Delta/a^2} \left[
       {1 \over 4} \dot n_1 + {1 \over 4} \left( 2 H + {\Delta \over
       3 H a^2} \right) n_1 - H n_3 \right]
       - {1 \over 3} a^3 n_0
       \right\}^\cdot
   \nonumber \\
   & & \qquad
       = - {1- c_s^2 \Delta/(a^2 \dot H) \over
       3 + \Delta/(a^2 \dot H)}
       {\Delta \over a^2} \varphi_\delta
       + {1 \over \mu + p} {\Delta \over 3 a^2}
       \left[ e + {2 \over 3} \left( 3 \dot H + {\Delta \over a^2}
       \right) \Pi \right]
   \nonumber \\
   & & \qquad
       + {\Delta \over 9 \dot H a^2} \left[
       {1 \over 4 a^2} \left( {a^2 \over H} n_1 \right)^\cdot
       - n_3
       - \left( 3 \dot H + {\Delta \over a^2} \right) n_4 \right],
   \label{ddot-varphi_delta-eq-NL} \\
   & & {1 \over a^3} \left\{ a^3 \left[ \dot \varphi_\kappa
       + {H \over 3 \dot H + \Delta/a^2} \left(
       \left( 1 + 3 c_s^2 \right) \left( {\Delta \over a^2}
       \varphi_\kappa - {1 \over 4} n_1 \right)
       - 12 \pi G e - n_3 \right) - {1 \over 3} n_0 \right] \right\}^\cdot
   \nonumber \\
   & & \qquad
       = {\Delta \over 3 a^2} \left\{
       - \varphi_\kappa + {1 \over 3 \dot H + \Delta/a^2} \left[
       \left( 1 + 3 c_s^2 \right) \left( {\Delta \over a^2}
       \varphi_\kappa - {1 \over 4} n_1 \right)
       - 12 \pi G e - n_3 \right] - 8 \pi G \Pi - n_4 \right\}.
   \label{ddot-varphi_kappa-eq-NL}
\eea Equations
(\ref{ddot-varphi_chi-eq-NL}),(\ref{ddot-varphi_v-eq-NL}) follow by
combining Eqs.\
(\ref{dot-varphi_chi-eq-NL}),(\ref{dot-varphi_v-eq-NL}). Equation
(\ref{ddot-varphi_delta-eq-NL}) follows from Eqs.\
(\ref{eq1}),(\ref{eq2}),(\ref{eq4}),(\ref{eq5}). Equation
(\ref{ddot-varphi_kappa-eq-NL}) follows from Eqs.\
(\ref{eq1}),(\ref{eq2}),(\ref{eq4}),(\ref{eq5}). The perturbed order
variables in Eqs.\
(\ref{ddot-varphi_delta-eq-NL}),(\ref{ddot-varphi_kappa-eq-NL}) are
evaluated in the uniform-density gauge ($\delta \equiv 0$), and the
uniform-expansion gauge ($\kappa \equiv 0$), respectively.

%%%%%%%%%%%%%%%%%%%%%%%%%%%%%%%%%%%%%%%%%%%%%%%%%%%%%%%%%%%%%%%
\subsection{Large-scale solutions}

Now, we assume an ideal fluid, thus set $e = 0 = \Pi$. In the
large-scale limit, thus ignoring the $\Delta/(aH)^2$-order higher
terms, Eqs.\
(\ref{ddot-varphi_v-eq-NL})-(\ref{ddot-varphi_kappa-eq-NL}) give
\bea
   & &
       \dot \varphi_v
       - {1 \over 3} \left( n_0 - n_2 \right) + a H n_6
       + {H c_s^2 \over 4 \pi G \left( \mu + p \right)} \left( - {1 \over 4} n_1 + H n_2 \right)
       \propto {H^2 c_s^2 \over 4 \pi G \left( \mu + p \right) a^3},
   \label{dot-varphi_v-eq2} \\
   & & \dot \varphi_\delta
       + {1 \over 3 \dot H} \left(
       {1 \over 4} \dot n_1 + {1 \over 2} H n_1 - H n_3 \right)
       - {1 \over 3} n_0
       \propto {1 \over a^3},
   \\
   & & \dot \varphi_\kappa
       - {H \over 3 \dot H} \left(
       {1 + 3 c_s^2 \over 4} n_1 + n_3 \right) - {1 \over 3} n_0
       \propto {1 \over a^3},
\eea where the perturbed order variables in Eq.\
(\ref{dot-varphi_v-eq2}) are evaluated in the comoving gauge ($v
\equiv 0$). We already used the behavior of linear order solutions
in Eqs.\ (\ref{linear-solutions-C}),(\ref{linear-solutions-d}) in
order to show that the right-hand-side of Eqs.\
(\ref{ddot-varphi_v-eq-NL})-(\ref{ddot-varphi_kappa-eq-NL}) vanish.
Using the solutions in Eqs.\
(\ref{linear-solutions-C}),(\ref{linear-solutions-d}) we can show
that \bea
   & & \left( \varphi_v - \varphi_v^2 \right)^\cdot
       + {\cal O} (\Delta C^2, \Delta^2 d^2)
       \propto {H^2 c_s^2 \over \left( \mu + p \right) a^3},
   \label{dot-varphi_v-eq-LS-NL} \\
   & & \left( \varphi_\delta - \varphi_\delta^2 \right)^\cdot
       + {\cal O} (\Delta C^2, \Delta^2 d^2)
       \propto \left( \varphi_\kappa - \varphi_\kappa^2
       \right)^\cdot
       + {\cal O} (\Delta C^2, \Delta^2 d^2)
       \propto {1 \over a^3}.
   \label{dot-varphi_delta-eq-LS-NL}
\eea Thus, we have general large-scale asymptotic solutions \bea
   & & \varphi_v - \varphi_v^2
       = C ({\bf x}) + {\Delta \over 4 \pi G} d ({\bf x})
       \int^t {c_s^2 H^2 \over \left( \mu + p \right) a^3} dt,
   \label{varphi_v-sol-NL} \\
   & & \varphi_\delta - \varphi_\delta^2
       = \varphi_\kappa - \varphi_\kappa^2
       = C ({\bf x})
       + {\Delta \over 3} d ({\bf x}) \int^t {dt \over a^3},
   \label{varphi_delta-sol-NL}
\eea where $C ({\bf x})$ and $d ({\bf x})$ are integration constants
now including the second-order contributions, i.e., $C = C^{(1)} +
C^{(2)}$, etc. Ignoring the transient solutions in an expanding
phase we have \bea
   & & \varphi_v = \varphi_\delta = \varphi_\kappa = C ({\bf x}),
\eea even to the second order in perturbations in the large-scale
limit.

%%%%%%%%%%%%%%%%%%%%%%%%%%%%%%%%%%%%%%%%%%%%%%%%%%%%%%%%%%%%%%%
%
% Discussion
%
%%%%%%%%%%%%%%%%%%%%%%%%%%%%%%%%%%%%%%%%%%%%%%%%%%%%%%%%%%%%%%%
\section{Discussion}
                                      \label{sec:Discussion}

In this work we presented pure general relativistic effects of
second-order perturbations in Friedmann cosmological world model. In
our previous work we have shown that to the second-order
perturbations, the density and velocity perturbation equations of
general relativistic zero-pressure, irrotational, single-component
fluid in a flat background {\it coincide exactly} with the ones
known in Newton's theory, \cite{second-order}. We also have shown
the effect of gravitational waves to the second-order, and pure
general relativistic correction terms appearing in the third-order
perturbations, \cite{second-order,third-order-PRD}. Here, we
presented results of second-order perturbations relaxing all the
assumptions made in our previous work in \cite{second-order}. We
derived the general relativistic correction terms arising due to (i)
pressure, (ii) multi-component, (iii) background curvature, and (iv)
rotation. We also presented a general proof of large-scale conserved
behaviors of curvature perturbation variable in several gauge
conditions, now to the second order.

Effects of pressure can be found in Eqs.\
(\ref{E-conserv-eq4})-(\ref{ddot-delta-eq4}). As we emphasized, the
effect of pressure is generically relativistic even in the
background world model and the linear order perturbations. Still,
our equations show the pure general relativistic effects of pressure
(including stresses) appearing in the second-order perturbations.
Effects of multi-component fluids can be found in Eqs.\
(\ref{ddot-delta-eq-N})-(\ref{ddot-delta-eq-i}). Although these
equations apparently show deviations from Newtonian situation, in
Sec.\ \ref{sec:multi-Newtonian} we showed that if we ignore purely
decaying terms in an expanding phase the equations are effectively
{\it the same} as in the Newtonian situation. Effects of background
spatial curvature $K$ can be read from Eqs.\
(\ref{density-pert-K-N}),(\ref{velocity-pert-K-N}) or Eqs.\
(\ref{E-conservation-zK})-(\ref{Mom-constraint-K}). Effects of
vector-type perturbation can be read from Eqs.\
(\ref{rot-second-order-eq1})-(\ref{rot-second-order-eq3}). In the
small-scale limit we showed that, if we ignore the tensor-type
perturbation, the equations {\it coincide} with the Newtonian ones.

Our results may have important practical implications in cosmology
and the large-scale structure formation. Our new result showing
relativistic/Newtonian correspondence in the zero-pressure
irrotational multi-component fluids is practically relevant in
currently favored cosmology where baryon and dark matter are two
important ingredients of the current matter content in addition to
the cosmological constant. All equations in our work are valid in
the presence of the cosmological constant. A related important
result is the relativistic/Newtonian correspondence valid in the
presence of rotational perturbation far inside horizon. Thus, inside
the horizon scale, even in the presence of rotational perturbations
we can still rely on the Newtonian equations to handle quasi-linear
evolution of large-scale structures. As the spatial curvature in the
present cosmological era is known to be small \cite{Boomerang}, the
possible presence of small spatial curvature may not be important in
the second-order perturbations. Still, while the Newtonian equation
is exactly valid to the linear order even in the presence of the
spatial curvature, we have nontrivial general relativistic
correction terms present to the second order in perturbations. Our
second-order perturbation equations in the presence of pressure may
have an interesting role as we approach early stage of universe
where the effect of radiation becomes important. The importance of
pressure to the second-order perturbations, of course, depends on
whether nonlinear effects are significant in the early evolution
stage of the large-scale structure during the radiation era and in
the early matter dominated era. %\textbf{[R+D???]}
Realistic estimations of the diverse pure general relativistic
contributions using the complete set of equations presented in this
work are left for future investigations.

In an accompanying paper we will investigate the effects of
third-order perturbations of zero-pressure irrotational
multi-component fluids in a flat background. This is one obvious
remaining issue in our series of investigation of nonlinear
cosmological perturbations where nontrivial general relativistic
effects are expected. In the case of a single fluid  we presented
the pure general relativistic effects appearing in the third order
in \cite{third-order-PRD}. Corresponding results in the case of
multi-component will be presented in \cite{third-order-multi}.

%%%%%%%%%%%%%%%%%%%%%%%%%%%%%%%%%%%%%%%%%%%%%%%%%%%%%%%%%%%%%%%
%
% Acknowledgments
%
%%%%%%%%%%%%%%%%%%%%%%%%%%%%%%%%%%%%%%%%%%%%%%%%%%%%%%%%%%%%%%%
\subsection*{Acknowledgments}

H.N. was supported by grant No.\ C00022 from the Korea Research
Foundation.

%%%%%%%%%%%%%%%%%%%%%%%%%%%%%%%%%%%%%%%%%%%%%%%%%%%%%%%%%%%%%%%
%
% References
%
%%%%%%%%%%%%%%%%%%%%%%%%%%%%%%%%%%%%%%%%%%%%%%%%%%%%%%%%%%%%%%%

%%%%%%%%%%%%%%%%%%%%%%%%%%%%%%%%%%%%%%%%%%%%%%%%%%%%%%%%%%%%%%%
\end{document}